\documentclass[10pt]{article}
\usepackage[disable]{todonotes}
\usepackage{enumerate}

\usepackage{geometry}
\usepackage{amsmath}
\usepackage{caption, booktabs}
\usepackage{caption}
\usepackage{subcaption}
\usepackage{hyperref}

\usepackage{tabularx}
\usepackage{array}

\title{Resampling Methods that Generate Time Series Data to Enable Sensitivity and Model Analysis in Energy Modeling}

\author{Kelly Wang \\
 University of Pennsylvania \\ 
 {\underline{kwang10@wharton.upenn.edu} }\\ \and
Steven O. Kimbrough\\
 University of Pennsylvania \\
 {\underline{sok@upenn.edu}} \\
 }

\begin{document}
\maketitle
\begin{abstract}
Energy systems modeling frequently relies on time series data, whether observed or forecast. This is particularly the case, for example, in capacity planning models that use  hourly production and load data forecast to occur over the coming several decades. This paper addresses the attendant problem of performing sensitivity, robustness, and other post-solution analyses using time series data. We explore two efficient and relatively simple, non-parametric, bootstrapping methods for generating arbitrary numbers of time series from a single observed or forecast series. The paper presents and assesses each method. We find that the generated series are both visually and by statistical summary measures close to the original observational data. In consequence these series are credibly taken as stochastic instances from a common distribution, that of the original series of observations. With climate change in mind, the paper further proposes and explores two general techniques for systematically altering (increasing or decreasing) time series. Both for the perturbed and unperturbed synthetic series data, we find  that the generated series induce variability in properties of the series that are important for energy modeling, in particular periods of under- and over-production, and periods of increased ramping rates. In consequence, series produced in this way are apt for use in robustness, sensitivity, and in general post-solution analysis of energy planning models. These validity factors auger well for applications beyond energy modeling.

\todo[inline]{To be done:
\vskip 9 pt
Convert to have the image files in the same folder but absent from the document, for submission purposes.
\vskip 9 pt
Read again and add to the paper:
{\bf Turowski, M., Heidrich, B., Weingärtner, L., Springer, L., Phipps, K., Schäfer, B., Mikut, R., \& Hagenmeyer, V. (2024). Generating synthetic energy time series: A review. Renewable and Sustainable Energy Reviews, 206, 114842. https://doi.org/10.1016/j.rser.2024.114842} 

Add to paper 

Kelly's finds: 
}
\end{abstract}

\vskip 12 pt
\centerline{{\bf Highlights}}

\begin{itemize}
    \item Energy system models typically incorporate hourly time series data on an annual basis, e.g., for representing load and generation.
    \item The time series data may be observed or forecast. In either case, the imperatives of model analysis and model-based decision making require that alternatives to assumed parameter values be explored, a task largely unsupported with current model analysis methods.
    \item This study identifies and explores two non-parametric bootstrap re-sampling methods that can be effective in producing synthetic time series data for model analysis.
    \item The study proposes and explores methods for generating systematically displaced time series compared to a basis series, as well as for broadly replicating an original series.
    \item The study uses recent data from PJM to illustrate and explore the properties of the several methods it investigates.
\end{itemize}

\newpage






\tableofcontents


\newpage
\section{Introduction\label{sec_introduction}}











Time series data are used ubiquitously. They are present in engineering design applications,   in economics, finance, and  sales, in applications in pattern recognition, bench marking, and quality control, among other areas. They are essential in energy modeling, the application that occasioned this study,  where the vagaries of solar photovoltaic (PV) and wind energy production loom large over time.  

Our point of departure is the fact that a single observational series  does not afford sensitivity analysis, robustness analysis, and other forms of post-solution analysis of models (model analysis or robustness analysis for short) \cite{kimbrough_business_2016}. For that, either multiple observational series
or else a generating model is needed in order to  have sufficient variability for probing the performance of a designed object or system. We explore the latter approach in this paper. Our interest has been motivated by modeling related to the energy transition, in which it is standard to require hourly load (alias demand)  and production data for decades into the future. In the presence of inevitably high uncertainty, having alternate series to use for model analysis is imperative. We emphasize that improved support for model analysis, rather than improved forecasting, is the subject of this study.

Energy system modeling presents demands for data that often exceed those of other applications, such as for building and infrastructure design, discussed in \S\ref{sec_related_applications}. There are two principal reasons for this. First, electricity grids are operated and need to be controlled in nearly real time. This leads to temporal requirements on modeling data to be granular at the hour or shorter duration. Second, while maxima and minima of supply and demand certainly matter for electricity systems (as they do for infrastructure generally), \emph{duration} of under supply and over supply of generation and load  also matter greatly, as does rapidity of change in generation. This is notably the case for the problem of integrating variable renewable energy (VRE, principally wind and solar PV systems) into electric power systems (alias  grids). There, the concern arises of prolonged under generation and how it can be accommodated, given the high expense of longer duration storage, conventionally defined as 10 hours or longer. Varieties of forecast plus reserve margin planning have been, and continue to be, used operationally in the electricity industry. Finding richer forms of data to support planning would surely be useful. For planning decades into the future, with the attendant high uncertainties, it is requisite. An extensive literature exists that proposes such long horizon models, which typically look to 2050 as the end planning date. A recent review \cite{hansen_status_2019} identifies more than 180 such ``grid deep decarbonization'' studies. A newer study \cite{yilmaz_power--gas_2022} identifies and discusses another two dozen or so modeling studies published since \cite{hansen_status_2019}. Due to absence of means to obtain alternative series, these studies generally do not report robustness analysis based on the forecast of production and demand data series, which may be  30 years or more into the future.

Our motivating prototypical use case may be described as follows. There is an energy system model that relies on hourly time series data that has been forecast by an agreed method,  e.g.,  hourly production of solar PV or wind energy 20 years hence.
In order to perform a model robustness analysis involving the time series data, we seek multiple alternative credible time series in order to have some indication of the variability of the data and its affects on the performance of the model. In short, instead of seeking a better forecast, we seek to make better use of the forecasts that we have. How do variations on our existing forecasts affect the behavior of the model?

This core use  case---and variations on it---sets the key requirement for the synthetic time series methods we seek. That requirement is for production of significant numbers of credible alternative time series in order to gain insight into the distribution of possibilities for model analysis. The case is quite similar to that of model analysis for scalar parameters. In the  case of a scalar parameter of unknown distribution, it is standard practice to obtain multiple observations of the parameter and then use bootstrapping to estimate distributional properties of the parameter for the sake of testing model robustness across the distribution of realized values. We extend this well-accepted methodology to time series. Given time series data used in a model, we use bootstrapping to estimate distributional properties of that time series for the sake of testing model robustness across the distribution of realized values. Thus, the problem we address is one of generating a  plurality  of time series  that credibly represent draws from a common causal mechanism. Bootstrapping approaches are {\it prima facie} plausible for this purpose. That is what they were conceived for doing. 


The need for multiple production and demand time series in the electricity sector can be seen through by two main analyses. First, a plurality of time series is necessary to analyse generation adequacy. The load in the electricity grid has to be balanced over short intervals, therefore the system has to be prepared for different load and electricity production situations. This task becomes more difficult as the share of VRE increases, because they are not dispatchable and because they create additional difficulties in balancing electricity demand and production. To prepare a robust electricity system, the system should ideally be tested using a large number of time series, reaching at least into the hundreds, of electricity demand and production scenarios. In these studies, analysts create scenarios that probe the energy system, such as very low electricity production from renewable energy sources (RES, including VRE as well as storage and other services) and high electricity demand. If a design is not robust in these situations, there is concomitant risk of system outages.

Second,  time series are necessary to analyse the uncertainties, e.g., to quantify the flexibility requirements of a system. Due to fluctuations in electricity generation from RES, short-term flexibility options such as batteries and pumped storage power plants are used to compensate for  fluctuations over short periods of time, e.g., one day. For this reason, the data generated should consider both aspects and include different extreme cases, while maintaining appropriate resemblance to 
original time series. 

In this study, we propose and explore methods that address these challenges and are non-parametric, specifically, bootstrap methods, which we explain in the subsequent sections. A key motivation for our approach is the aim of generating time series data suitable for sensitivity analysis and other forms of model analysis, rather than for making point predictions. Again: our purpose is model analysis, not prediction. This goal in turn has two aspects. First, the methods should be capable of generating arbitrary numbers of new series that appropriately resemble the original series. The challenge with this requirement is that the generating distributions of solar and wind supply are obscure, and existing time series methods using distributional assumptions have been unable to predict well  for any but very short periods of time (see \S\ref{sec_lit_review} for a discussion). This is in conflict with the needs of energy systems studies, which need hourly or finer data for periods of a year or more.

The second aspect of our overall goal is that the methods should be capable of generating arbitrary numbers of new series that \emph{deviate} in prescribed ways from the original series. Droughts and deluges are a central concern in energy system design and planning. Electrical grids have to be kept in balance, while wind and solar, and to some extent hydro, are not readily dispatchable. It is in consequence necessary to plan for times of unusual shortage, as well as surplus, knowing full well that there are years of solar, wind, and hydro drought and that climate change may well be affecting this behavior. Short of being able to predict changes in production distributions, being able to explore directional changes from today's observations---in model analysis mode for decision support---is a useful capability. It is one we address with our proposed methods.

We have been able to identify two bootstrapping methods for time series generation that are suitable to the first aspect of our goal: to mimic original series in energy systems. In what follows we present and discuss these two methods. We then explore directional  perturbations on time series data using bootstrap methods. Throughout, we use data from PJM \cite{pjm_pjm_2024}
to illustrate our methods. 

In a nutshell, the principal contributions are as follows.
\begin{itemize}
    \item We propose the problem of undertaking model analysis (sensitivity analysis and post-solution analysis in multiple forms) using synthetic time series data. 
    \item The problem presents requirements that are quite distinct from well-established synthetic time series applications. Specifically, traditional synthetic time series methods aim to learn the underlying data to generate series that minimally deviate from the statistical properties of original series to allow for point predictions.  
    \item For the purposes of model analysis, however, we propose to find a way to generate large ensembles of synthetic time series that can plausibly capture random deviations from the observed time series.  For example, given observed solar PV production for a given year (and installed capacity), how might that production vary in coming years?  Instead of asking what the most likely single series is for a future period, we aim to characterize the plausible deviations from the observed series.  This affords robustness and sensitivity analysis, rather than point predictions.
    \item 
    Bootstrapping methods are routinely used with scalar data to generate synthetic distributions of the data. We propose to extend this capability to synthetic time series data, yielding \emph{distributions} of synthetic time series. Alternative time series may then be substituted into a model to probe its behavior, just as is standardly done with alternative scalar values when doing sensitivity and model analysis.
    \item We identify, demonstrate, and assess the behavior of two bootstrapping methods for generating time series data.
    \item We introduce, demonstrate, and investigate two additions to the bootstrapping methods that afford modeling of time series systematically displaced from the original series. This affords, for example, modeling of systematically  higher (lower) loads and production levels in future years. 
    \item We make available our code for replication studies and use by others.
\end{itemize}


The paper is organized as follows. \S1, ``Introduction,'' adumbrates the applied problem addressed by this paper (doing sensitivity and other forms of model analysis on energy models) and provides motivation for the contribution approach to be taken.  \S2, ``Related Art,'' presents necessary background material, on similar applications for which the approach developed here may be useful (\S2.1), and on prior literature that explores generation of time series data in the broadly energy context (\S2.2). \S3, ``Bootstrap Methods for Generating Similar Time Series,'' begins by laying out  the core ideas of bootstrap estimation as they pertain to the present instance (\S3.1). It then  presents and compares two bootstrap methods suitable for generating corpora (large numbers) of new time series that are similar to and distributed about an original time series. We find that the method introduced here (the Symmetric Block Bootstrap, SBB) is superior in some ways to the only previously introduced method we have found (the Nearest Neighbors Lagged Bootstrap, NNLB). \S4, ``Creating Displaced Time Series,'' extends the SBB method in order to create corpora of synthetic time series that are are systematically more extreme (higher or lower) than a given source time series. This capability is needed for many purposes for which time series data are needed but observations are unavailable, such as designing for  extreme solar or wind droughts or for large deviations in demand for electricity. \S4 also includes a case study with PJM data using the SBB method. \S5, ``Discussion and Conclusion,'' concludes the paper with a summary and discussion.

\section{Related Art\label{sec_related_art}}

\subsection{Related Applications\label{sec_related_applications}}


It is standard practice to configure buildings to a specified level of performance using historical weather data. What is typically used are dry bulb and wet bulb ambient temperatures that are higher (for cooling, lower for heating) than a majority (typically 99.6\%) of the historical temperature observations over the period of a year. Put otherwise, threshold temperatures are used that have a probability of exceedance of not more than 0.004. Then, a safety or over-sizing factor of 15--25\%\ is typically used to size heating and cooling equipment. To illustrate, if $X$ tonnes of air cooling per hour are needed for the threshold temperature of the historical data (e.g., hotter than 99.6\%\ of the observed temperatures), then the space cooling system will be designed with a capacity of, say, $1.2\times X$ tonnes per hour. (See the {\it ASHRAE Fundamentals} for one widely-used source of data and standards \cite{ashrae_handbook_2024}.)


Building energy performance is commonly modeled (simulated) during design and afterwards using 
\emph{typical meteorological year} (TMY) data.
``TMYs contain one year of hourly data that best represents median weather conditions over a multiyear period'' \cite{nrel_tmy_2024}. The obvious problem with modeling a building to a TMY, when the building will last for more than 50 years, is that the building's performance is being assessed over a range of conditions (the median) that is less varied than is likely to occur. Also obvious is that a reserve buffer is  a blunt instrument for dealing with stochastic variability and may often result in over-configuration of the building with attendant extra costs. In both cases, it would be desirable  to have sufficient data for an accurate estimate of the variability that is likely to occur over, say, the next 50 years. Such data are normally unavailable, thus the recourse to TMYs.

Infrastructure projects, such as bridges and roadways, are typically built to withstand a 100-year flood (storm, etc.) event, a flood (storm, etc.) level expected to be exceeded only once in 100 years \cite{usgs_100-year_2024}). When the requisite data are available, the 100--year (generally, $n$--year) event criterion supports probability assessments that can be used to balance risks and outcome consequences. Even so, these risks would potentially be better evaluated were replications of $n$--year data series available. The occurrence of the most severe event in the last 100 years is a random variable for which we have at best one observation, making it impossible to estimate its distribution absent further information. Alas, this is rarely if ever possible with observational data. Worse yet, with climate change in evidence, adjustment of historical time series may be needed to explore and analyze modeling results.


\subsection{Prior art on generating time series for energy systems modeling\label{sec_lit_review}}



We note with pleasure an excellent recent review of 169 papers, published over the past 40 years, on synthetic time series data generation methods for energy applications \cite{turowski_generating_2024}. This is a valuable resource. We limit ourselves here to discussion in somewhat greater detail of the most pertinent methods for the model analysis challenges identified (above) to be addressed in this study. Interestingly, \cite{turowski_generating_2024} identifies in the literature seven use cases for synthetic energy time series: customer profiling, event analysis, load segregation, analysis of power consumption, privacy, and network analysis. Model analysis--the subject of this study---is not among them. The authors of \cite{turowski_generating_2024} worry the limited number of use cases addressed ``[C]ould unnecessarily limit the benefits of using synthetic energy time series although they are very valuable for research such as the generation of scenarios for the design of future energy systems'' \cite[page 13]{turowski_generating_2024}. We agree entirely; our framing of the model analysis use case should be seen as a generalization of scenario analysis. In either case, the emphasis is off of prediction and forecasting and on representing circumstances of interest.

We now comment briefly on representative related efforts to generate time series for modeling loads and energy production.

Markov chains \cite{papaefthymiou_mcmc_2008, ettoumi_statistical_2003} are one of the most popular methods of generating wind power data  as input for energy system models. Brokish et al. \cite{brokish_pitfalls_2009} used an energy system model to assess the quality of the time series generated by Markov chains. The results show that synthetic wind energy time series using Markov chains lead to lower storage requirements. The study concludes that Markov models correctly reflect the probability density function of wind data, but are not necessarily suitable for quantifying storage requirements. They can be used to a limited extent, especially for the generation of wind time series with time steps of less than 40 minutes \cite{pesch_new_2015}. Unfortunately, they find that neither the level of fluctuation nor characteristic trends can be correctly presented in the generated time series. 

ARMA or ARIMA models \cite{bertolina_exploring_2024,billinton_time-series_1996,chen_synthetic_2017,chen_arima-based_2009} are another widely used approach in this field to generate electricity production time series from wind. However, it has been reported that these models  do not reliably maintain the probability distribution of the original data \cite{brokish_pitfalls_2009}.



Usaola \cite{usaola_synthesis_2014} uses a random block bootstrap method to generate time series for electricity generation from wind for the Spanish peninsula. The study mainly generates time series to carry out sequential Monte Carlo-based studies on generation adequacy. A new annual series is synthesised by taking random blocks of successive values, each sample block containing a few days. The transition between successive blocks is then smoothed, so that the difference between the last value of one block and the first value of the next block is not  large. Smoothing is necessary because random block methods assume stationarity of the time series sampled. However, we cannot assume that energy time series data are stationary; indeed, it is plain that they are not in our case. As shown in their respective auto-correlation graphs, 
 found in our supplemental materials (\S1.1), variable renewable energy sources and energy demand follow seasonal trends.

Even if seasonality was accounted for, time-dependent extreme events, like a wind drought, are frequent enough that even assuming a distribution for every season does not accurately capture the data. Because it becomes more complicated to transform our data into a stationary time series, we explore bootstrap methods that do not require this assumption. Finally, there are energy modeling studies that use forecasting but make no attempt to generate time series, for example \cite{veall_boostrapping_1987} and \cite{al-sahlawi_forecasting_1990}.

 We would emphasize, again, that our use case calls for generating an ensemble of time series for the sake of undertaking model analysis. The methods in the prior art cited are largely directed at a different use case, that of predicting a system state. The two use cases are not entirely distinct, but they are different. Going forward, it will be important to make advances on both fronts and the multiple extant methods constitute a valuable resource for such developments.





\section{Bootstrap Methods for Generating Similar Time Series\label{sec_bootstrap_similar}}

\subsection{Bootstrap Estimation}
\label{sec_Bootstrap}

Bootstrap estimation is an established and widely used family of statistical methods that use resampling of observational data to support statistical inference \cite{efron_introduction_1993,lahiri_bootstrap_2006}.  
We summarize a base case here for the sake of appreciating the challenges of  time series estimation and for how we formulate our bootstrapping approach.

In a prototypical case, there is an unknown population distribution function, $F(X)$, for which we wish to estimate $\theta$, a population characteristic or statistic of interest, such as the mean, median, or variance. To do this, we draw a sample of size $n$ from the population  $X = x_1, x_2, \ldots, x_n$. In the prototypical case we assume that the $x_i$s are independent and drawn from a common distribution (i.i.d.), $F$. Using the sample, $X$, we can then calculate an estimate of $\theta$, which we designate $\hat\theta$. For example, if $\theta$ is the population mean, then we might estimate $\hat\theta = \bar{X} = \sum_{i=1}^n x_i/n$.

At this point, bootstrapping estimation begins to diverge from traditional parametric statistical inference. Instead of assuming a parametric distribution for $\hat\theta$, we construct the bootstrapped empirical probability distribution (EPD), $\hat{F^*}(\hat \theta^*)$, as follows. We create $B$ new samples by sampling (resampling) $X$ with replacement for each new sample of $n$ items, leading to $X_1, X_2, \ldots, X_b, \ldots, X_B$. For each $X_b$ we calculate the statistic of interest, say $\hat\theta_b = \bar{X}_b = \sum_{i=1}^n x^b_i/n$.
We then  place a probability of $1/B$ on each observation $\hat\theta_b$ to get $\hat{F^*}(\hat \theta^*)$. (We are assuming here i.i.d. $B$ needs to be a reasonably large number, say 1000 or more in the prototypical case.)
This sampling distribution estimate for $\hat\theta$, which is  the sample estimate of the population statistic of interest $\theta$, is denoted $\hat{F}^*(\hat{\theta}^*)$. We may now use $\hat{F}^*(\hat{\theta}^*)$ to undertake statistical inferences pertaining to $\hat\theta$, e.g., to construct confidence intervals for $\hat\theta$. The process here is similar to how we would use the $z$ or $t$ distributions in standard parametric statistical inference.

The challenge in obtaining bootstrap estimators, the $\hat{\theta}_b$ values, of time series data is that the elements of $X$ are not independently identically distributed in real world observational series, such as those encountered in energy systems, specifically wind and solar production, and demand. There are essentially two approaches that can be taken to meet this challenge. The first is to impute a distributional structure to the time series, e.g., that it is ARMA, ARIMA, Markov, stationary or some other known time series structure. This in hand, bootstrapping approaches have been developed for statistical inference \cite{hardle_bootstrap_2003,hongyi_li_bootstrapping_1996,kreiss_bootstrap_2012}. 
 The problem, as we have noted, is that energy series do not prima facie have these structures, and the results of assuming they do have not, as noted above, been satisfactory for long periods such as the 8760 hours in a year.

The second approach is to develop bootstrapping estimators without making parametric assumptions about the originating series. It is this approach that we are exploring in the present study. We have been able to identify two published estimators that meet this criterion, the Nearest Neighbor Lagged Bootstap (NNLB) estimator \cite{lall_nearest_1996} and the Symmetric Block Bootstrap (SBB) 
\cite{kimbrough_symmetric_2021}. We explore them and extensions of them. 

We now define notation and the statistics (distribution properties) $\theta$s for our study. Our  aim  is  to capture the ``extreme'' periods of a time series compared to the given time series.

\begin{enumerate}
    \item $X = \langle x_1, x_2, \ldots, x_n\rangle$
    
    $X$ is a time series of (normally but not necessarily original) observations or forecasts, $x_i$. For the series we are most interested in, for energy modeling, we cannot assume $X$ is i.i.d., nor do we assume that the generating system is anything but indeterminate.\footnote{In the case of a forecast series we can in principle recover the generation method, but the point of the exercise is to found the analysis on actual observations.}
    
    \item $X^b = \langle x^b_1, x^b_2, \ldots, x^b_n\rangle$ is the $b$\textsuperscript{th} bootstrap sample from $X$. How this is done is addressed in the sequel. 
    
    \item We sequence the time series into periods of length $l$. $X_l$ is $X$ ``chunked'' sequentially into periods of length $l$, wrapping at the end to ensure an integral number of periods. For example, if $l=24$, we are interested in the sequential 24-hour of periods of the series so that
    \newline $X_l  = \langle\langle x_1, x_2, \ldots, x_{24}\rangle , \langle x_{25}, \ldots x_{48}\rangle, \ldots \langle x_{8736}, \ldots x_{8760}\rangle\rangle$.
    
    \item $X^b_l$ is $X^b$ ``chunked'' sequentially into periods of length $l$, wrapping at the end to ensure an integral number of periods. 
    
    \item $X_{l,i}$ is the $i$\textsuperscript{th} element of $X_l$.
    
    \item $X^b_{l,i}$ is the $i$\textsuperscript{th} element of $X^b_l$.
    
    \item $\theta^+$ and $\theta^-$ are the  population statistics  to be estimated as $\hat\theta^+$ and $\hat\theta^-$. We seek to find their distributions, that is $\hat{F}^{+*}(\hat{\theta}^{+*})$ and 
    $\hat{F}^{-*}(\hat{\theta}^{-*})$, in order to make statistical inferences. In our case, the statistics of interest are really functions, which we define as follows. 
    
    
    \begin{equation}
    \hat\theta_b^{-*}(l,e) = \sum_{i=1}^l (X_{l,i} - X^b_{l,i})\ \ \mbox{for}\ \ X_{l,i} - X^b_{l,i} \ge e
    \end{equation}
    $\hat\theta^-$ sums the under production chunks that are in deficit of at least $e$.
    \begin{equation}
        \hat\theta_b^{+*}(l,e) = \sum_{i=1}^l ( X^b_{l,i} - X_{l,i})\ \ \mbox{for}\ \ X^b_{l,i} - X_{l,i} \ge e
    \end{equation}
    $\hat\theta^+$ sums the over production chunks that are in surplus of at least $e$. 
    
    It should be noted the comparison variable $e$ can be a scalar quantity, or a vector of some proportion of the original series. For example, if we were looking for $\theta_b^{-*}(l,e)$ for solar energy, we could set $l = 24$ hours and $e = 20$ mW to find the number of days that our generated solar series fall below $20$ mW of the original solar energy generation. We could also, instead, declare $e = \alpha \times X_l$ where $\alpha$ is a parameter that scales the given series to a certain proportion. That means if we set $\alpha = 0.05$, then we are trying to find the number of days in our generated solar series that falls below $95\%$ of the original solar series. In the other direction, if we were looking for $\theta_b^{+*}(l,e)$ with $l = 24$ hours and $e = 0.05 \times X_l$, then we are counting the number of days in our generated solar series that lie above $105\%$ of the original solar series.
    
    \item The empirical bootstrap distribution for underage assigns a probability of $1/B$ to each 
    $\hat\theta_b^{-*}(l,e)$ for $b = 1, 2, \ldots B$, where $B$ is the number of bootstrap samples taken. Similarly, the empirical bootstrap distribution for overage assigns a probability of $1/B$ to each $\hat\theta_b^{+*}(l,e)$ for $b = 1, 2, \ldots B$, where $B$ is the number of bootstrap samples taken.
    
\end{enumerate}
Put simply in words, for each bootstrap sample we collect two totals: the total underage compared to the matching segment in $X_l$ beyond the threshold $e$ and the total overage compared to the matching segment in $X_l$ beyond the threshold $e$. Taking many samples, we form two empirical distributions, one for underage and one for overage.\footnote{The $e$ values associated with $\hat\theta_b^{+*}(l,e)$ and $\hat\theta_b^{-*}(l,e)$ need not be identical. We treat them as such for the sake of simplicity of exposition.}


We now turn to a discussion of how the bootstrap samples may be created and, once created, how they may be used to undertake statistical inference regarding under and over production. Note that in declaring our statistics of interest, our $\theta$s, we have defined two statistics for every permitted combination of $l$ and $e$.  This complicates the analysis while enriching the potential results.

\subsection{Nearest Neighbors Lagged Bootstrap (NNLB) Method \label{sec_NNLB}}

    
    


\subsubsection{Background Information}

Lall and Sharma \cite{lall_nearest_1996} note that autoregressive moving average (ARMA) models have long been used to model hydrologic (streamflow and weather sequences) data. Because the resulting series were judged not to look like the observed sequence (``failed to pass the ocular test'' as the folk saying has it) a number of other methods have been tried for generating such series, including bootstrapping (resampling) approaches and specifically various moving block bootstrap (MBB) methods, all without fully satisfactory upshot, as noted above. With this as background, 
Lall and Sharma \cite{lall_nearest_1996}  proposed a $k$-nearest neighbors lagged bootstrap (NNLB) algorithm for generating hydrologic and similar series.

\subsubsection{Description of the Procedure}
The NNLB has two key parameters: $l$ is the size of the lag and $k$ is the number of nearest neighbors. We will show how to obtain predictions for each point in a given series as follows.
\begin{enumerate}

\item We are first given a time series with $t$ observations $\mathbf{X} = x_1, x_2, \ldots, x_{t-1}, x_{t}$ from which we want to form a lag vector 
$\mathbf{D}$ for each point in $\mathbf{X}$. The lag vector $i$ that we associate with point $x_i \in \mathbf{X}$, which we denote as $\mathbf{D}_i$, has the length of the lag $l$, and is composed of the $l$ observations \emph{preceding} the target point $x_i$ that we wish to predict. Therefore, we set $\mathbf{D}_i = \langle x_{i-l}, x_{i-(l-1)}, \ldots, x_{i-2}, x_{i-1}\rangle$. We loop the time series observations so that the successor of the last point is the first point in the series (and the predecessor of the first point is the last point). Therefore, if we were to find the lag vector for point $x_2$ in a series $\mathbf{X}$ with 8760 observations and a lag of $l=4$, our lag vector would be $\mathbf{D}_2 = \langle x_{8758}, x_{8759}, x_{8760}, x_{1} \rangle$. 

\item Next, for each observation $x_i$, we find the $k$ nearest lag vectors to $\mathbf{D}_i$ and compile them together into an array $P_i = \langle \mathbf{D}_{t_1}, \mathbf{D}_{t_2}, \ldots, \mathbf{D}_{t_k} \rangle$ of size $k \times l$. The subscripts $t$ indicate the original indices in the series $\mathbf{X}$ for each respective lag vector. The $k$ nearest lag vectors are found by calculating the Euclidean distances between all lag vectors and $\mathbf{D}_i$ and taking the $k$ vectors with the shortest distances. Our code allows for us to choose whether we include $\mathbf{D}_i$ in this calculation or not. The default is to include $\mathbf{D}_i$, meaning in $P_i$, $\mathbf{D}_i$ has the smallest Euclidean distance to $\mathbf{D}_i$ since the distance is equal to 0. This is important to note because we plan on sorting $P_i$ from shortest to longest distance from $\mathbf{D}_i$ to get the ordered set $P_{j(i)} = \langle \mathbf{D}_{1(t)}, \mathbf{D}_{2(t)}, \ldots, \mathbf{D}_{k(t)} \rangle$. This notation means $\mathbf{D}_{1(t)}$ is the closest lag vector and $\mathbf{D}_{k(t)}$ is the $k^{\text{th}}$ closest lag vector to $\mathbf{D}_{i}$. Note that when we include $\mathbf{D}_i$ in our calculations, $\mathbf{D}_{1(t)}$ will always equal $\mathbf{D}_i$.

\item Because the lag vector $\mathbf{D}_i$ does not contain the observation $x_i$, we need another vector, which we will call the index vector $\mathbf{N}_i$ with length $k$, to keep track which $k$ observations in $\mathbf{X}$ are the nearest to $x_i$ based on their respective lag vectors. We named $\mathbf{N}$ the index vector because the observations $x \in \mathbf{N}$ depend on the initial indices of $\mathbf{D} \in P_i$ before we sorted $P_i$. This means $\mathbf{N}_i$ is tied with $P_i$ and we can define $\mathbf{N}_i = \langle x_{t_1}, x_{t_2}, \ldots, x_{t_k} \rangle$ where $x_{t} \in \mathbf{N}_i$ is the observation succeeding the points in the lag vector $\mathbf{D}_{t} \in P_i$. This step is done in tandem with the previous step because we will compile $P_i$ and $\mathbf{N}_i$ at the same time, and then sort them in the same way to get $\mathbf{N}_{j(i)} = \langle x_{1(t)}, x_{2(t)}, \ldots, x_{k(t)} \rangle$. so that each observation in $\mathbf{N}_{j(i)}$ stays with their respective lag vector ordered in $P_{j(i)}$. Therefore, the point $x_{j(t)} \in \mathbf{N}_{j(i)}$ is the $j^{\text{th}}$ closest to point $x_i$ in terms of its lag vector. 


\item Under a suitably defined randomization scheme, pick a point $x_{j(t)} \in \mathbf{N}_{j(i)}$ and use it as the predicted value, $\hat{x}_i$.

\begin{enumerate}
    \item In our analysis, we follow Lall and Sharma's \cite{lall_nearest_1996} definition of a resampling kernel, also known as a probability mass function, for choosing the new point in the synthetic series. \\
    Once we have $\mathbf{N}_{j(i)}$, we create the probability mass function as follows.
    \[ K(j(i)) = \frac{1/j}{\sum_{i=1}^{k} 1/j}\]
    $K(j(i))$ is the probability point $x_{j(i)}$ is picked from $\mathbf{N}_{j(i)}$ to be a part of the new series. That means this probability mass function is the same for any observation at time $i$, so it can be calculated and stored before actively resampling for each new point in the series. 
    
    \item It should be noted that any reasonable resampling kernel for $\hat{x}_i$ from $\mathbf{N}_{j(i)}$ can be used in this step. It depends on the user's intentions.

\end{enumerate}

\item Repeat this process for every point $x_i$ in the given series $\mathbf{X}$ to generate a new series.

\end{enumerate}

\subsubsection{Examining the Generated Series}

We have run the $k$-nearest neighbors lagged bootstrapping (NNLB) method on PJM's 2021 generation data 1000 times for solar energy, wind energy, and energy load using a lag size of $5$ and $20$ of each point's nearest neighbors. We also repeated the 1000 trials with a lag size of $9$ and nearest neighbors of $100$ using the same original dataset. These parameters do not have any privileged meaning; they were arrived at through a parameter sweep in which we took care that the generated series not deviate too much from the original series, while still having room to show variability. They were also chosen to match the same distance vector sizes that are used in \S\ref{sec_SBB}. 

Each type of dataset will showcase a 48 hour snapshot during a  time period of the year, chosen for illustration and  comprising  the 20 most extreme series (generated series with the 10 highest and 10 lowest means), one with a mean in the middle of the 1000 generated series, and the original series graphed for reference. Based on a snapshot of each dataset, we can see that the new generated time series capture  the main features of the original time series. 

However, a slight bias can be found in the generated series depending on the characteristics of the original series used. This could be due to the second step in the NNLB procedure where we are finding the lag vector for every point in the given time series. The notable aspect of this vector is that it does not contain the original point itself in the vector. This is because in \cite{lall_nearest_1996}, Lall and Sharma's intention was to use the NNLB method for prediction, whereas we are attempting to emulate the original series. Therefore, two lag vectors may have close Euclidean distances, but the respective points that those vectors represent may not be as similar. We note that it is well known that bootstrap estimators may be biased.

\begin{figure}[h!]
        \centering
       \includegraphics[width=\textwidth]{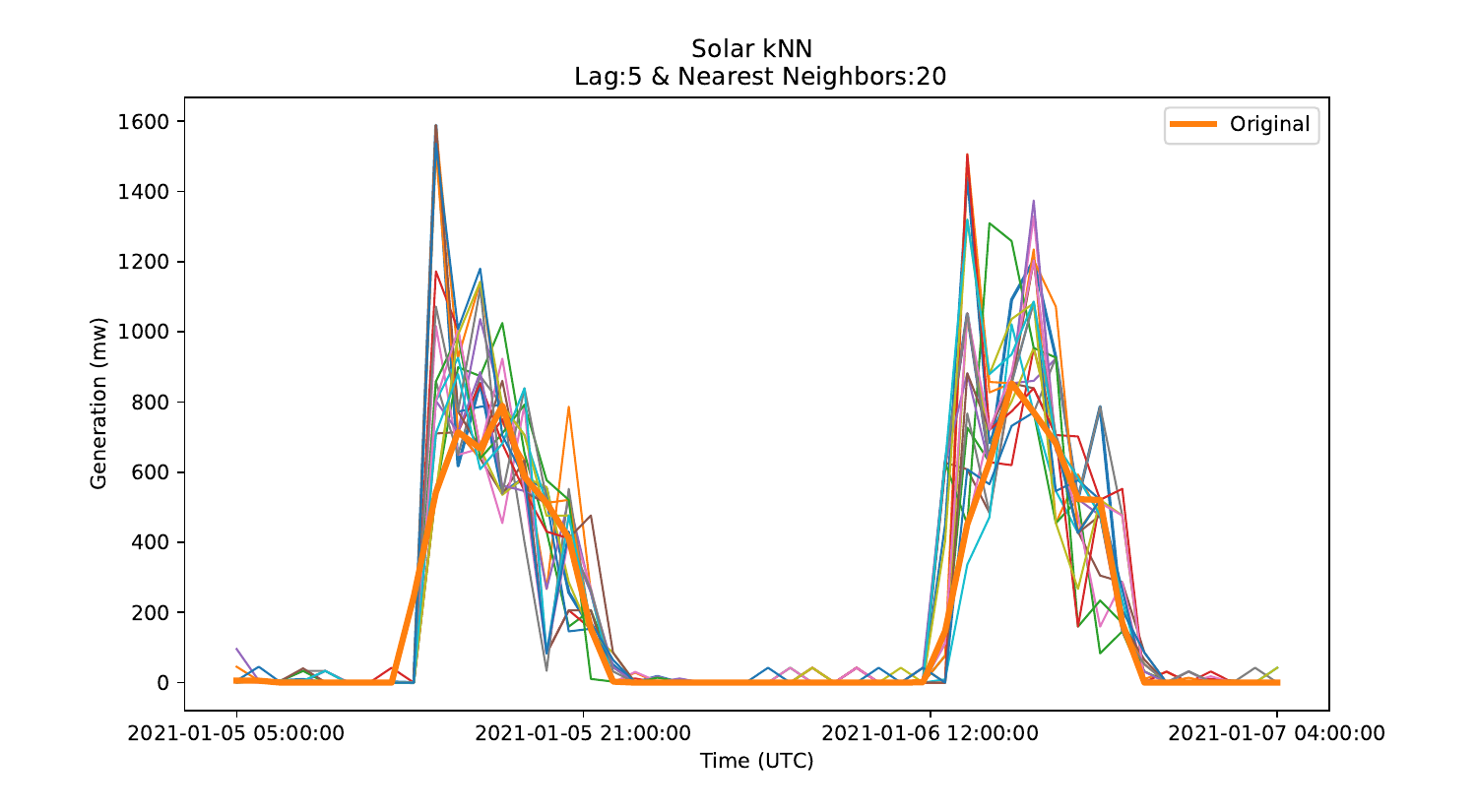}
        \caption{Demonstration of 21 generated time series for solar energy using the NNLB method. Lag=5. Nearest neighbors=20.}
        \label{fig_solar_knn}
    \end{figure}

\begin{figure}[h!]
        \centering
         \includegraphics[width=\textwidth]{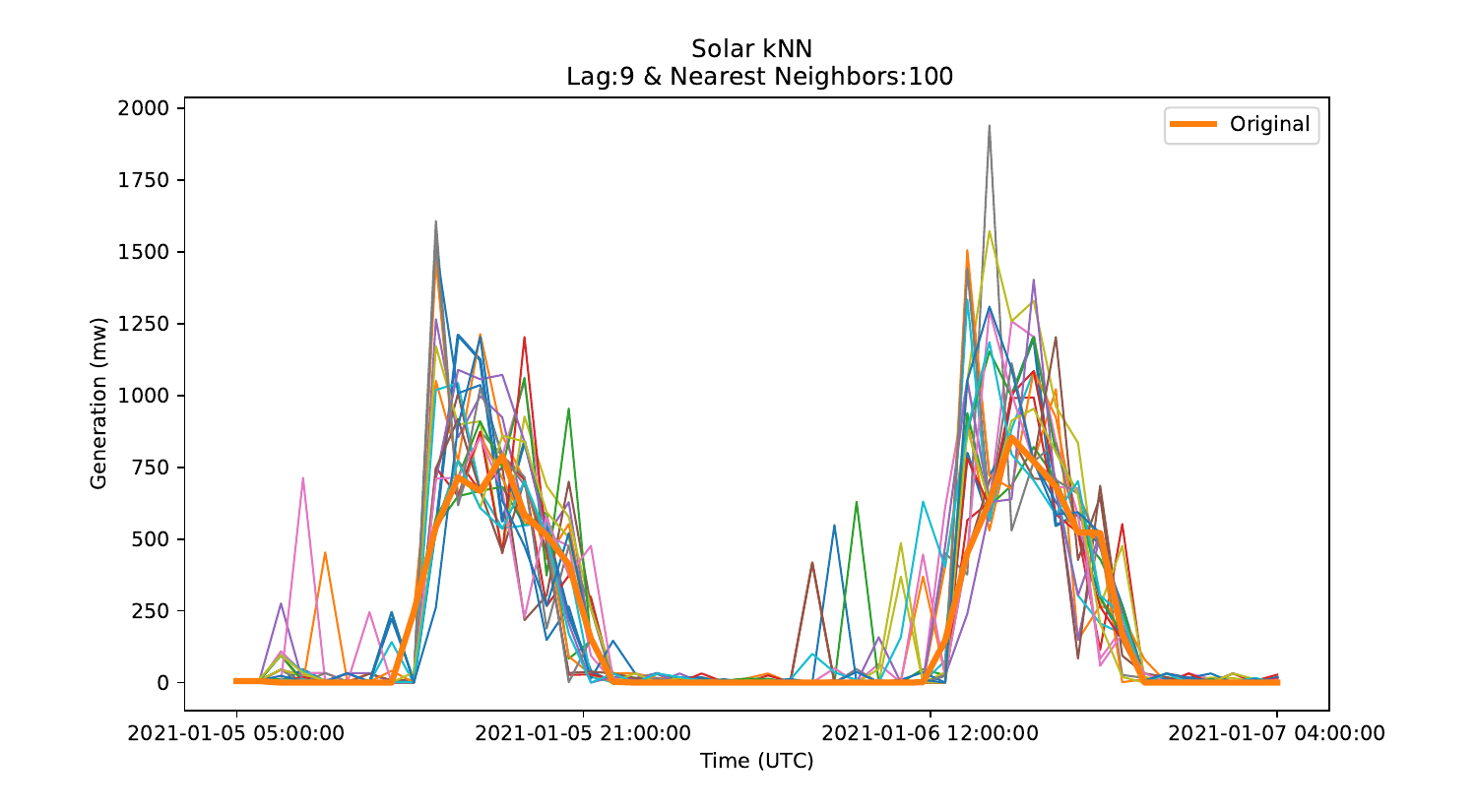}
        \caption{Demonstration of 21 generated time series for solar energy using the NNLB method. Lag=9. Nearest neighbors=100.}
        \label{fig_solar_knn_2}
    \end{figure}

    \paragraph{Solar}
        As seen in Figures \ref{fig_solar_knn} and \ref{fig_solar_knn_2}, the NNLB method encapsulates the general sense of the original solar time series from PJM. However, it is notable that at night, when solar energy generation should be 0 mW, the NNLB method has the possibility of generating series exhibiting energy production at night. This means the NNLB method does not capture the original series to the fullest extent desired. 
    
\begin{figure}[h!]
        \centering
         \includegraphics[width=0.45\textwidth]{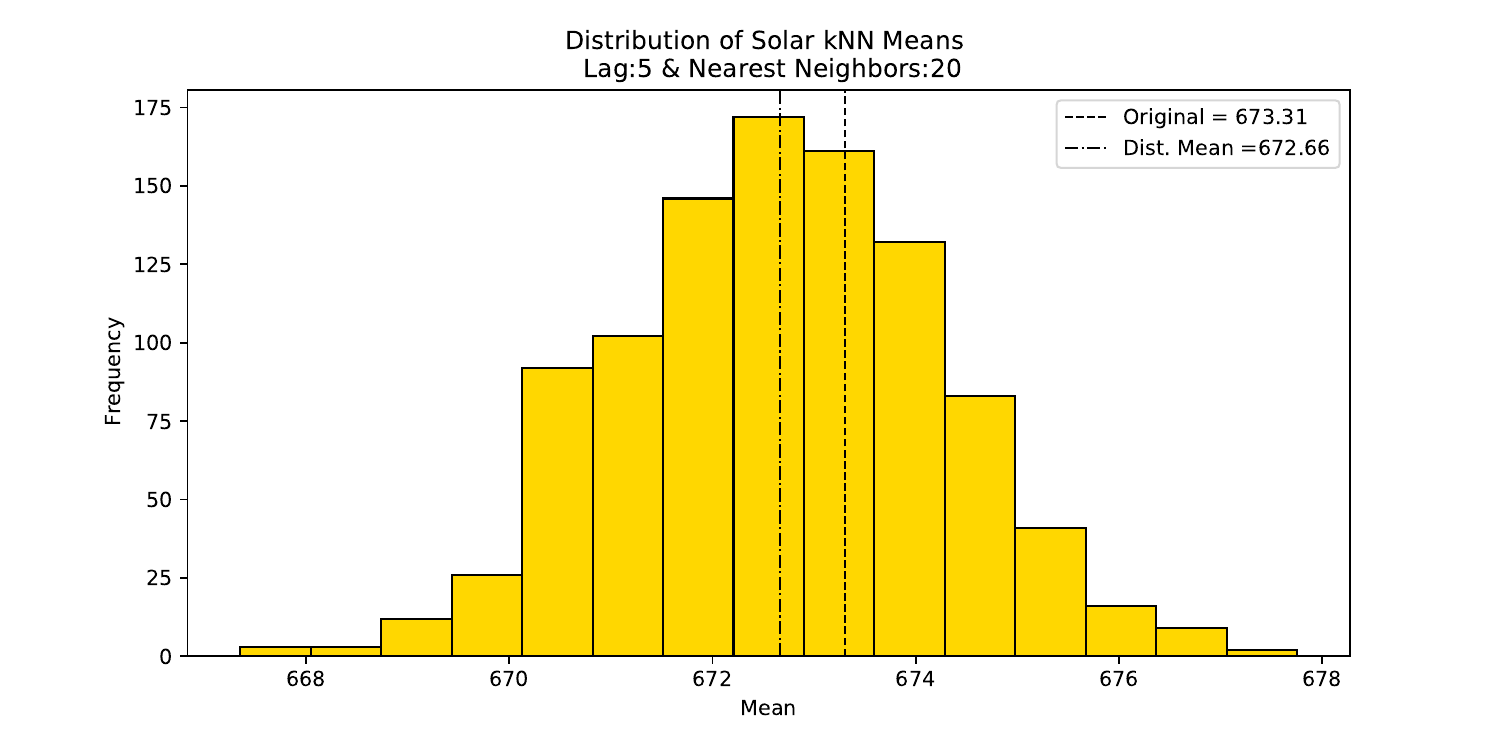}
        \includegraphics[width=0.45\textwidth]{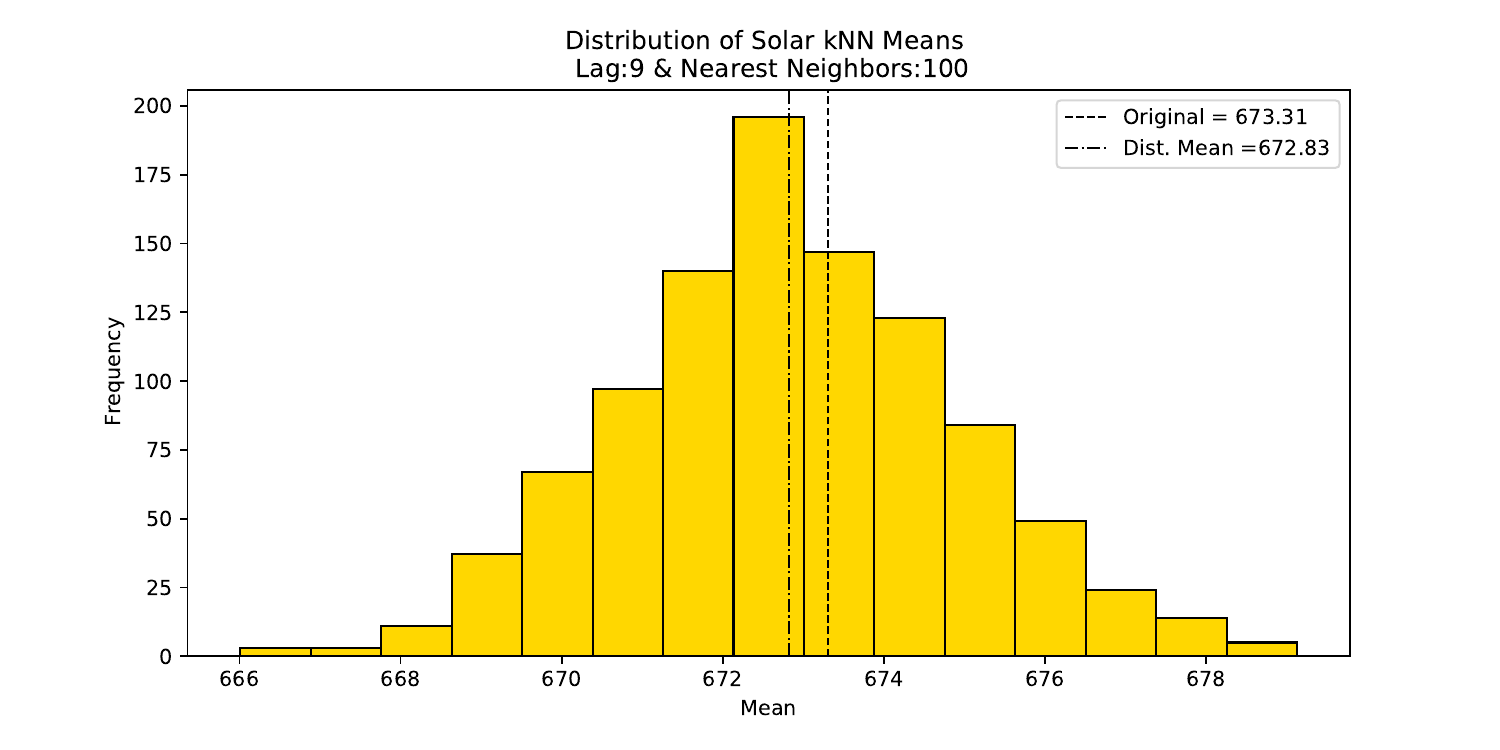}
        \caption{Distribution of NNLB means for solar.}
        \label{fig_solar_knn_mean_dist}
    \end{figure}

    Figure \ref{fig_solar_knn_mean_dist} shows the bias in the generated series is minimal under both   lag and nearest neighbors size configurations exhibited. However, it is noticeable that the NNLB method has the capability of over-reaching certain points by more than double the original energy generation of that hour. This means the averages appear to be evened out by both the over- and under-reaching of points.
    
    The statistics shown in the two tables in Table \ref{tab_solar_knn_stats}   showcase the amount of variability these 1000 series have from each other. The  first columns list the statistics we used to measure each individual generated series. The table first rows list the statistics used on each of the of the 1000 individual statistics we calculated for each generated NNLB series. For example, the number $0.867$ in the bottom left corner of Table \ref{tab_solar_knn_stats} for a lag of 5 and using the 20 nearest neighbors represents the average of all 1000 daily autocorrelations calculated for the 1000 NNLB time series that we generated.
    
    \begin{table}[h!]
    \centering
    Lag: 5 \& Nearest Neighbors: 20
    \begin{tabular}{|l|r|r|r|r|r|r|r|r|}
    \toprule
    Description       &    mean &    std &      min &     25\% &     50\% &     75\% &      max &  Original \\
    \midrule
    Min               &    0.00 &  0.000 &    0.00 &    0.00 &    0.00 &    0.00 &    0.00 &   
    0.00 \\
    First Quartile    &    2.85 &  0.090 &    2.60 &    2.79 &    2.90 &    2.90 &    3.00 &      2.90 \\
    Median            &   22.06 &  1.949 &   19.10 &   20.15 &   21.80 &   23.40 &   25.30 &     24.80 \\
    Third Quartile    & 1444.28 & 13.517 & 1402.05 & 1436.35 & 1446.90 & 1454.10 & 1474.80 &   1436.78 \\
    Max               & 2910.44 & 15.699 & 2852.70 & 2899.50 & 2922.50 & 2922.50 & 2922.50 &   2922.50 \\
    Mean              &  672.66 &  1.577 &  667.35 &  671.61 &  672.70 &  673.79 &  677.75 &    673.31 \\
    Standard Dev.     &  901.00 &  1.382 &  896.84 &  900.09 &  900.99 &  901.96 &  905.52 &    900.21 \\
    Coeff. of Var.    &   1.339 &  0.003 &   1.331 &   1.338 &   1.340 &   1.341 &   1.348 &     1.337 \\
    Autocorr. Lag: 24 &   0.867 &  0.002 &   0.860 &   0.866 &   0.867 &   0.869 &   0.874 &     0.895 \\
    \bottomrule
    \end{tabular}
    
    Lag: 9 \& Nearest Neighbors: 100
    \begin{tabular}{|l|r|r|r|r|r|r|r|r|}
    \toprule
    Description       &    mean &    std &     min &    25\% &    50\% &    75\% &     max &  Original \\
    \midrule
    Min               &    0.00 &  0.000 &    0.00 &    0.00 &    0.00 &    0.00 &    0.00 &      0.00 \\
    First Quartile    &    2.96 &  0.098 &    2.60 &    2.90 &    2.90 &    3.00 &    3.40 &      2.90 \\
    Median            &   19.41 &  0.632 &   17.75 &   19.10 &   19.30 &   19.90 &   24.10 &     24.80 \\
    Third Quartile    & 1451.03 & 13.392 & 1393.42 & 1445.90 & 1453.05 & 1459.88 & 1483.70 &   1436.78 \\
    Max               & 2909.24 & 17.045 & 2840.30 & 2899.50 & 2922.50 & 2922.50 & 2922.50 &   2922.50 \\
    Mean              &  672.83 &  2.082 &  666.01 &  671.50 &  672.74 &  674.15 &  679.13 &    673.31 \\
    Standard Dev.     &  900.46 &  1.790 &  894.65 &  899.25 &  900.47 &  901.65 &  906.05 &    900.21 \\
    Coeff. of Var.    &   1.338 &  0.003 &   1.329 &   1.336 &   1.338 &   1.340 &   1.348 &     1.337 \\
    Autocorr. Lag: 24 &   0.861 &  0.003 &   0.853 &   0.859 &   0.861 &   0.862 &   0.869 &     0.895 \\
    \bottomrule
    \end{tabular}
    \caption{Distribution of all the statistics calculated for the solar NNLB series.}
    \label{tab_solar_knn_stats}
    
    \end{table}
    
    Although there does not appear to be bias among the means, the distributions of the medians show that the NNLB method still has some underlying bias.
    
\begin{figure}[h!]
        \centering
        \includegraphics[width=0.45\textwidth]{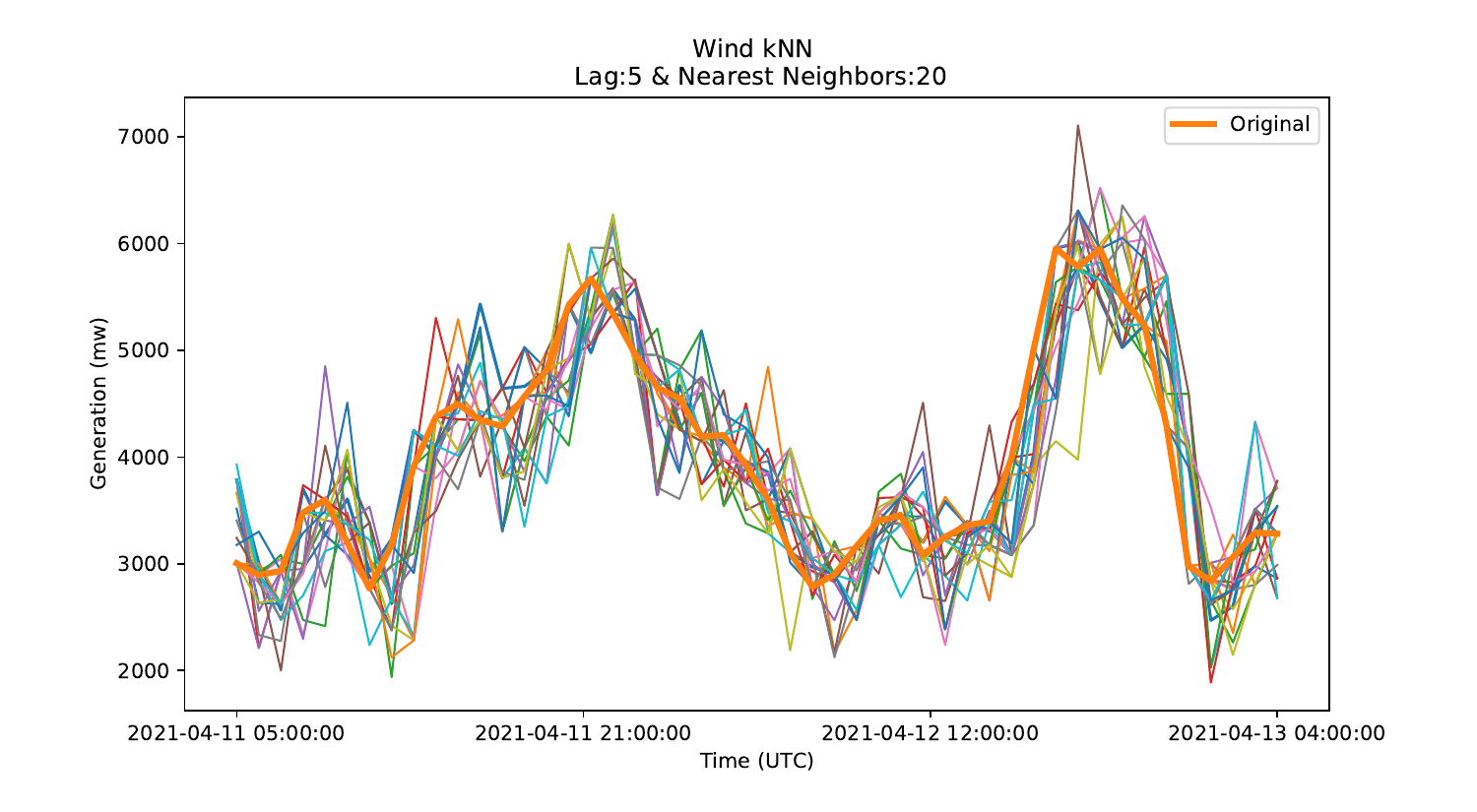}
        \includegraphics[width=0.45\textwidth]{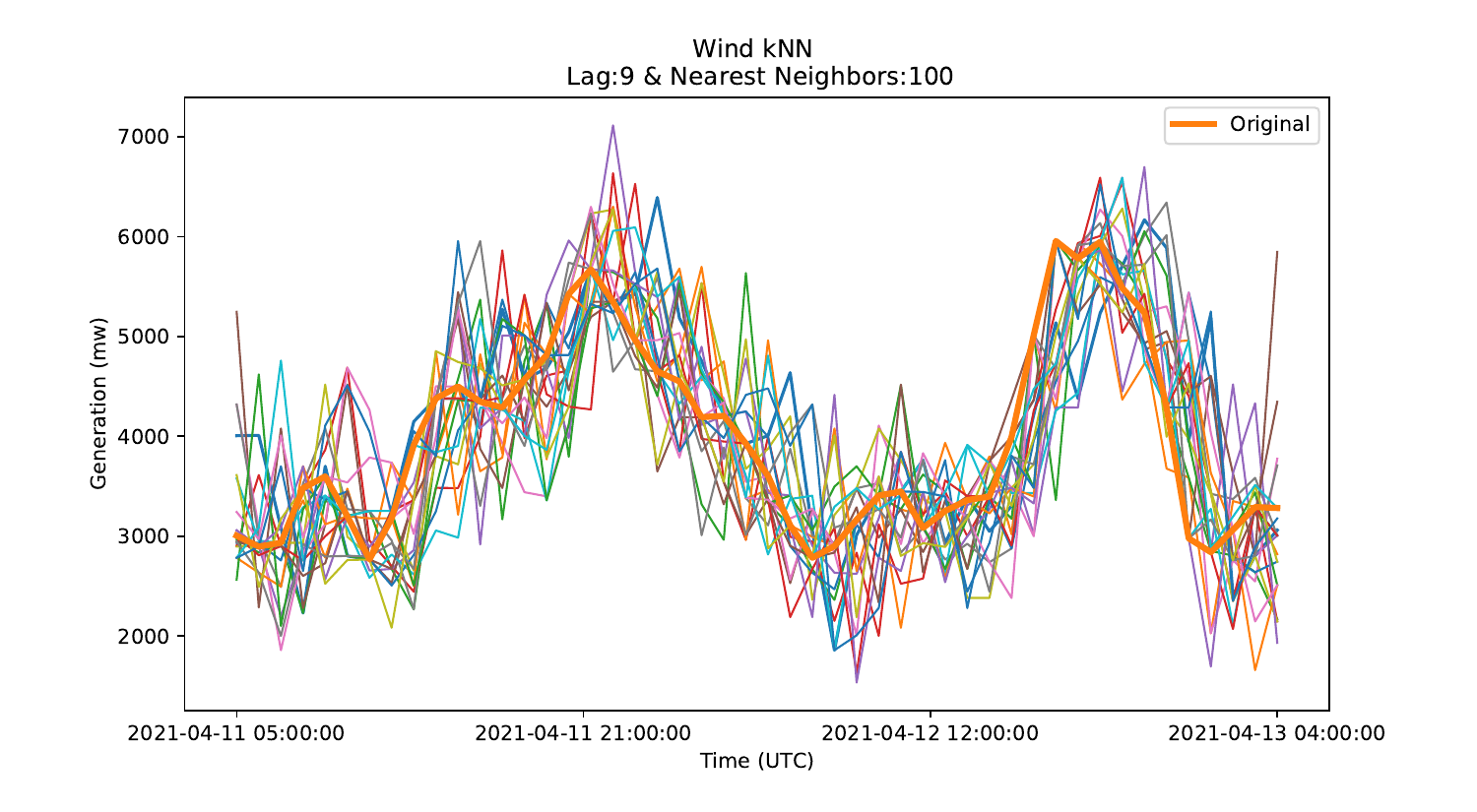}
        \caption{Demonstration of 21 generated time series for wind energy using the NNLB method.}
        \label{fig_wind_knn}
    \end{figure}

    \paragraph{Wind} 
    From Figure \ref{fig_wind_knn}, we see that the NNLB method captures the main trends of wind energy, but with more variation in the series when we increase the lag size and number of nearest neighbors used. 
    
    However, there appears to be even more deviation from the original series mean than we saw in the case of the generated solar series. Though this bias appears to be more prominent than the bias in the solar generated series, when we look at the distribution of medians for wind in Table \ref{tab_wind_knn_stats}, there is less deviation from the original series than seen with solar.
    
    \begin{figure}[ht!]
        \centering
        \includegraphics[width=0.45\textwidth]{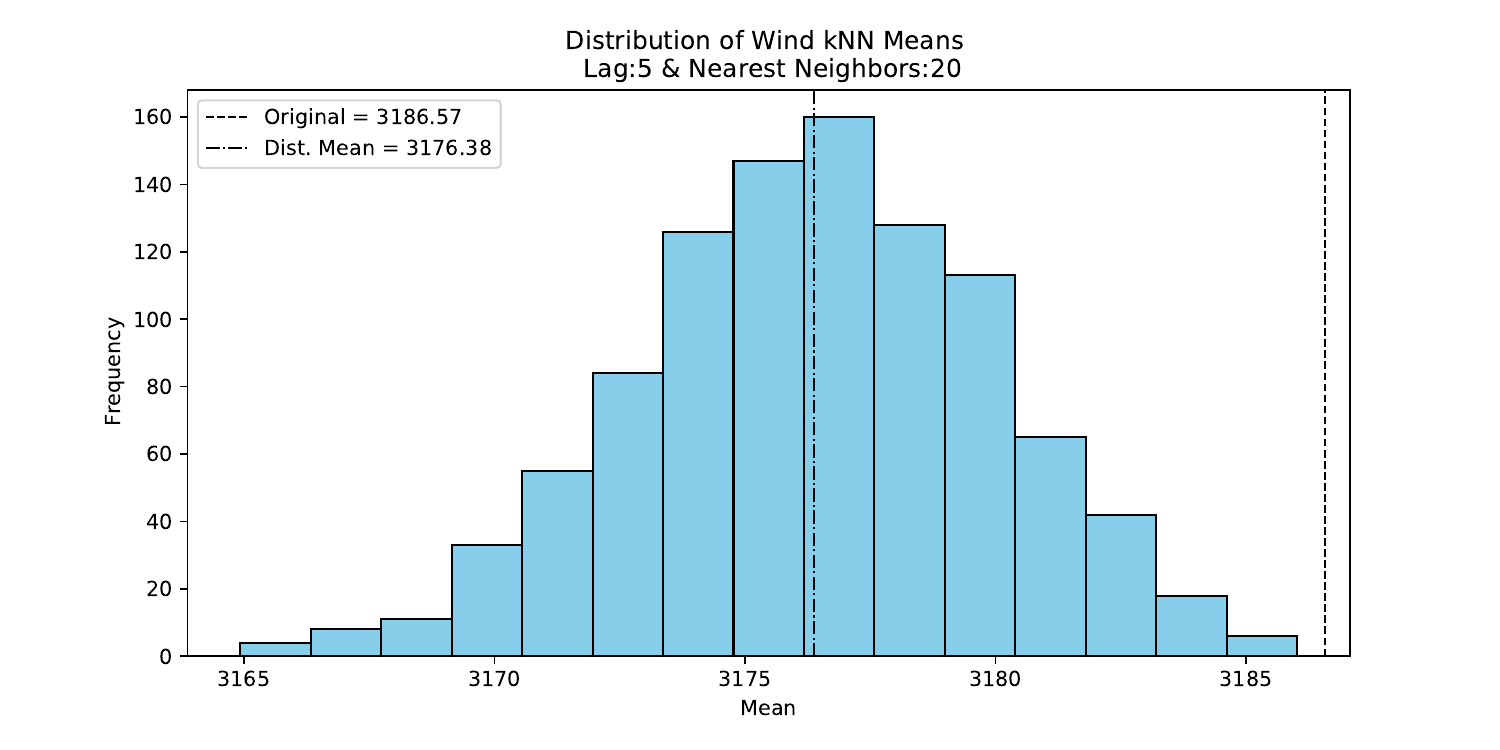}
        \includegraphics[width=0.45\textwidth]{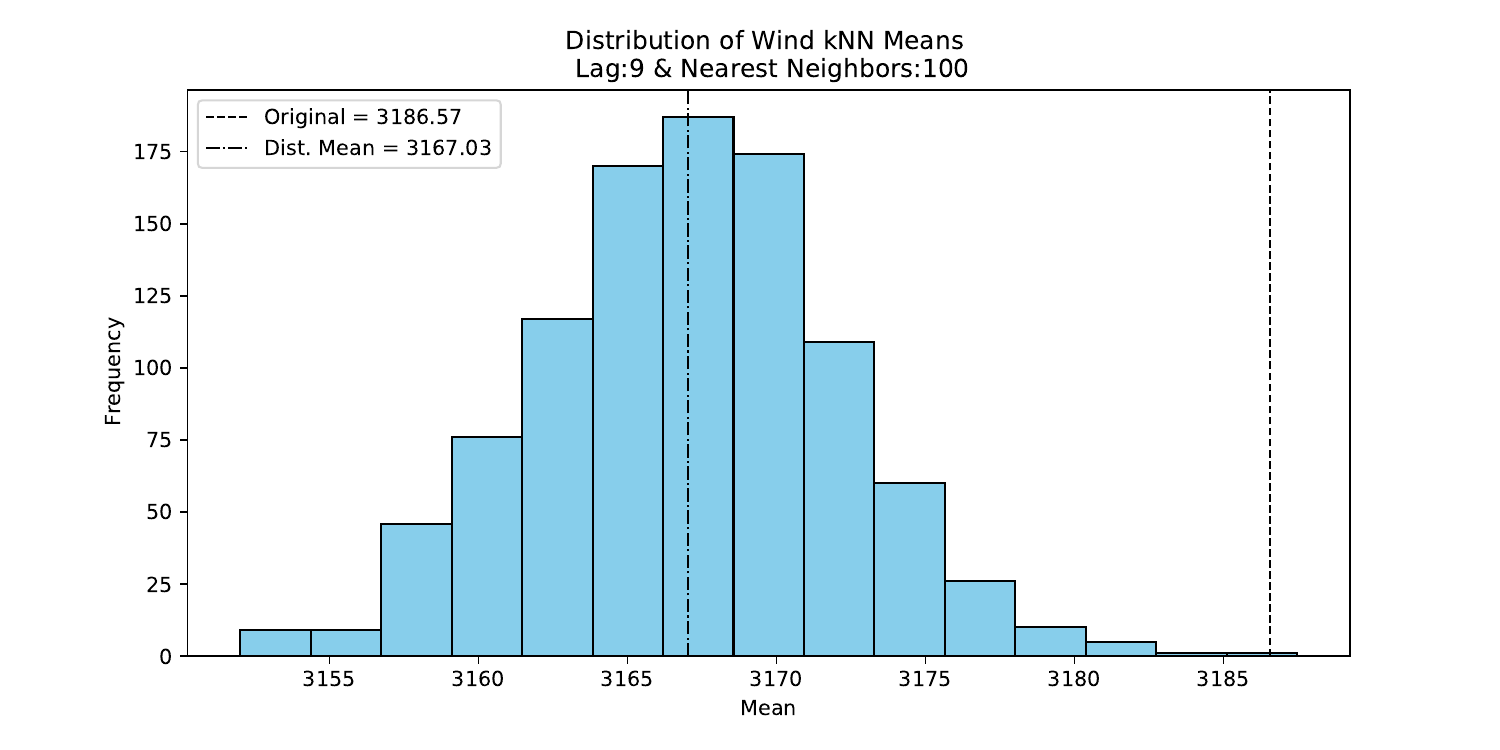}
        \caption{Distribution of NNLB block means for wind.}
        \label{fig_wind_knn_mean_dist}
    \end{figure}
    
    The more extreme bias in the means is associated with the lower autocorrelation that wind energy has compared to solar energy. Hence, with solar energy generation, there are more points (hours) that could have been selected into the pooled samples that are close in distance with the original point's generation. Wind energy, on the other hand, is not as consistent day-to-day, so it is harder for points to match up well with each other based on the Euclidean distance with their lag vectors.
    
    \begin{table}[ht!]
    \centering
    
    Lag: 5 \& Nearest Neighbors: 20
    \begin{tabular}{|l|r|r|r|r|r|r|r|r|}
    \toprule
    Description       &    mean &    std &      min &     25\% &     50\% &     75\% &      max & Original \\
    \midrule
    Min               &   76.84 & 14.742 &   63.30 &   63.30 &   76.60 &   81.10 &  123.40 &     63.30 \\
    First Quartile    & 1370.47 &  8.434 & 1341.20 & 1367.17 & 1370.42 & 1376.50 & 1394.20 &   1372.90 \\
    Median            & 2725.87 & 12.143 & 2680.60 & 2720.00 & 2727.15 & 2734.35 & 2759.90 &   2739.50 \\
    Third Quartile    & 4712.83 & 19.462 & 4660.30 & 4697.40 & 4712.25 & 4729.60 & 4765.93 &   4743.15 \\
    Max               & 8973.98 & 27.787 & 8746.20 & 8973.70 & 8990.00 & 8990.00 & 8990.00 &   8990.00 \\
    Mean              & 3176.38 &  3.584 & 3164.93 & 3174.04 & 3176.45 & 3178.93 & 3186.02 &   3186.57 \\
    Standard Dev.     & 2129.21 &  3.160 & 2119.70 & 2127.05 & 2129.20 & 2131.33 & 2140.18 &   2138.40 \\
    Coeff. of Var.    &   0.670 &  0.001 &   0.667 &   0.670 &   0.670 &   0.671 &   0.674 &     0.671 \\
    Autocorr. Lag: 24 &   0.425 &  0.002 &   0.416 &   0.423 &   0.425 &   0.426 &   0.433 &     0.437 \\
    \bottomrule
    \end{tabular}
    
    Lag: 9 \& Nearest Neighbors: 100
    
    \begin{tabular}{|l|r|r|r|r|r|r|r|r|}
    \toprule
    Description       &    mean &    std &     min &    25\% &    50\% &    75\% &     max &  Original \\
    \midrule
    Min               &   83.54 & 18.370 &   63.30 &   63.30 &   76.60 &  102.10 &  131.00 &     63.30 \\
    First Quartile    & 1368.46 & 10.950 & 1332.25 & 1362.15 & 1368.92 & 1376.50 & 1402.90 &   1372.90 \\
    Median            & 2720.06 & 15.250 & 2670.90 & 2708.50 & 2721.60 & 2730.50 & 2766.45 &   2739.50 \\
    Third Quartile    & 4703.40 & 21.311 & 4645.70 & 4687.45 & 4703.80 & 4717.95 & 4765.70 &   4743.15 \\
    Max               & 8977.79 & 23.790 & 8733.60 & 8973.70 & 8990.00 & 8990.00 & 8990.00 &   8990.00 \\
    Mean              & 3167.03 &  5.083 & 3152.01 & 3163.78 & 3167.09 & 3170.42 & 3187.47 &   3186.57 \\
    Standard Dev.     & 2120.58 &  4.529 & 2105.65 & 2117.32 & 2120.70 & 2123.61 & 2138.45 &   2138.40 \\
    Coeff. of Var.    &   0.670 &  0.001 &   0.664 &   0.669 &   0.670 &   0.671 &   0.674 &     0.671 \\
    Autocorr. Lag: 24 &   0.422 &  0.003 &   0.412 &   0.419 &   0.422 &   0.424 &   0.436 &     0.437 \\
    \bottomrule
    \end{tabular}
    
    \caption{Distribution of all the statistics calculated for the wind NNLB series.}
    \label{tab_wind_knn_stats}
    \end{table}
    
    \paragraph{Load} 
    
    Energy demand approximately follows the same patterns every day, which corroborates the high daily autocorrelations in the statistics; see Table \ref{tab_load_knn_stats}.
    
    \begin{figure}[h!]
        \centering
        \includegraphics[width=0.45\textwidth]{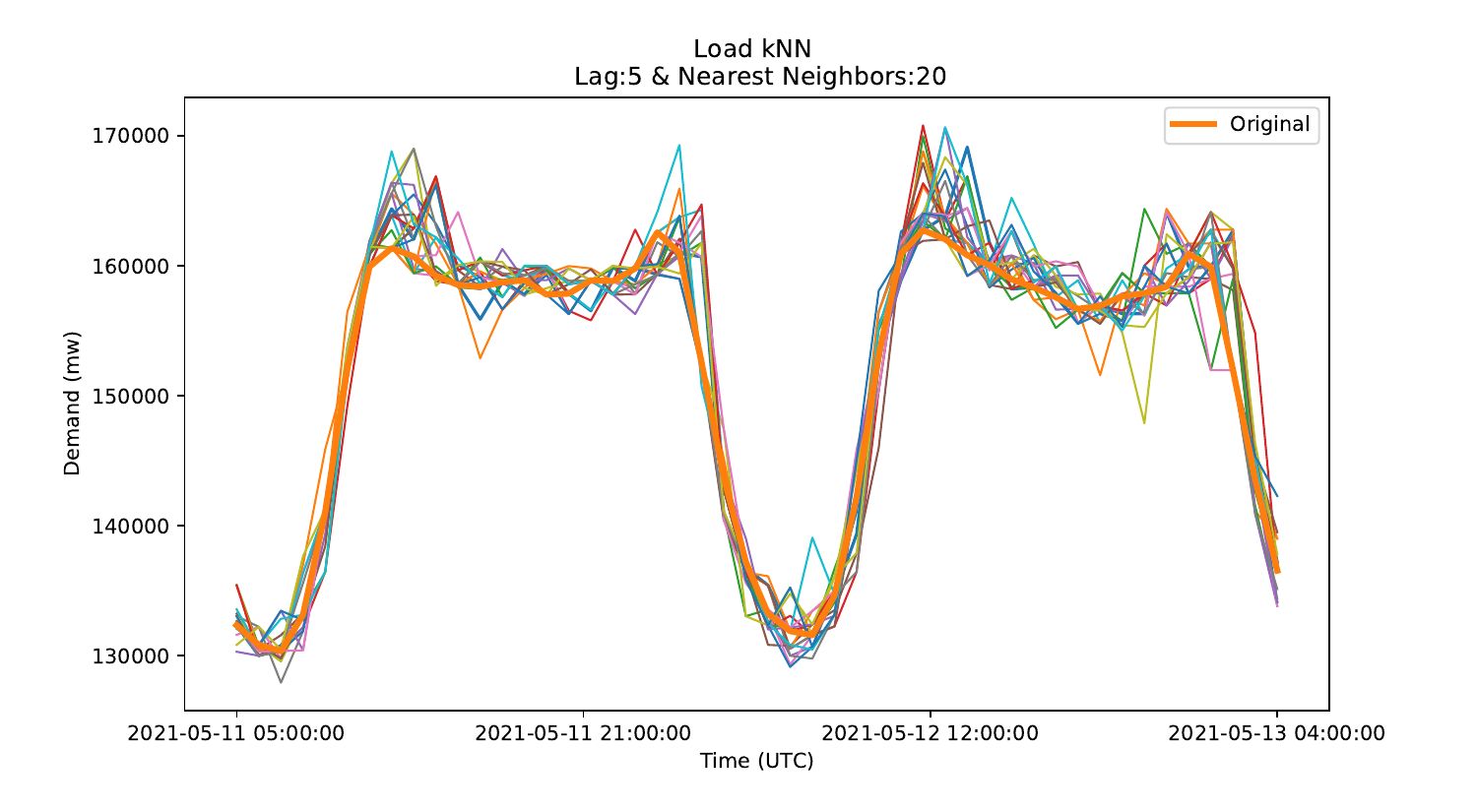}
        \includegraphics[width=0.45\textwidth]{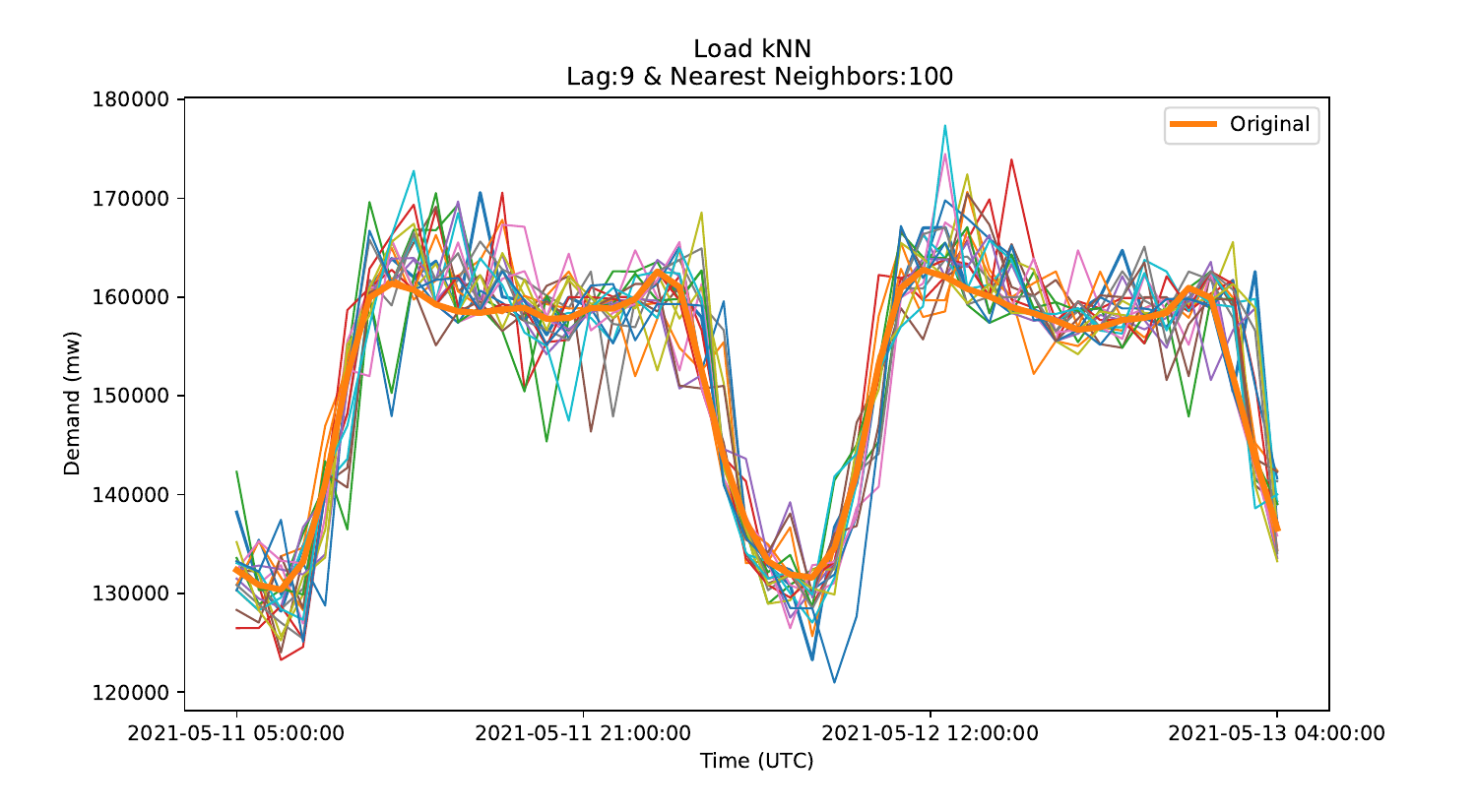}
        \caption{Demonstration of 21 generated time series for energy load using the NNLB method.}
        \label{fig_load_knn}
    \end{figure}
    
    Once again, the bias is negligible with energy demand since a day's expected load is significantly larger than the difference in means between the generated and original load time series. 
    
    \begin{figure}[h!]
        \centering
        \includegraphics[width=0.45\textwidth]{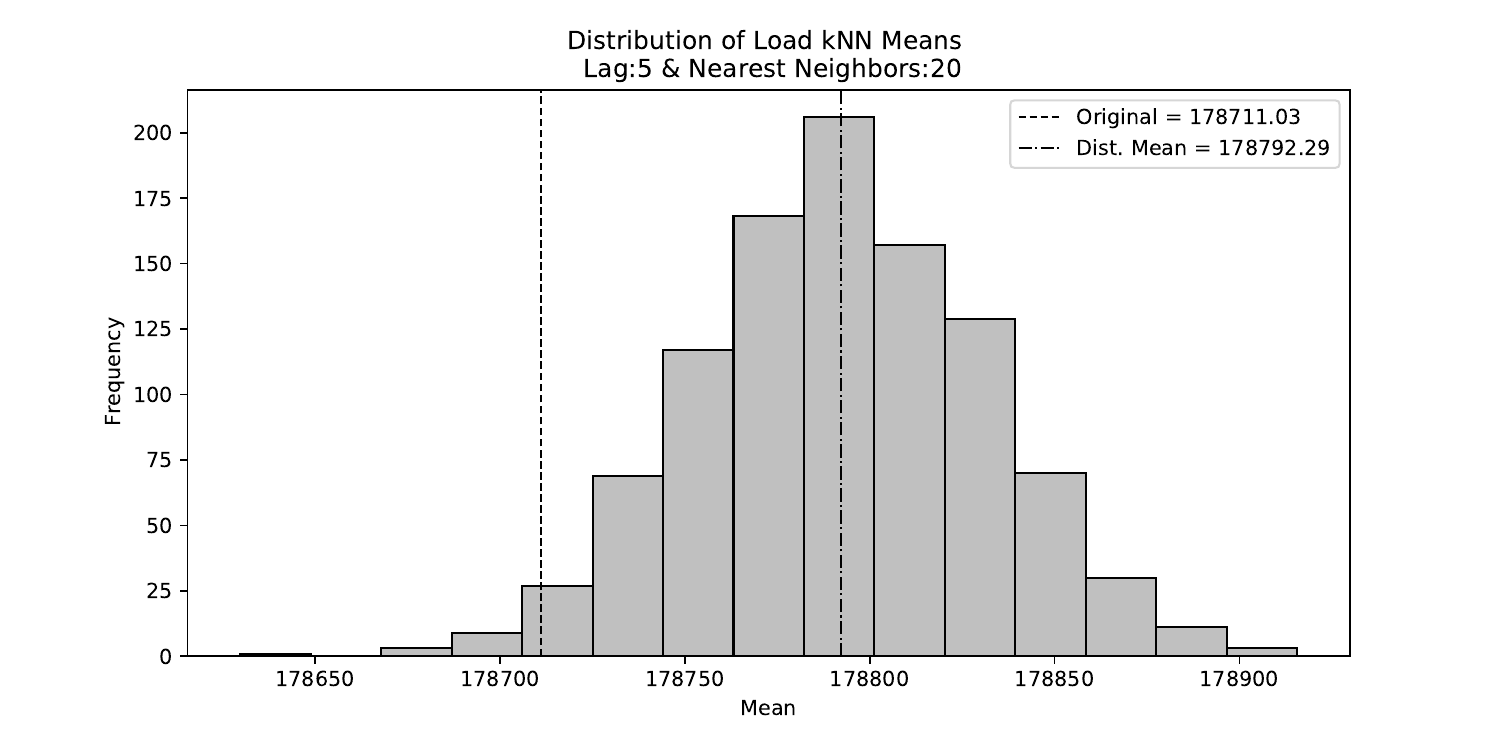}
        \includegraphics[width=0.45\textwidth]{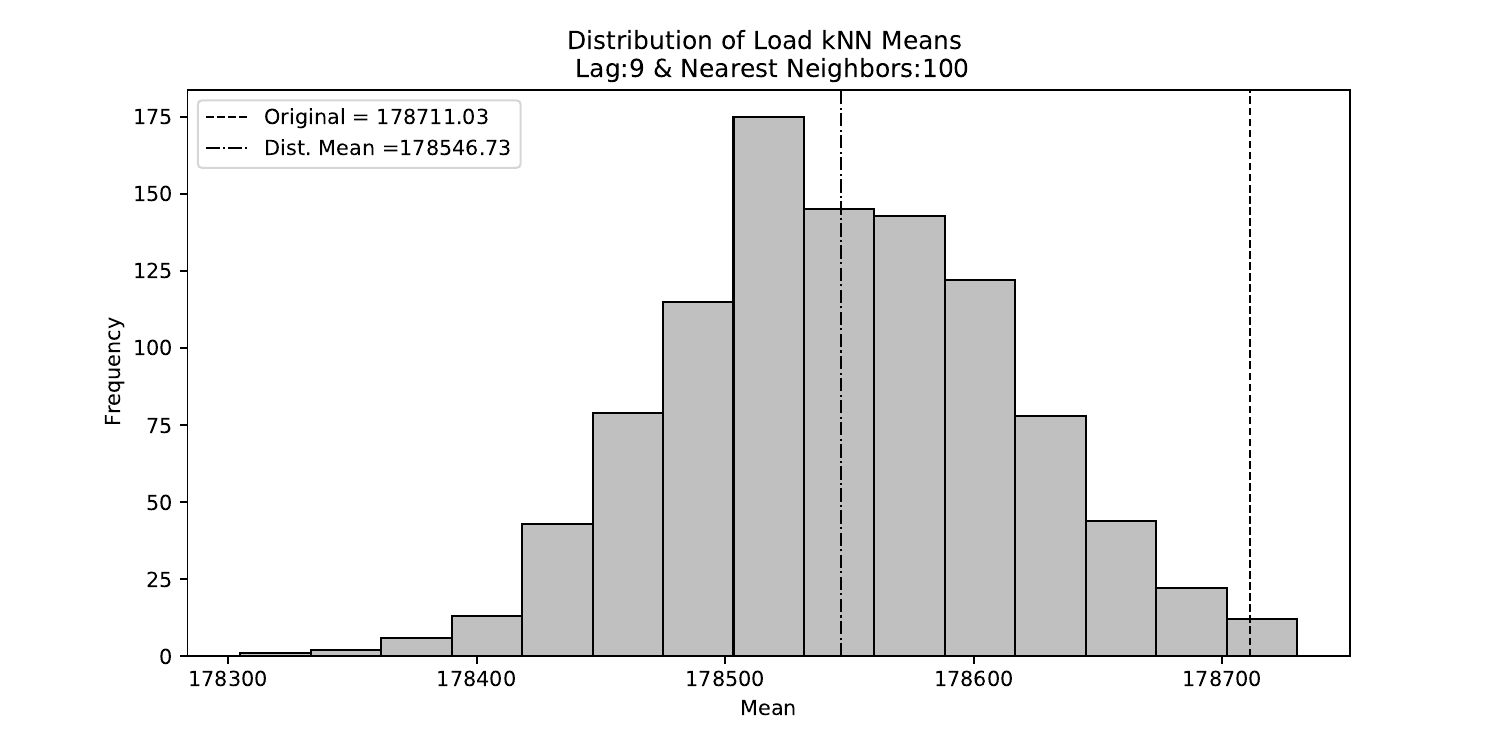}
        \caption{Distribution of NNLB means for load.}
        \label{fig_load_knn_mean_dist}
    \end{figure}
    
    What is notable is that the distribution of the generated series' means increases as lag size and the number of nearest neighbors used increases. This trend can also be seen in the solar and wind sections.
    
    \begin{table}[b!]
        \centering
        
    Lag: 5 \& Nearest Neighbors: 20
    \begin{tabular}{|l|r|r|r|r|r|r|r|r|}
    \toprule
    Description       &   mean &  std &    min &    25\% &    50\% &    75\% &    max &  Original \\
    \midrule
    Min               & 117721 &  540.405 & 117370 & 117370 & 117428 & 118554 & 118884 &    117370 \\
    First Quartile    & 156607 &  125.507 & 156161 & 156542 & 156597 & 156644 & 157152 &    156350 \\
    Median            & 174116 &  130.208 & 173656 & 174039 & 174124 & 174207 & 174577 &    174063 \\
    Third Quartile    & 197102 &  160.884 & 196538 & 197004 & 197087 & 197226 & 197614 &    197085 \\
    Max               & 297198 &  783.151 & 293709 & 297495 & 297541 & 297541 & 297541 &    297541 \\
    Mean              & 178792 &   39.246 & 178630 & 178765 & 178791 & 178819 & 178916 &    178711 \\
    Standard Dev.     &  32108 &   45.471 &  31934 &  32076 &  32108 &  32140 &  32259 &     32242 \\
    Coeff. of Var.    &  0.180 &    0.000 &  0.179 &  0.179 &  0.180 &  0.180 &  0.180 &  0.180 \\
    Autocorr. Lag: 24 &  0.899 &    0.002 &  0.890 &  0.898 &  0.899 &  0.900 &  0.903 &  0.910 \\
    \bottomrule
    \end{tabular}
 
    Lag: 9 \& Nearest Neighbors: 100
    \begin{tabular}{|l|r|r|r|r|r|r|r|r|}
    \toprule
    Description       &   mean &      std &    min &   25\% &   50\% &   75\% &    max &  Original \\
    \midrule
    Min               & 117667 &  516.724 & 117370 & 117370 & 117370 & 117428 & 119183 &   117370 \\
    First Quartile    & 156617 &  203.465 & 155928 & 156517 & 156598 & 156738 & 157326 &   156350 \\
    Median            & 173977 &  175.391 & 173420 & 173874 & 173949 & 174096 & 174623 &   174063 \\
    Third Quartile    & 196712 &  208.389 & 196028 & 196561 & 196725 & 196840 & 197480 &   197085 \\
    Max               & 297218 &  764.019 & 292916 & 297495 & 297541 & 297541 & 297541 &   297541 \\
    Mean              & 178547 &   67.813 & 178305 & 178502 & 178545 & 178595 & 178730 &   178711 \\
    Standard Dev.     &  31789 &   82.965 &  31523 &  31732 &  31790 &  31845 &  32047 &    32242 \\
    Coeff. of Var.    &  0.178 &    0.000 &  0.177 &  0.178 &  0.178 &  0.178 &  0.179 &    0.180 \\
    Autocorr. Lag: 24 &  0.872 &    0.003 &  0.861 &  0.870 &  0.872 &  0.873 &  0.879 &    0.910 \\
    \bottomrule
    \end{tabular}
    
    \caption{Distribution of all the statistics calculated for the load NNLB series.}
    \label{tab_load_knn_stats}
    \end{table}

\subsubsection{Under-Performance Analysis}

Fixing solar or wind physical capacity may result in variable production on a seasonal or annual basis, due to seasonal and annual variation in realized wind and insolation. Of particular concern are wind and solar ``droughts'' which temporarily result in much reduced energy production, a phenomenon well known in practice. Energy systems need to be designed not just for a few recent years of data; they need to anticipate reasonable levels of variability---and possible droughts---decades in advance. This impinges on a central motivation for this study. For example, given a history of wind or solar availability, what sorts of risks of production shortage exist even assuming that the underlying causal generation process remains unchanged?

We will now be using the statistics we laid out in \S\ref{sec_Bootstrap} to describe our generated series. For our purposes, we will only be showcasing the underage of each generated series, but the overage is provided in the supplemental materials.

We report the number of days in our generated series that fall below 95\% of the original PJM dataset. Solar generated series using the NNLB method has the highest average number of days that fall below 95\% of the original series. Energy demand has nearly none of its generated series fall below the 5\% threshold. 

\begin{figure}[h!]
    \centering
    \includegraphics[width=0.45\textwidth]{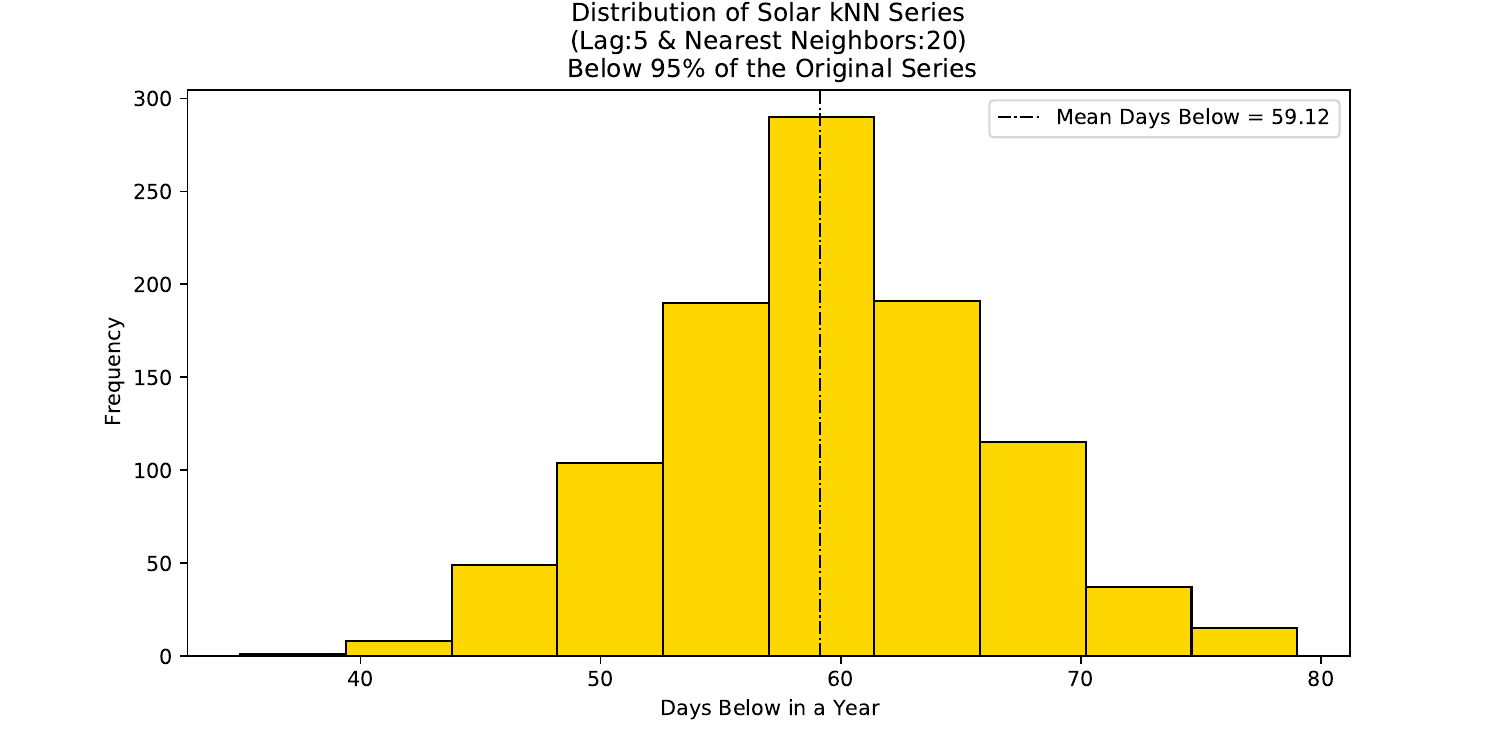}
    \includegraphics[width=0.45\textwidth]{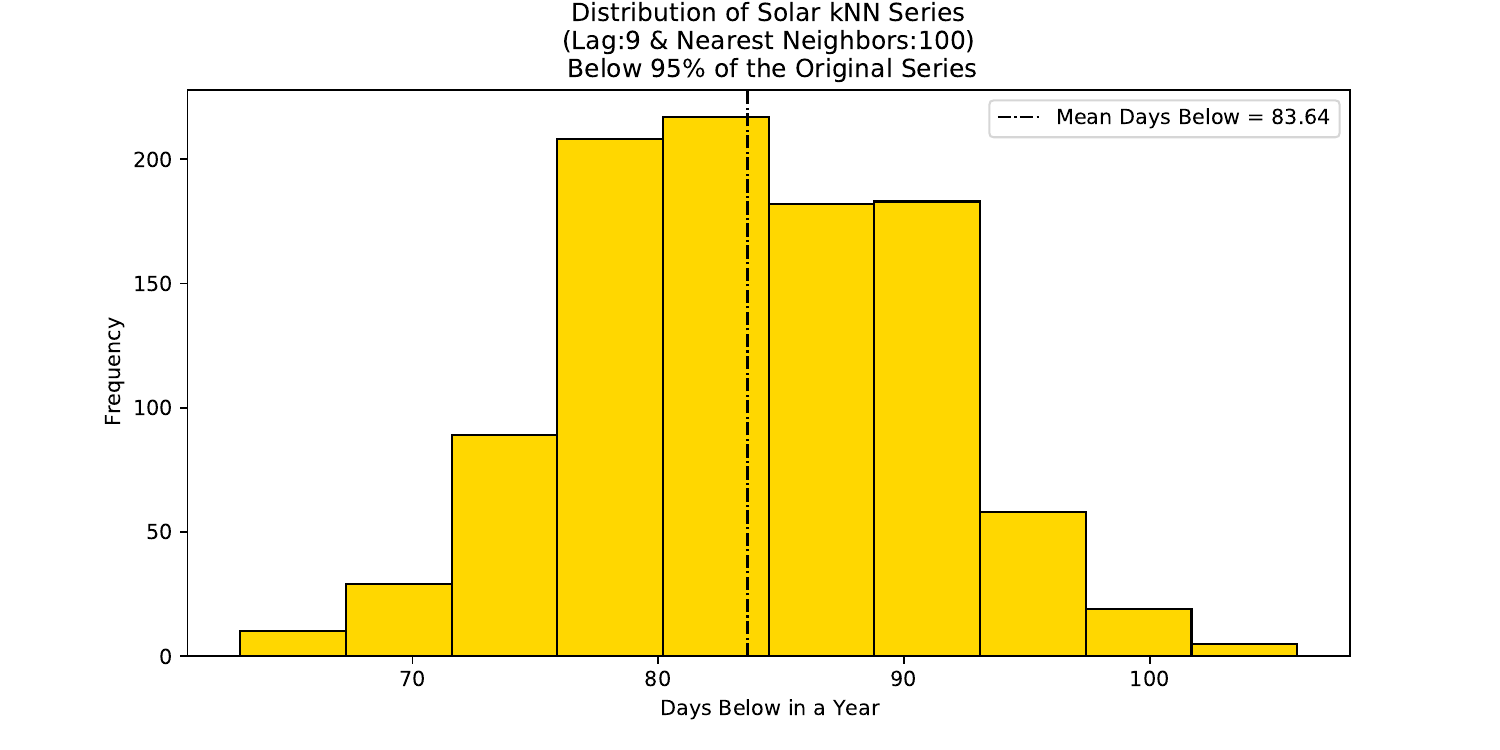}
    \caption{Distribution of the number of days from 1000 solar NNLB series that fall below 95\% of the original PJM data for solar energy.}
    \label{fig_solar_knn_extreme_below}
\end{figure}

These statistics on  series extreme periods cohere with the coefficients of variation for each dataset type. Solar has the highest coefficient of variation, whereas load's coefficient of variation is significantly lower than wind's and solar's. 

\begin{figure}[h!]
    \centering
    \includegraphics[width=0.45\textwidth]{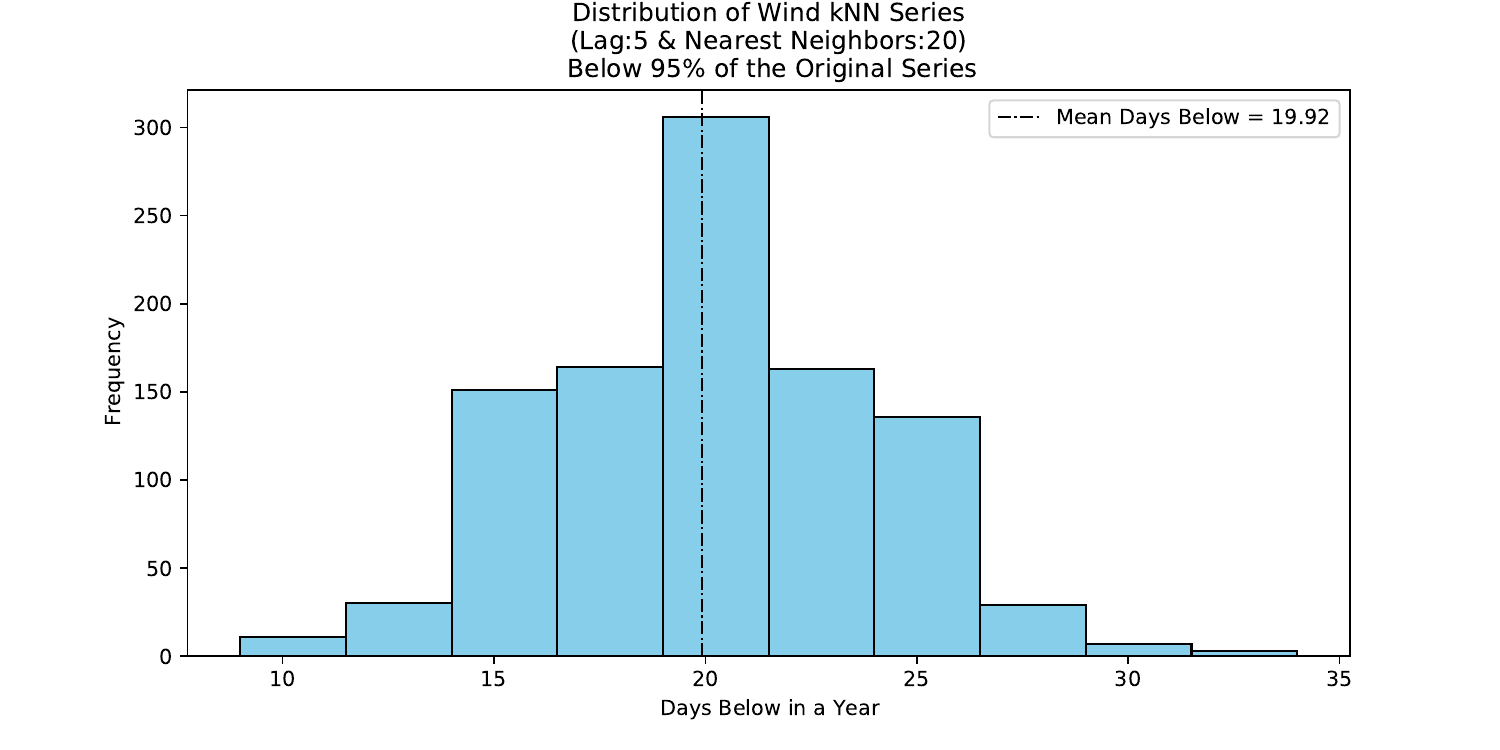}
    \includegraphics[width=0.45\textwidth]{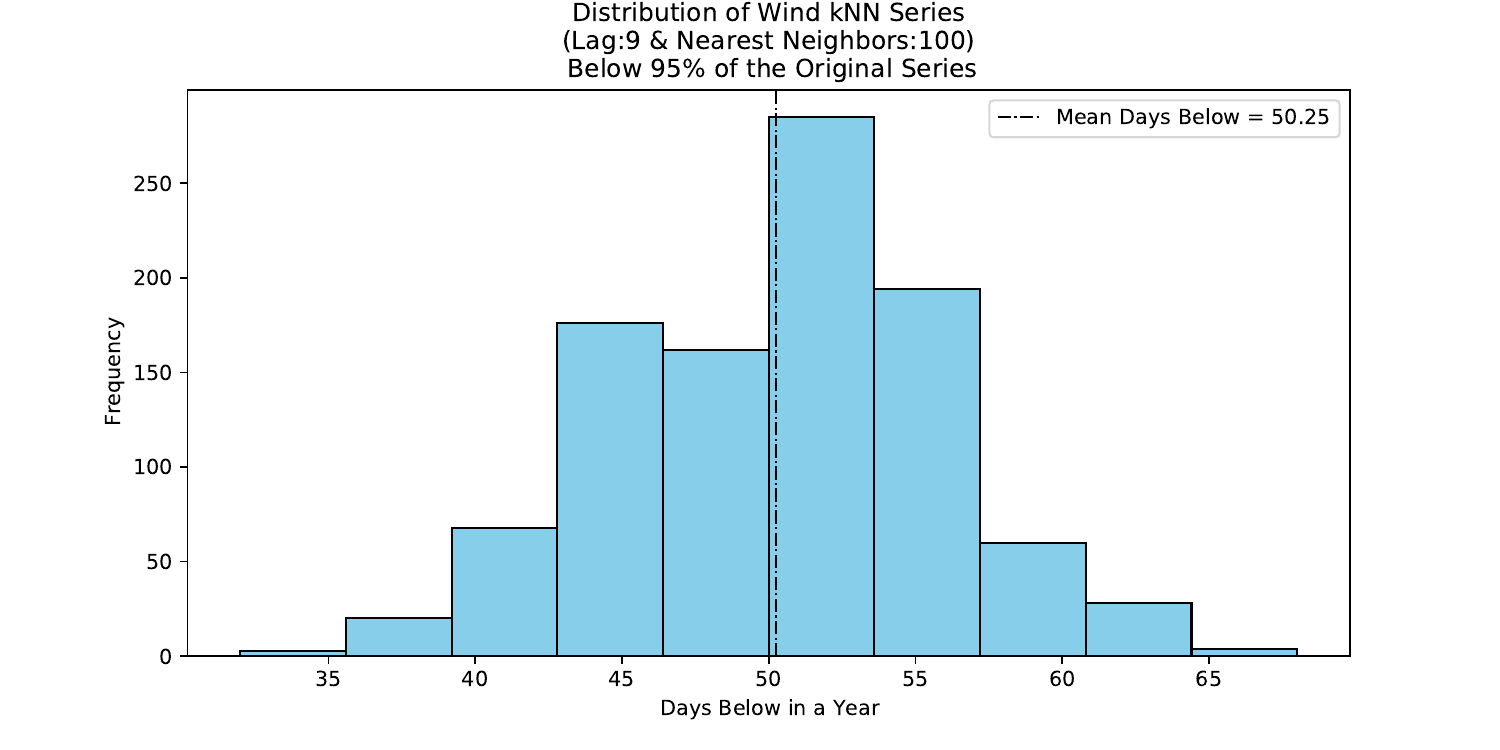}
    \caption{Distribution of the number of days from 1000 wind NNLB series that fall below 95\% of the original PJM data for wind energy.}
    \label{fig_wind_knn_extreme_below}
\end{figure}

A high daily auto-correlation and low coefficient of variation is what leads to load  having only two generated series fall below or threshold, and this only occurs when we use a lag of 9 and the 100 nearest neighbors.

\begin{figure}[h!]
    \centering
    \includegraphics[width=0.45\textwidth]{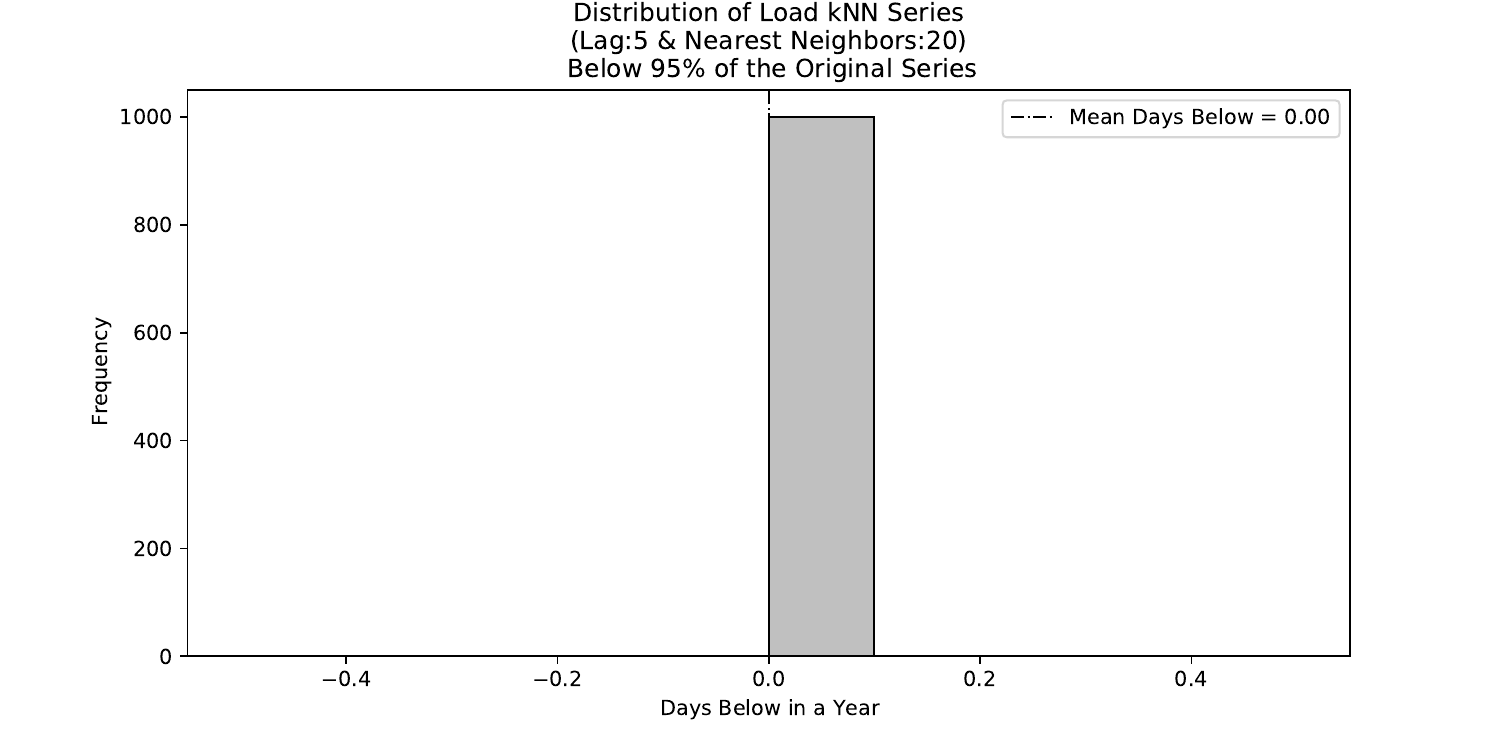}
    \includegraphics[width=0.45\textwidth]{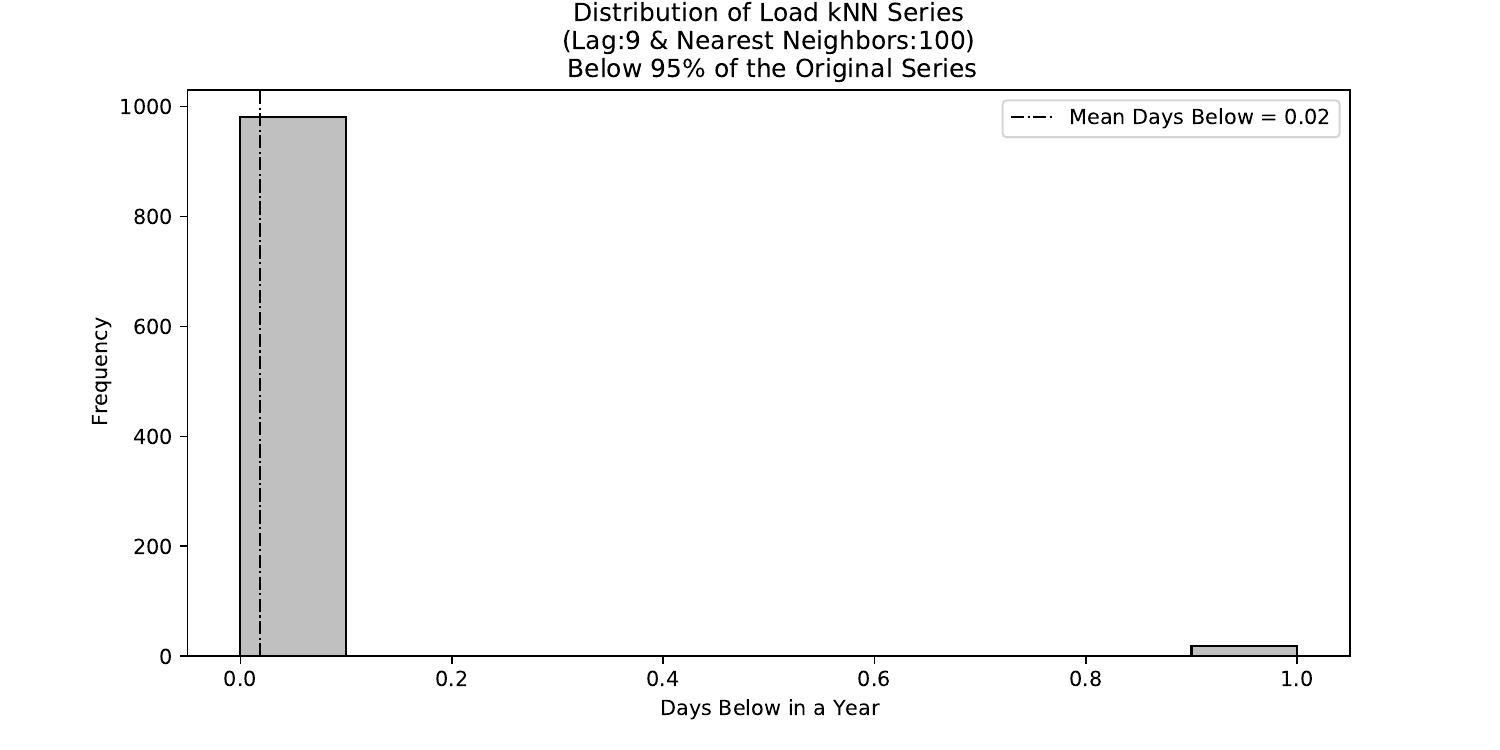}
    \caption{Distribution of the number of days from 1000 load NNLB series that fall below 95\% of the original PJM data for load.}
    \label{fig_load_knn_extreme_below}
\end{figure}

\subsection{Symmetric Block Bootstrap (SBB) Method \label{sec_SBB}}

\subsubsection{Background Information}


The symmetric block bootstrap (SBB) method was introduced in
\cite{kimbrough_symmetric_2021} with minimal development.
This section describes the SBB method and summarizes our re-implementation. In addition, we present new analyses, visuals, and statistical summaries to illustrate findings about this method. These go well beyond what was reported in \cite{kimbrough_symmetric_2021}.  We remind the reader that this is the second of two non-parametric methods---NNLB is the first---for generating alternative time series by bootstrapping that broadly mimic the source distribution, which is presumed to be non-stationary. In subsequent sections, we explore principled directional deviations from the original series as a way of generating data for model analysis. These explorations draw upon and extend both the NNLB and SBB generated series.

As mentioned in \S\ref{sec_NNLB}, the NNLB method was designed for making short-range (point) predictions. Our purpose in this paper is instead is to capture plausible overall distributions, which we perturb and use for purposes of model analysis (alias post-solution analysis). The SBB method of \cite{kimbrough_symmetric_2021} was originally created for that purpose. 
This design goal led to looking at data points on both sides of the focal point, instead of just data points preceding the focal point, as in NNLB.

\subsubsection{Description of the Procedure}

\begin{enumerate}
    \item We start with a given time series with $t$ observations $\mathbf{X} = x_1, x_2, \ldots, x_{t}$ and create a window $w$ around each observation in $\mathbf{X}$. Window $i$ is centered around the observation $x_i$, where $i$ is called the \emph{focal slot}. The window will contain $n$ observations, for some chosen $n$, taken from $\mathbf{X}$ before and after the observation $x_i$, hence the symmetric aspect of this method. Therefore, we define window $w_i = \langle x_{i-n}, \ldots, x_{i-1}, x_{i}, x_{i+1}, \ldots , x_{i+n} \rangle$ for every observation $x_i$ with a size of $1+2n$. We say the window has a size of $1+2n$ and a sash (size) of $n$. Note that as usual we treat each time series as effectively circular, with the last observation wrapping back around to the first observation in the time series.\footnote{This is appropriate for annual data and concatenations thereof. When the circularity assumption is not warranted, then the initial focal slot can be at the $(n+1)$\textsuperscript{th} position.} Thus, the immediate predecessor of an entity at time 0, coded as -1, has the maximum time value in the series, etc.
    
    \item Then, for each observation $x_i$, we collect a pool $\mathcal{P}_i$ of the $p$ windows most similar to $w_i$. The value of $p$ is a parameter in the method. We used values of 20 and 100 in this study; the software accommodates arbitrary values. The code also allows the user to choose whether  the focal point $x_i$ is included in the pool $\mathcal{P}_i$ of $w_i$s or not. Similarity is measured by taking the Euclidean distance between window $w_i$ with all other windows  and adding the $p$ windows with the smallest distance from window $w_i$.

    \item A new time series is generated by randomly selecting one window from each pool $\mathcal{P}_i$ and using the observation in the focal slot as a new observation $x_i'$. (The random selection is uniform in this section of the paper, meaning the closest point and the $20^{\text{th}}$ closest point have the same probability of being chosen. We invite other random schemes like the one used in Section \ref{sec_NNLB} or a new scheme that weights each point by their respective window's distance from the original point's window.) This generates a new time series with the elements $x'_1, x'_2, \ldots , x'_t$.
\end{enumerate}

\subsubsection{Examining the Generated Series}

We have run the symmetric block method on PJM's 2021 generation data 1000 times for solar energy, wind energy, and energy load using a window size of $1+2(2) = 5$ and pool size of 20. We also repeated the 1000 trials with a window size of $1+2(4) = 9$ and pool size of 100 using the same original dataset. Based on a snapshot of 48 hours for each energy type, we can see the new generated time series plausibly capture the patterns of the original time series. 

Through our analysis, we discovered that the symmetric block method is a slightly biased estimator of the mean. Time series generated by this method using generation energy data tend to have an approximately normal distribution of means, but the averages skew away from the more extreme values of the given original time series' distribution. We believe this is due to how the elements in the pool are selected for each observation. Since the pooled windows are most similar (in terms of shortest Euclidean distance) to the given observation's window, it excludes more extreme values from being considered. Therefore, the bias depends on the distribution of the given times series data. 

\begin{figure}[h!]
        \centering
        \includegraphics[width=\textwidth]{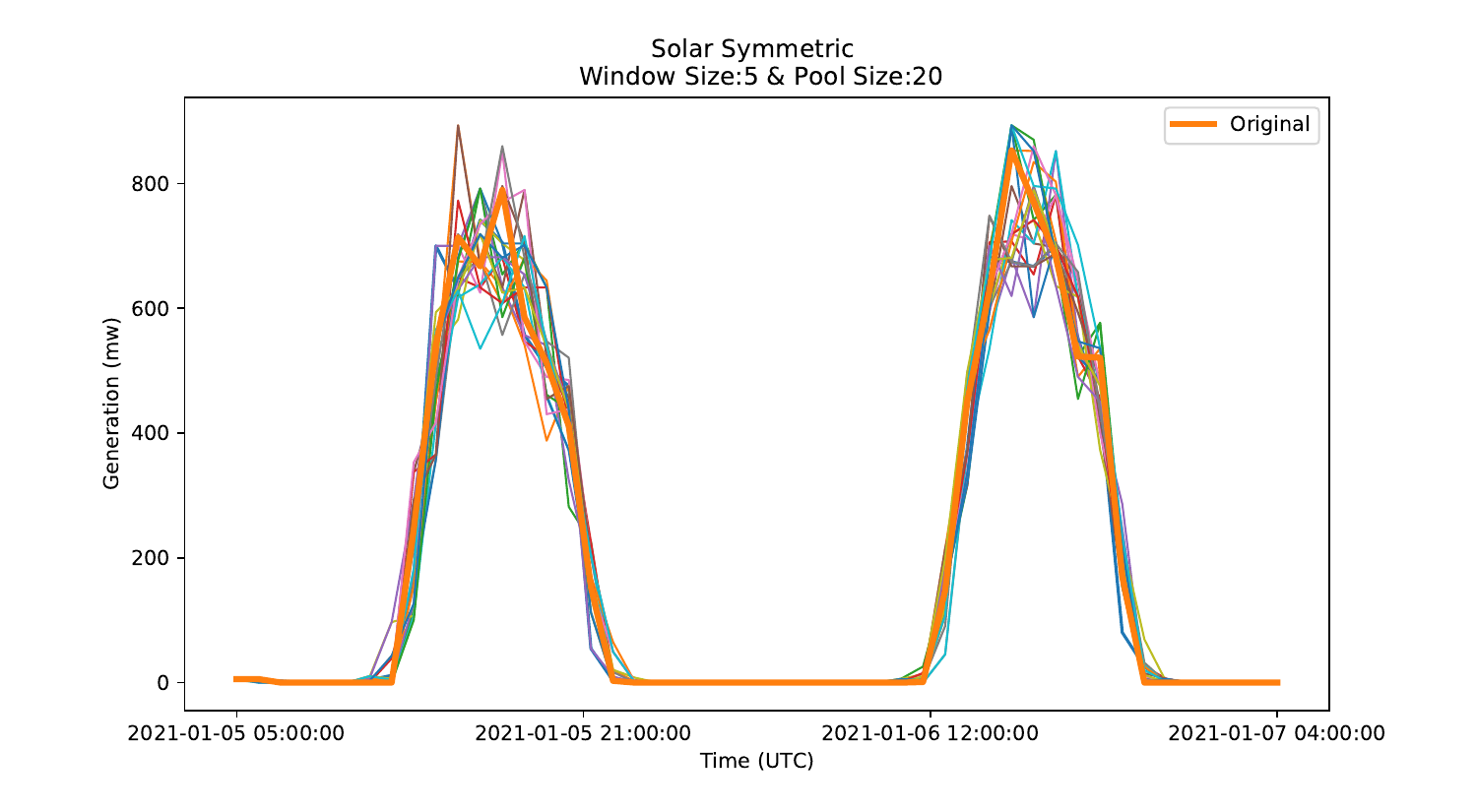}
        \caption{Demonstration of 21 generated time series for solar energy using the SBB method. Window size=5. Pool size=20.}
        \label{fig_solar_symmetric}
    \end{figure}

    \paragraph{Solar} 
    Recalling the NNLB generated series for solar energy, we pleased to report that even the most extreme SBB series for solar do not produce energy at nighttime. This improvement in better capturing the main attributes of the given series can even be seen when the window and pool size parameters are increased. There is  more variability when the parameters increase, but this does not affect the resampled versions' energy generation at night. See Figure \ref{fig_solar_symmetric}. 
    
     \begin{figure}[h!]
        \centering
        \includegraphics[width=\textwidth]{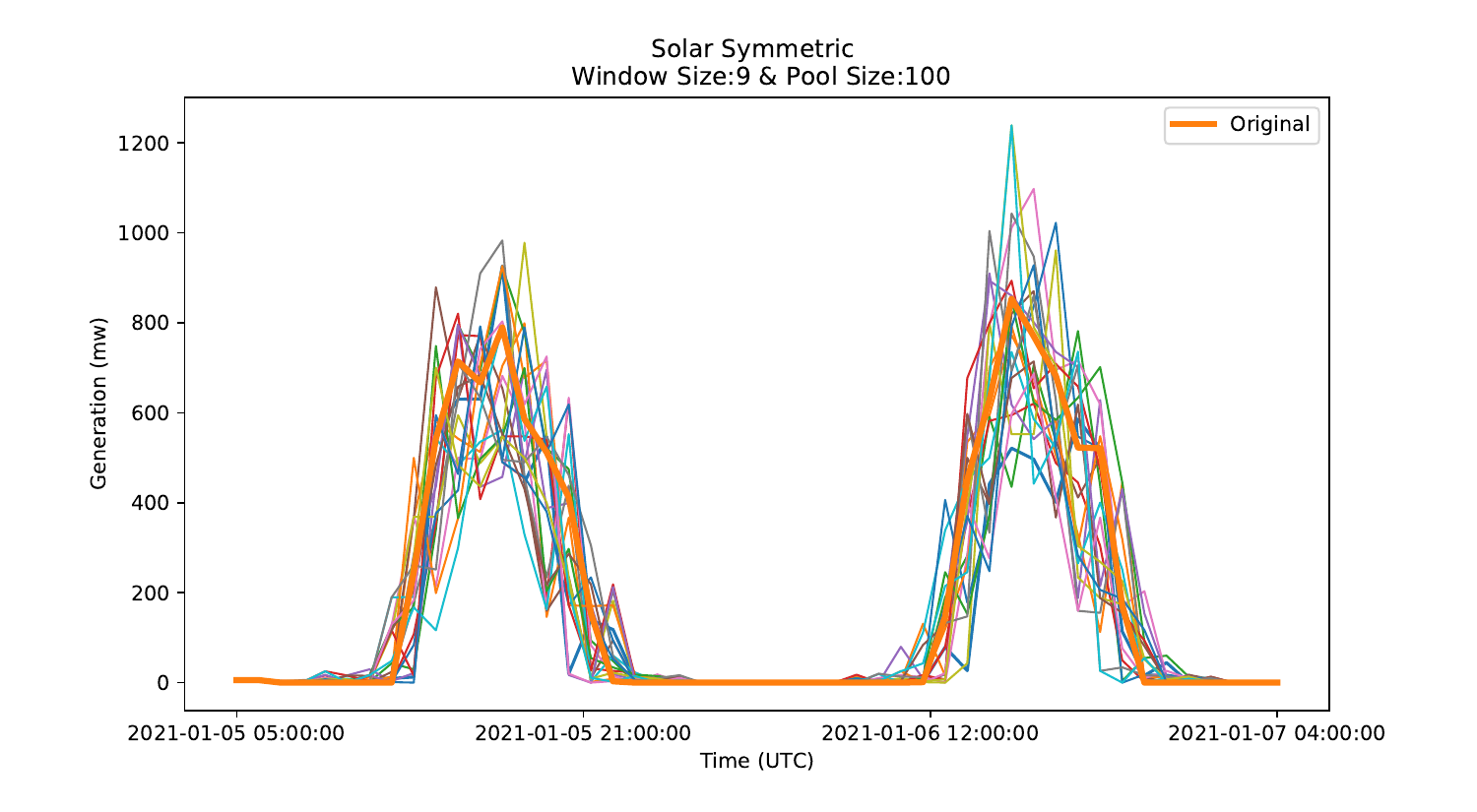}
        \caption{Demonstration of 21 generated time series for solar energy using the SBB method. Window size=9. Pool size=100.}
        \label{fig_solar_symmetric_2}
    \end{figure}
    
\begin{figure}[h!]
        \centering
        \includegraphics[width=0.75\textwidth]{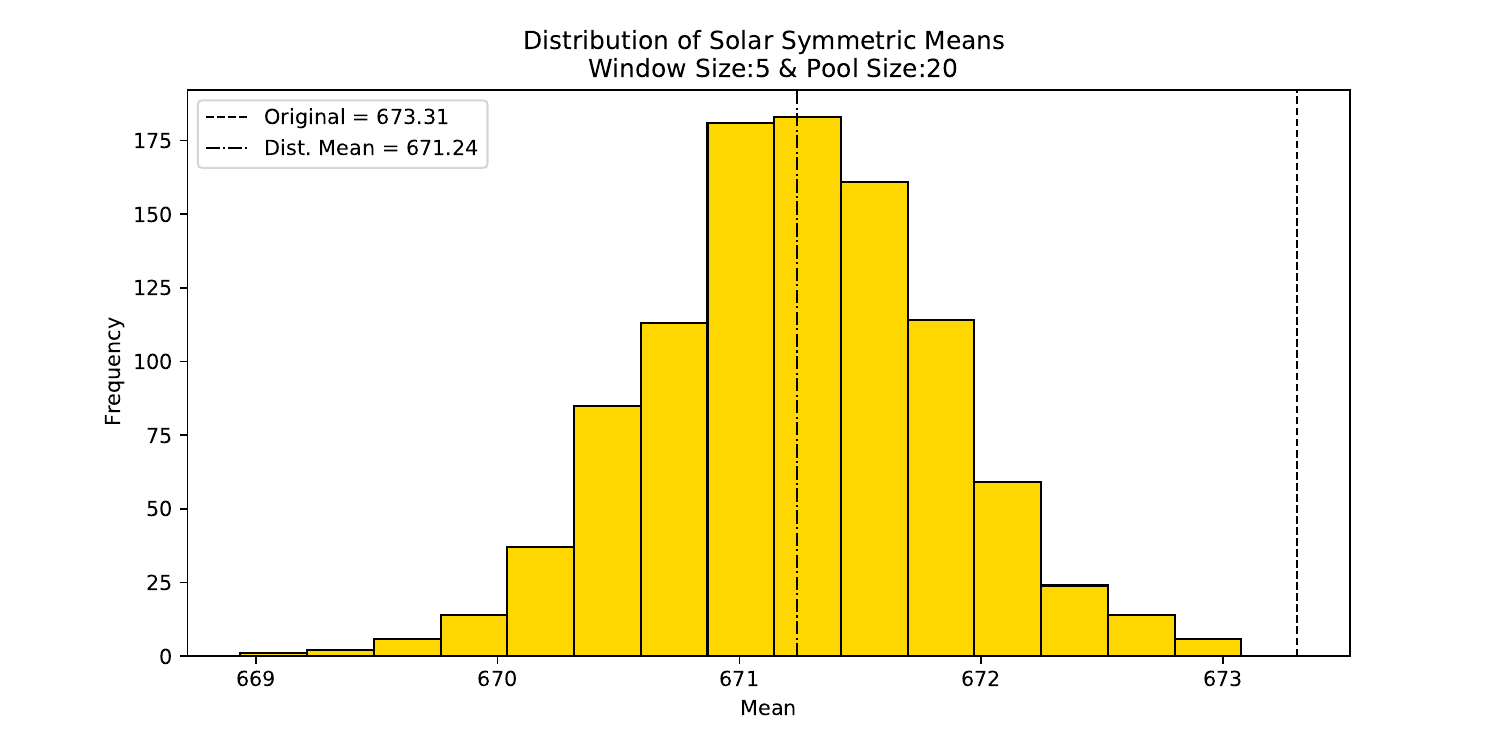}
        \includegraphics[width=0.75\textwidth]{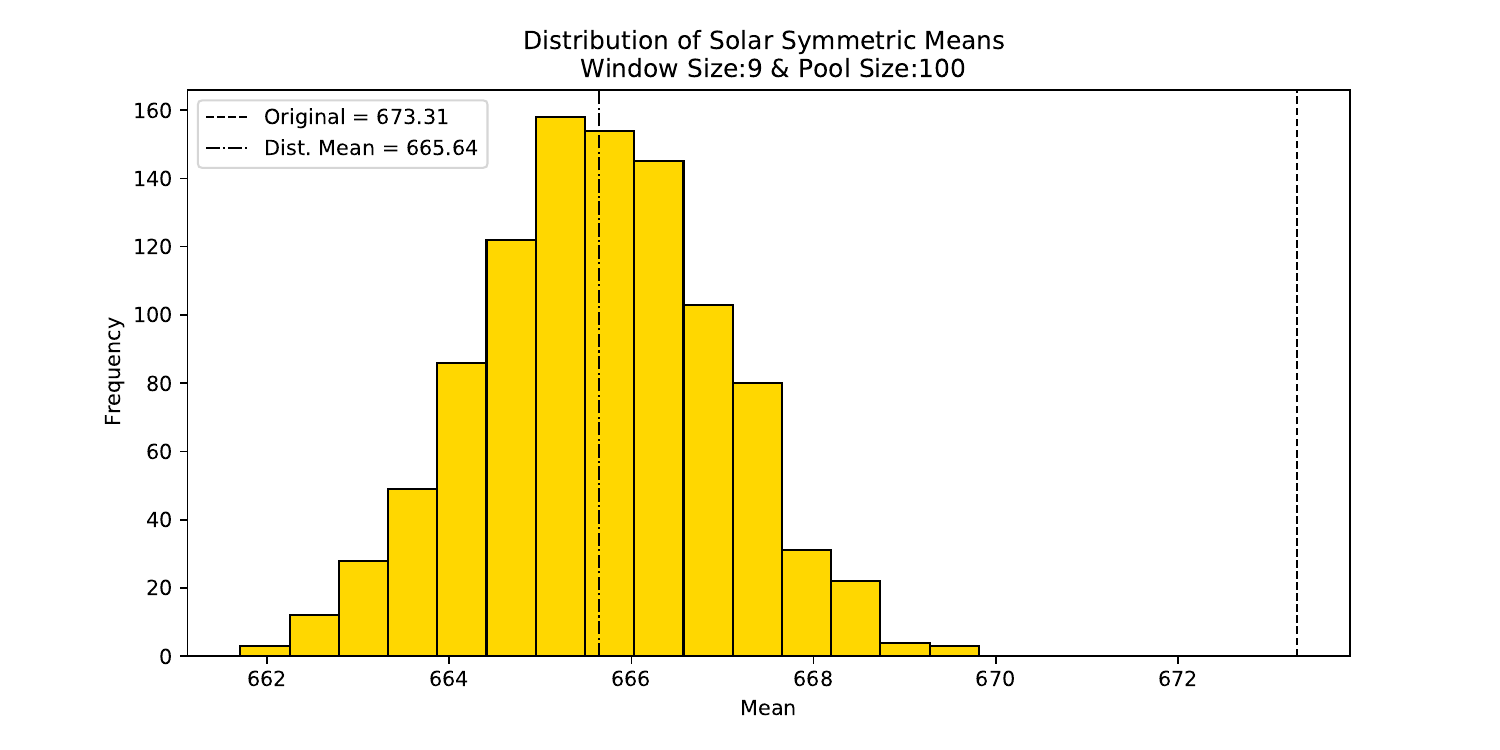}
        \caption{Distribution of symmetric block means for solar.}
        \label{fig_solar_sym_mean_dist}
    \end{figure}

    Even though the SBB was better at capturing the patterns of the original data, as mentioned above, the SBB method has a slight bias, numerically small but statistically significant (hence real). In consequence, the mean of the means of the generated series are consistently  below the original mean. See Figure \ref{fig_solar_sym_mean_dist}.

    Intuitively, this stands to reason because solar energy has quite extreme values. At night  solar energy will be 0 mW (megawatts) and in the daytime it can get to almost 3000 mW. The bias is statistically significant but small in magnitude as a percent of the mean output.
    
    
    \subsubsection*{Statistics Review}
    
    The tables in Table \ref{tab_solar_sym_stats} review the statistics for the 1000 synthetic series we created using the symmetric block method on PJM's 2021 solar energy generation time series. We note the general fidelity of the generated series with the recorded statistics in the tables of the original series.
    
    \begin{table}[h!]
    \centering
    
    Window Size: 5 \& Pool Size: 20
    \begin{tabular}{|l|r|r|r|r|r|r|r|r|}
    \toprule
    Description       &     mean &    std &      min &     25\% &     50\% &     75\% &      max &     Original \\
    \midrule
    Min               &    0.00 &  0.000 &    0.00 &    0.00 &    0.00 &    0.00 &    0.00 &      0.00 \\
    First Quartile    &    2.91 &  0.029 &    2.70 &    2.90 &    2.90 &    2.90 &    3.00 &      2.90 \\
    Median            &   20.72 &  0.956 &   19.30 &   20.00 &   20.20 &   21.25 &   24.45 &     24.80 \\
    Third Quartile    & 1439.56 &  9.774 & 1403.10 & 1432.58 & 1442.50 & 1446.90 & 1462.97 &   1436.78 \\
    Max               & 2903.18 & 21.481 & 2832.50 & 2889.80 & 2906.60 & 2922.50 & 2922.50 &   2922.50 \\
    Mean              &  671.24 &  0.602 &  668.94 &  670.85 &  671.24 &  671.63 &  673.08 &    673.31 \\
    Standard Dev.     &  898.86 &  0.662 &  897.01 &  898.38 &  898.88 &  899.32 &  900.94 &    900.21 \\
    Coeff. of Var.    &    1.339 &  0.001 &    1.336 &    1.339 &    1.339 &    1.340 &    1.342 &     1.337 \\
    Autocorr. Lag: 24 &    0.895 &  0.001 &    0.893 &    0.894 &    0.895 &    0.895 &    0.897 &     0.895 \\
    \bottomrule
    \end{tabular}
    
    Window Size: 9 \& Pool Size: 100
    \begin{tabular}{|l|r|r|r|r|r|r|r|r|}
    \toprule
    Description       &    mean &    std &     min &    25\% &    50\% &    75\% &     max &     Original \\
    \midrule
    Min               &    0.00 &  0.000 &    0.00 &    0.00 &    0.00 &    0.00 &    0.00 &      0.00 \\
    First Quartile    &    2.95 &  0.084 &    2.70 &    2.90 &    2.90 &    3.00 &    3.40 &      2.90 \\
    Median            &   18.72 &  0.449 &   17.60 &   18.30 &   18.65 &   19.10 &   19.90 &     24.80 \\
    Third Quartile    & 1421.79 & 12.746 & 1377.05 & 1412.65 & 1419.90 & 1431.40 & 1459.08 &   1436.78 \\
    Max               & 2906.83 & 19.269 & 2840.30 & 2899.50 & 2906.60 & 2922.50 & 2922.50 &   2922.50 \\
    Mean              &  665.64 &  1.323 &  661.71 &  664.73 &  665.66 &  666.56 &  669.81 &    673.31 \\
    Standard Dev.     &  897.43 &  1.397 &  893.06 &  896.51 &  897.44 &  898.41 &  901.79 &    900.21 \\
    Coeff. of Var.    &    1.348 &  0.002 &    1.340 &    1.347 &    1.348 &    1.350 &    1.355 &     1.337 \\
    Autocorr. Lag: 24 &    0.881 &  0.002 &    0.876 &    0.880 &    0.881 &    0.882 &    0.886 &     0.895 \\
    \bottomrule
    \end{tabular}
    \caption{Distribution of all the statistics calculated for the solar SBB series.}
    \label{tab_solar_sym_stats}
    \end{table}
    
    \newpage
    
\begin{figure}[h!]
        \centering
        \includegraphics[width=0.45\textwidth]{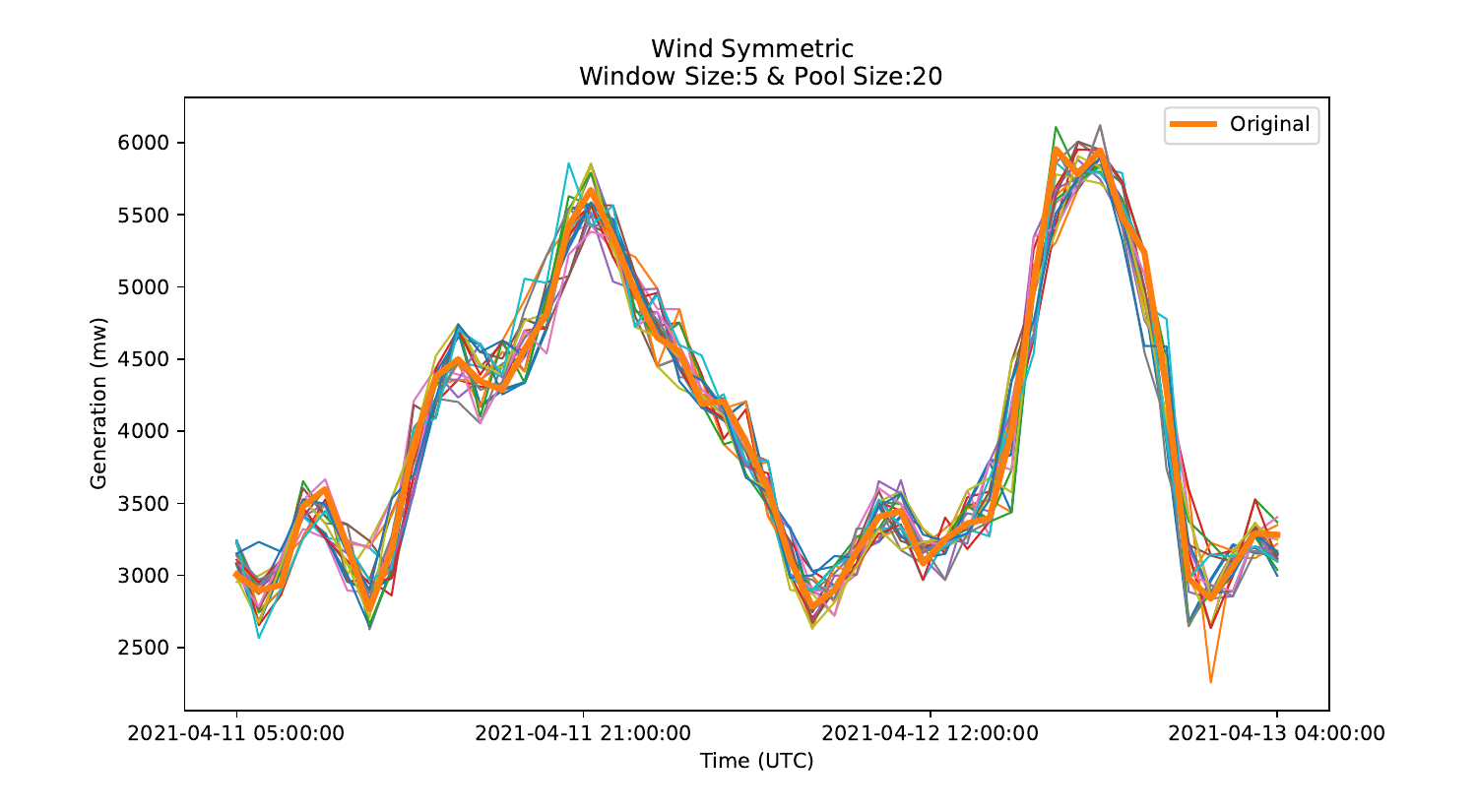}
        \includegraphics[width=0.45\textwidth]{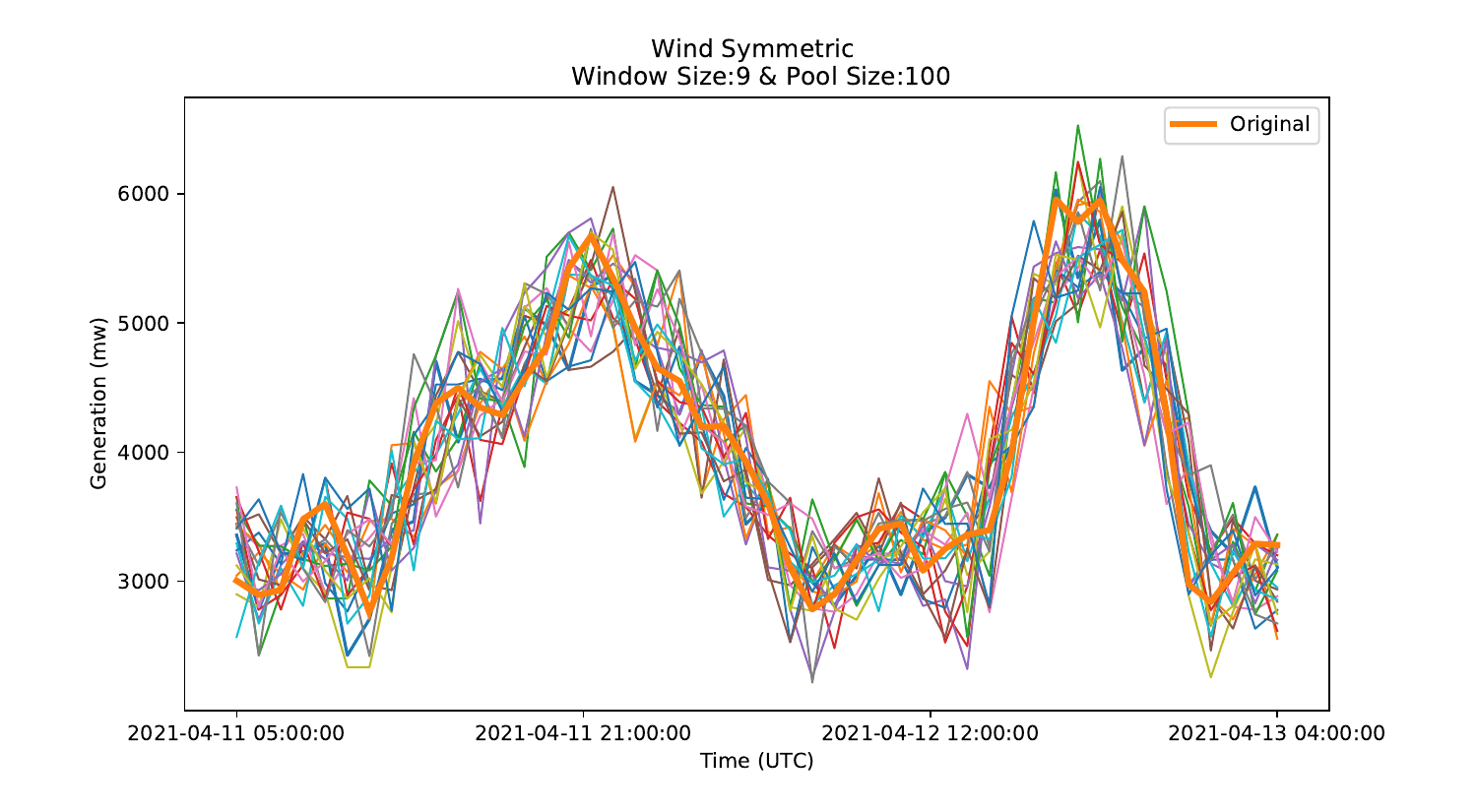}
        \caption{Demonstration of 21 generated time series for wind energy using the SBB method.}
        \label{fig_wind_symmetric}
    \end{figure}

    \paragraph{Wind}
    When we compare Figure \ref{fig_wind_symmetric} (SBB generated series) to Figure \ref{fig_wind_knn} (NNLB generated series), we see that even though both used the same vector of sizes of 5 and 9 when finding the most similar points, the symmetric block series has less variability compared to the NNLB series. 
    
\begin{figure}[ht!]
        \centering
        \includegraphics[width=0.45\textwidth]{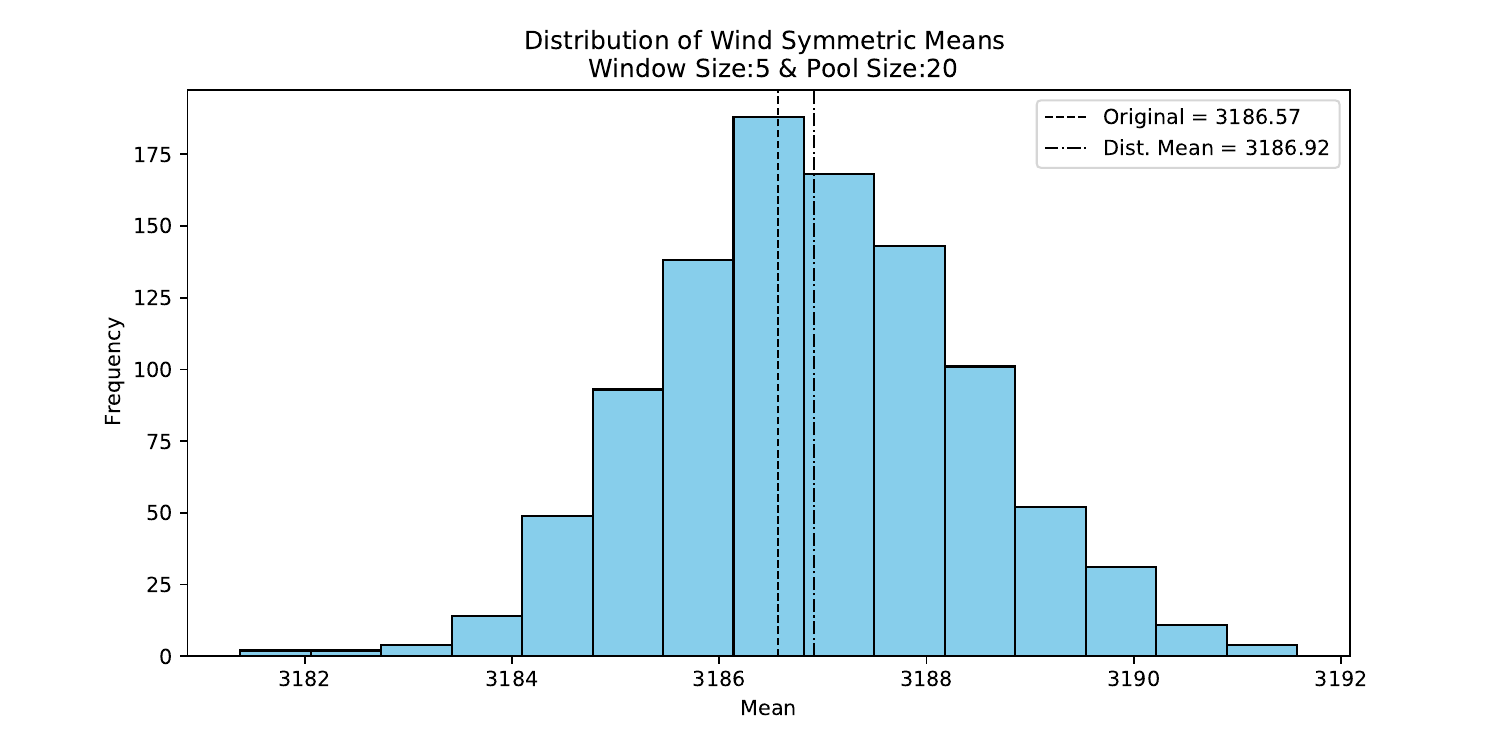}
        \includegraphics[width=0.45\textwidth]{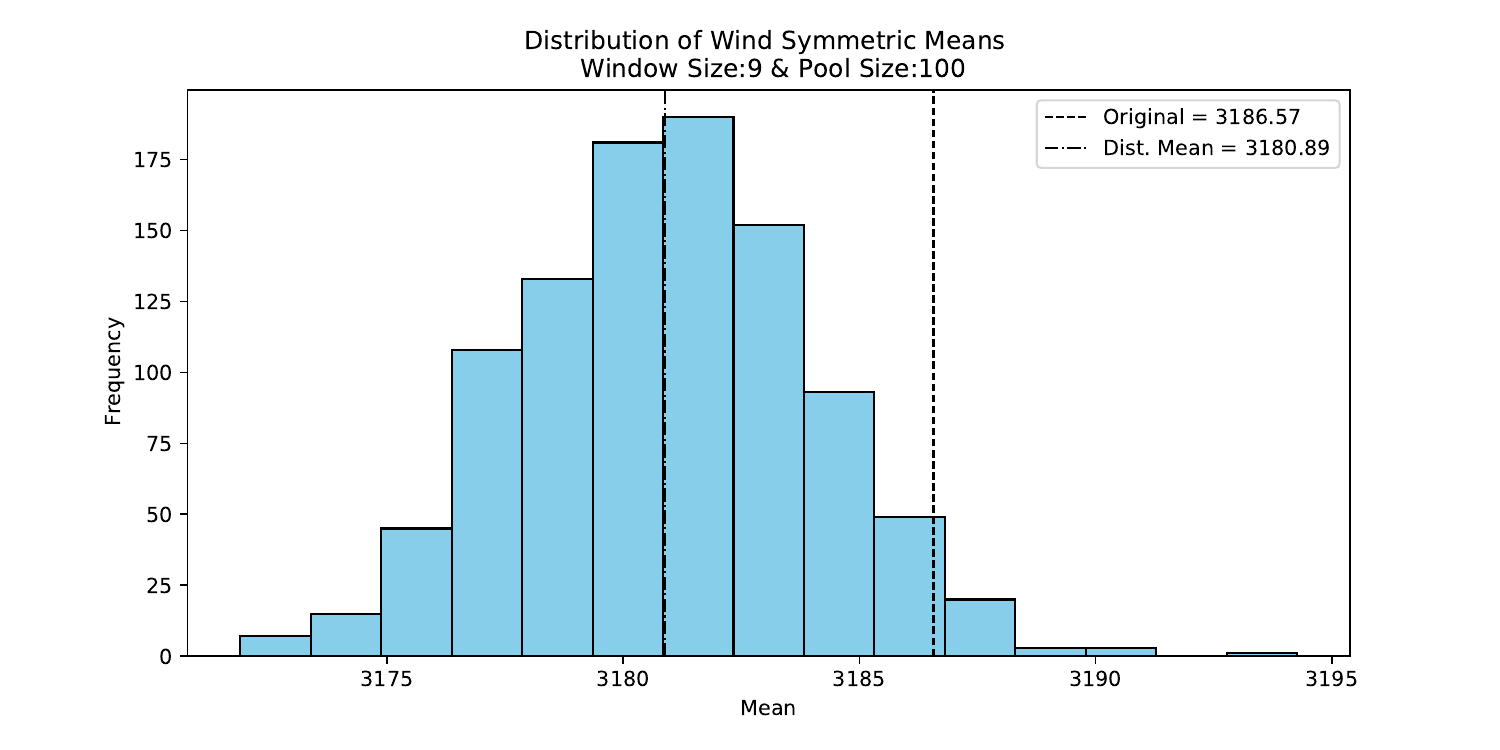}
        \caption{Distribution of symmetric block means for wind.}
        \label{fig_wind_sym_mean_dist}
    \end{figure}

    Also, unlike solar energy, symmetric blocks generated with a smaller window and pool size for wind energy yielded means that closely centered around the original dataset's mean. With a larger window and pool size, the symmetric block means begin to fall below the original mean. See Figure \ref{fig_wind_sym_mean_dist}.
    This is due to the median wind generation energy being less than the mean in the original time series. This means there are more extreme high values in wind generation. There are hours in the year where wind can generate close to 9000 MW. 
    
    
        \subsubsection*{Statistics Review}
    
    The tables in Table \ref{tab_wind_sym_stats} review the statistics for the 1000 synthetic series we created using the symmetric block method on PJM's 2021 wind energy generation time series. We note the general fidelity of the generated series with the recorded statistics in the tables of the original series.
    \begin{table}[h!]
    \centering
    
    Window Size: 5 \& Pool Size: 20
    
    \begin{tabular}{|l|r|r|r|r|r|r|r|r|}
    \toprule
    Description       &    mean &    std &     min &    25\% &    50\% &    75\% &     max &  Original \\
    \midrule
    Min               &   76.59 & 14.445 &   63.30 &   63.30 &   76.60 &   81.10 &  125.00 &     63.30 \\
    First Quartile    & 1375.60 &  4.197 & 1357.20 & 1373.67 & 1376.45 & 1377.30 & 1388.20 &   1372.90 \\
    Median            & 2746.18 &  7.691 & 2725.55 & 2740.35 & 2745.55 & 2751.15 & 2767.90 &   2739.50 \\
    Third Quartile    & 4738.10 & 11.509 & 4694.75 & 4730.05 & 4740.50 & 4746.10 & 4764.90 &   4743.15 \\
    Max               & 8965.83 & 34.865 & 8746.20 & 8973.70 & 8973.70 & 8990.00 & 8990.00 &   8990.00 \\
    Mean              & 3186.92 &  1.485 & 3181.38 & 3185.93 & 3186.87 & 3187.88 & 3191.57 &   3186.57 \\
    Standard Dev.     & 2132.66 &  1.451 & 2127.49 & 2131.72 & 2132.67 & 2133.64 & 2137.06 &   2138.40 \\
    Coeff. of Var.    &   0.669 &  0.000 &   0.668 &   0.669 &   0.669 &   0.669 &   0.670 &     0.671 \\
    Autocorr. Lag: 24 &   0.439 &  0.001 &   0.437 &   0.438 &   0.439 &   0.440 &   0.442 &     0.437 \\
    \bottomrule
    \end{tabular}
    
    Window Size: 9 \& Pool Size: 100
    
    \begin{tabular}{|l|r|r|r|r|r|r|r|r|}
    \toprule
    Description       &    mean &    std &     min &    25\% &    50\% &    75\% &     max &  Original \\
    \midrule
    Min               &   83.17 & 19.591 &   63.30 &   63.30 &   76.60 &  102.10 &  152.60 &     63.30 \\
    First Quartile    & 1375.21 &  7.302 & 1348.83 & 1369.10 & 1376.30 & 1380.58 & 1392.60 &   1372.90 \\
    Median            & 2751.75 & 10.671 & 2721.60 & 2743.20 & 2751.15 & 2759.61 & 2780.80 &   2739.50 \\
    Third Quartile    & 4735.99 & 15.582 & 4685.18 & 4725.30 & 4740.27 & 4747.00 & 4772.60 &   4743.15 \\
    Max               & 8942.20 & 64.301 & 8633.60 & 8906.00 & 8973.70 & 8990.00 & 8990.00 &   8990.00 \\
    Mean              & 3180.89 &  3.074 & 3171.89 & 3178.75 & 3180.92 & 3182.91 & 3194.27 &   3186.57 \\
    Standard Dev.     & 2119.73 &  2.876 & 2110.41 & 2117.88 & 2119.73 & 2121.64 & 2130.02 &   2138.40 \\
    Coeff. of Var.    &   0.666 &  0.001 &    0.663 &    0.666 &    0.666 &    0.667 &    0.670 &     0.671 \\
    Autocorr. Lag: 24 &   0.443 &  0.002 &    0.436 &    0.442 &    0.443 &    0.445 &    0.450 &     0.437 \\
    \bottomrule
    \end{tabular}
    
    \caption{Distribution of all the statistics calculated for the wind SBB series.}
    \label{tab_wind_sym_stats}
    \end{table}
    
    \paragraph{Load}  
    \begin{figure}[h!]
        \centering
        \includegraphics[width=0.45\textwidth]{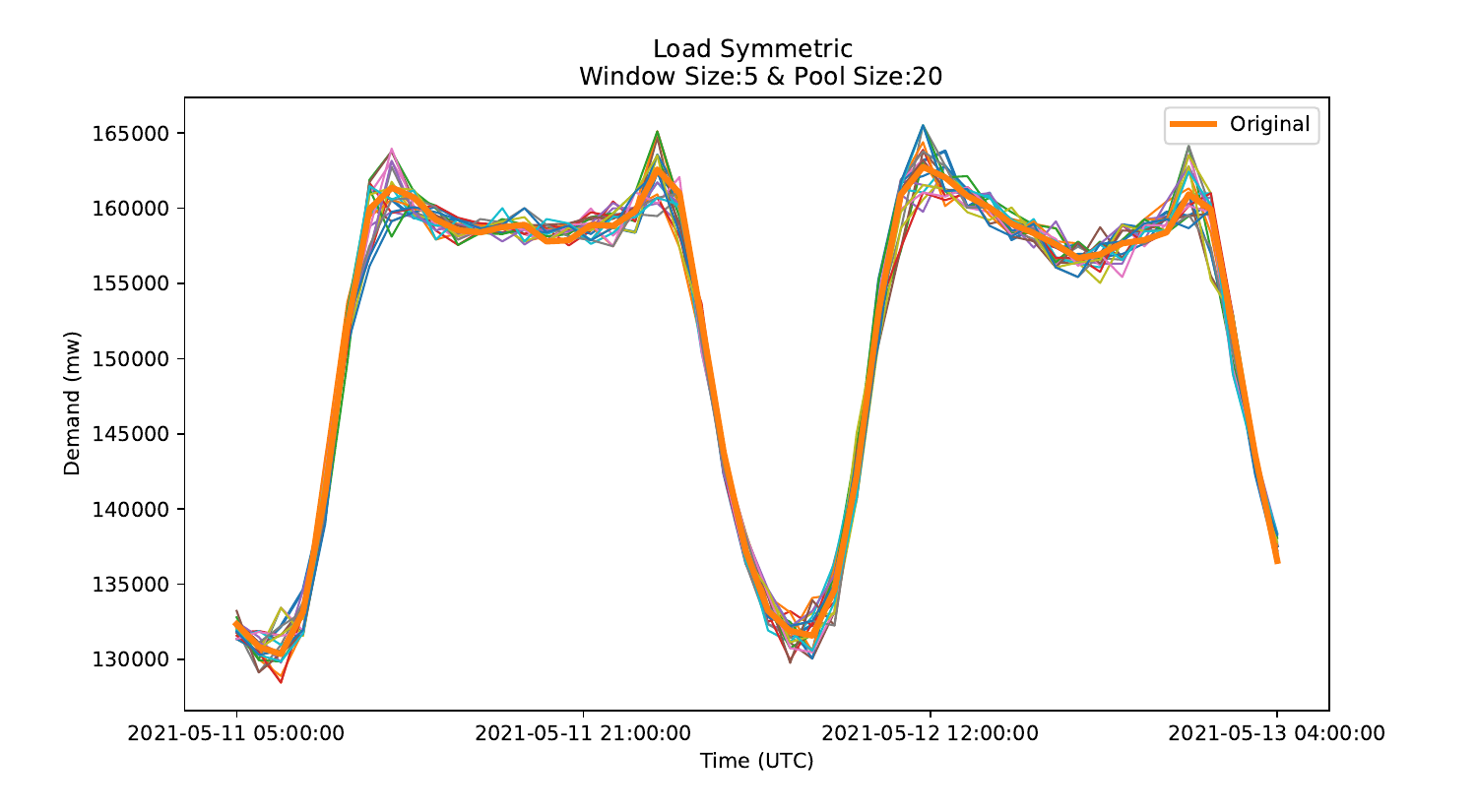}
        \includegraphics[width=0.45\textwidth]{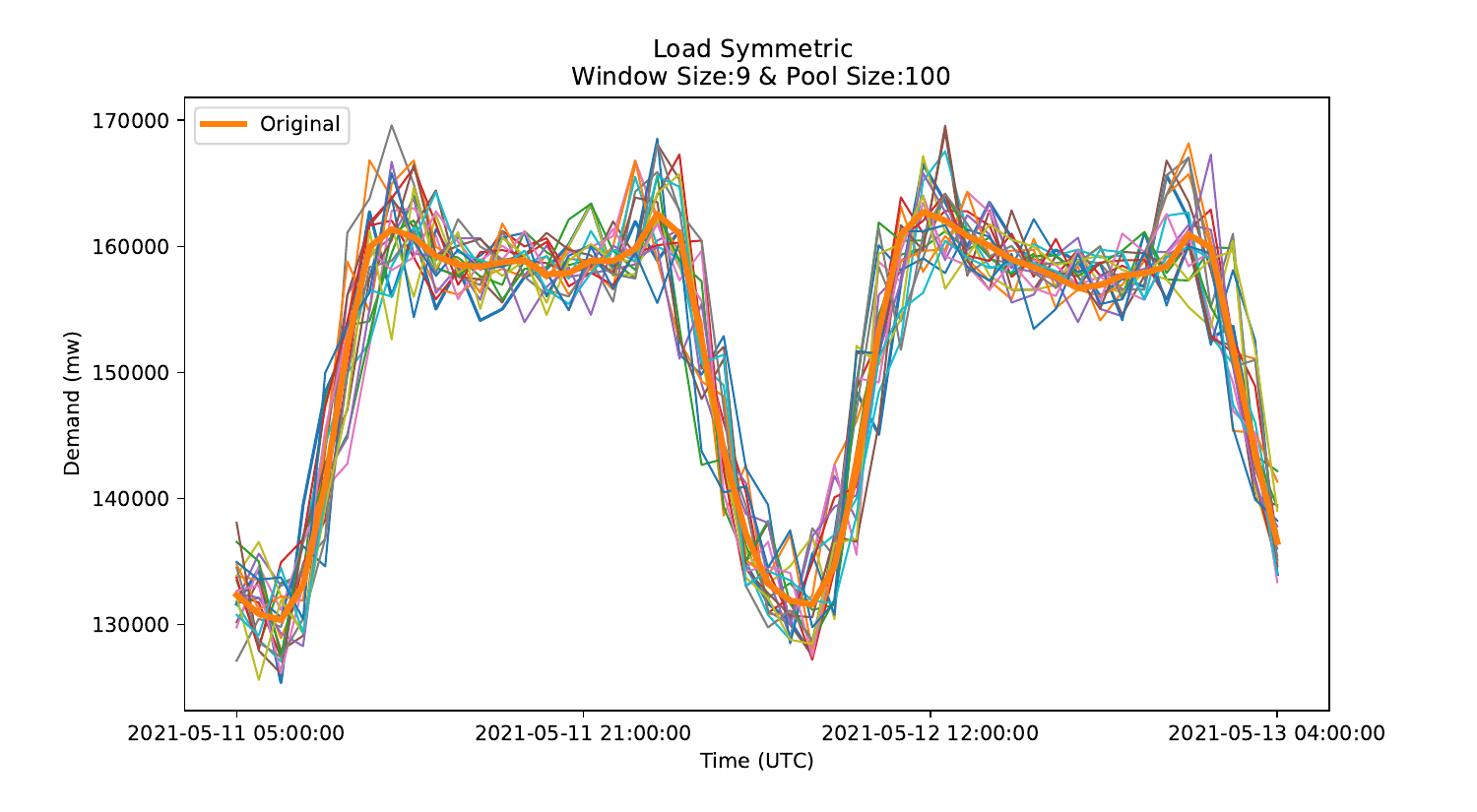}
         \caption{Demonstration of 21 generated time series for load energy using the SBB method.}
        \label{fig_load_symmetric}
    \end{figure}

\begin{figure}[h!]
        \centering
        \includegraphics[width=0.45\textwidth]{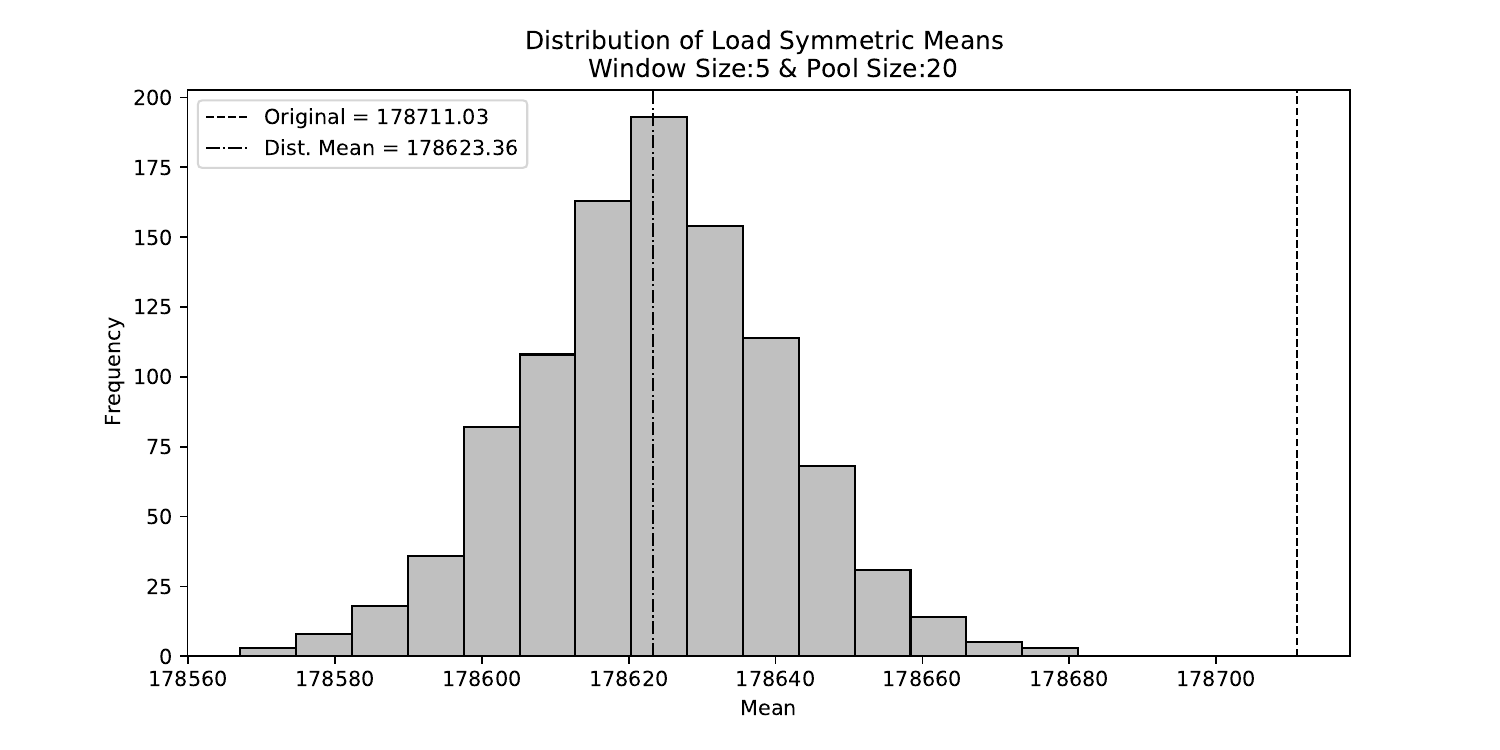}
        \includegraphics[width=0.45\textwidth]{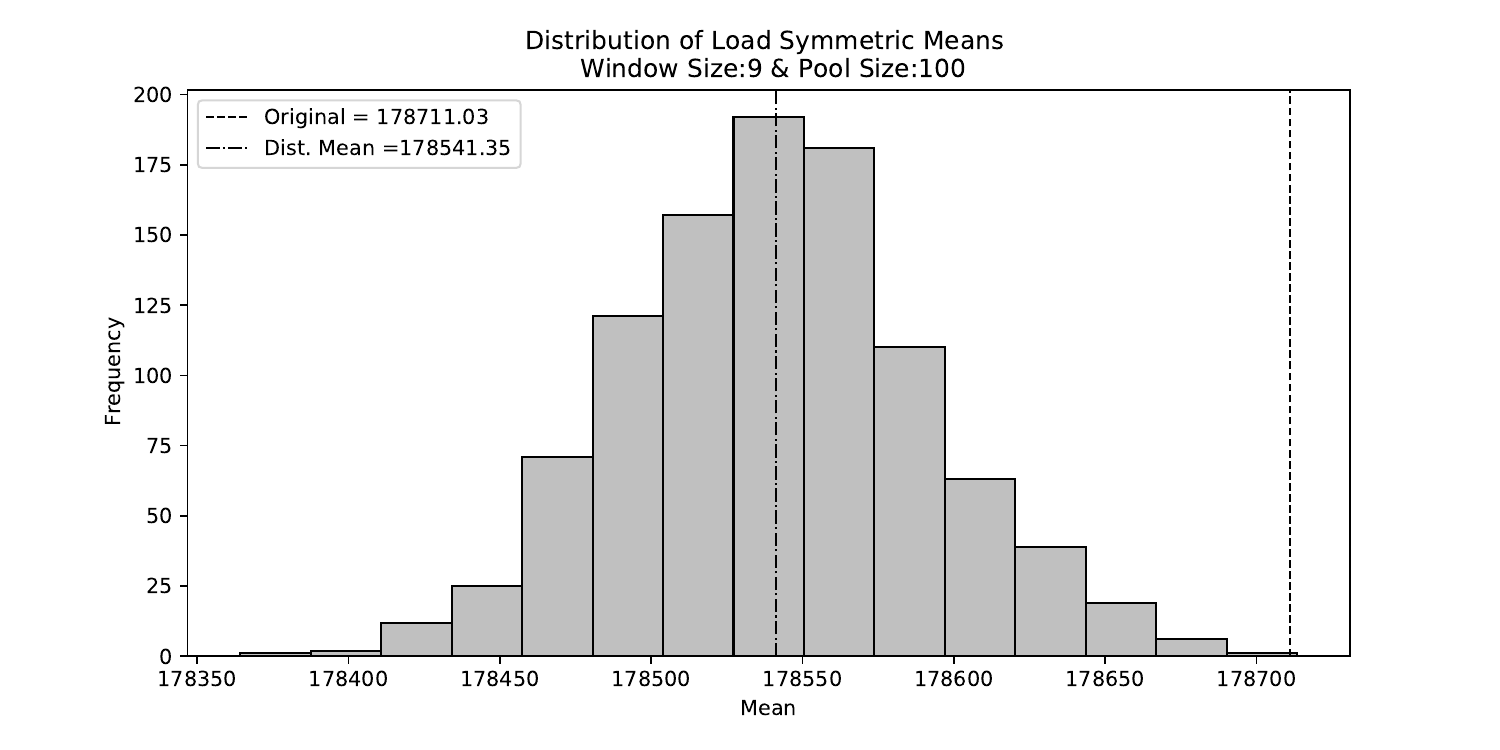}
        \caption{Distribution of symmetric block means for load.}
        \label{fig_load_sym_mean_dist}
    \end{figure}

    Figure \ref{fig_load_symmetric} plots generated series against the original PJM data for load. As we saw for wind and solar, the generated series do well at roughly tracking the original data, with greater variation evident for the larger window size.
     Time series for load  created using the symmetric block method have averages that fall below the mean of PJM's original time series, with  the magnitude of expected shortfall reminiscent of that for solar. See Figure  \ref{fig_load_sym_mean_dist}.

    We also still see the  symmetric block data created with the larger window and pool sizes having averages even further from the original's mean. It also appears that variation in the means is also wider with larger window and pool sizes as it is even possible for some of these series to achieve the original's mean, which was not possible with the window size of 5 and pool size of 20.
    
    With a larger pool size, there will be more points that are considered, leading to more variation among the symmetric blocks. This is demonstrated through the 48-hour period snapshots of each energy type.\footnote{Full year plots are not readable. Full datasets and visuals can be reproduced using the provided Python code in the supplemental material.}

Even with an indication of bias, the bias itself is relatively small compared to the scope of these series, so our purpose in this paper can still be accomplished without significant impact from the bias.
    
    \begin{table}[h!]
    \centering
    
    Window Size: 5 \& Pool Size: 20
    \begin{tabular}{|l|r|r|r|r|r|r|r|r|}
    \toprule
    Description       &   mean &      std &    min &   25\% &   50\% &   75\% &    max &  Original \\
    \midrule
    Min               & 117792 &  575.075 & 117370 & 117370 & 117428 & 118554 & 118884 &    117370 \\
    First Quartile    & 156356 &  101.776 & 156046 & 156294 & 156348 & 156427 & 156597 &    156350 \\
    Median            & 173966 &   95.897 & 173652 & 173889 & 173949 & 174047 & 174216 &    174063 \\
    Third Quartile    & 197011 &  108.330 & 196685 & 196926 & 197008 & 197084 & 197419 &    197085 \\
    Max               & 297073 &  924.985 & 292916 & 297495 & 297495 & 297541 & 297541 &    297541 \\
    Mean              & 178623 &   17.311 & 178567 & 178612 & 178624 & 178634 & 178681 &    178711 \\
    Standard Dev.     &  32166 &   22.880 &  32093 &  32151 &  32165 &  32181 &  32248 &     32242 \\
    Coeff. of Var.    &  0.180 &    0.000 &  0.179 &  0.179 &  0.180 &  0.180 &  0.180 &     0.180 \\
    Autocorr. Lag: 24 &  0.909 &    0.000 &  0.908 &  0.909 &  0.909 &  0.910 &  0.910 &     0.910 \\
    \bottomrule
    \end{tabular}
    
    Window Size: 9 \& Pool Size: 100
    \begin{tabular}{|l|r|r|r|r|r|r|r|r|}
    \toprule
    Description       &   mean &      std &    min &   25\% &   50\% &   75\% &    max &  Original \\
    \midrule
    Min               & 117909 &  625.407 & 117370 & 117370 & 117428 & 118554 & 119183 &  117370 \\
    First Quartile    & 156612 &  160.768 & 156143 & 156536 & 156596 & 156682 & 157206 &  156350 \\
    Median            & 173940 &  149.980 & 173442 & 173834 & 173915 & 174053 & 174358 &  174063 \\
    Third Quartile    & 196969 &  172.087 & 196404 & 196857 & 196950 & 197056 & 197564 &  197085 \\
    Max               & 296844 & 1229.443 & 290679 & 296432 & 297495 & 297541 & 297541 &  297541 \\
    Mean              & 178541 &   49.820 & 178364 & 178506 & 178542 & 178572 & 178713 &  178711 \\
    Standard Dev.     &  31829 &   67.785 &  31642 &  31781 &  31831 &  31873 &  32061 &   32242 \\
    Coeff. of Var.    &  0.178 &    0.000 &  0.177 &  0.178 &  0.178 &  0.179 &  0.180 &   0.180 \\
    Autocorr. Lag: 24 &  0.892 &    0.001 &  0.888 &  0.891 &  0.892 &  0.892 &  0.894 &   0.910 \\
    \bottomrule
    \end{tabular}
    
    \caption{Distribution of all the statistics calculated for the load SBB series.}
    \label{tab_load_sym_stats}
    \end{table}

\subsubsection*{Statistics Review}
    
    The tables in Table \ref{tab_load_sym_stats} review the statistics for the 1000 synthetic series we created using the symmetric block method on PJM's 2021 load  time series. We note the general fidelity of the generated series with the recorded statistics in the tables of the original series.

\subsubsection{Under-Performance Analysis}

The results above serve to demonstrate that the SBB method, like the NNLB method, plausibly replicates the distribution of the original series it samples. As seen, there is a small amount of bias, but overall the generated series have characteristic statistics that conform well to the original data and pass the ``ocular'' test: they look like the original series, but there is variation.

For our purposes the variation we are most interested in is under-performance that would need to be compensated in some way, whether with storage, load reduction or some other method. Such under-performance can be measured in a number of ways, depending on the exact application. Here, we focus on just one representative measure in order to illustrate what the methods can do. Specifically, we examine under-performance over a period of a year as the number of 24 hour periods (days) in which generation in the synthetic data is at least $X$\% (we use $X$=5) below that of the source data at the corresponding time. See \S\ref{sec_Bootstrap} for our formal description of underage and overage.

\begin{figure}[h!]
    \centering
    \includegraphics[width=0.45\textwidth]{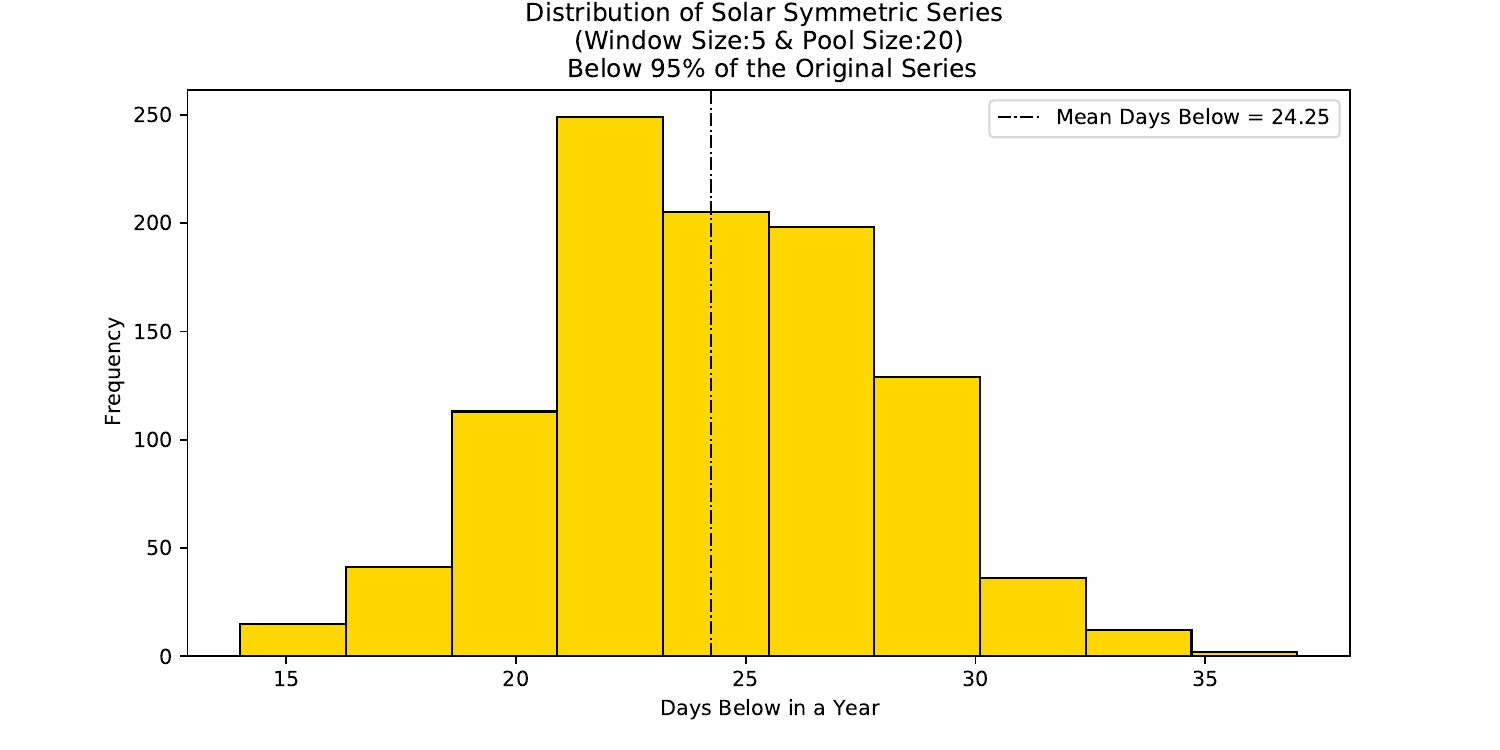}
    \includegraphics[width=0.45\textwidth]{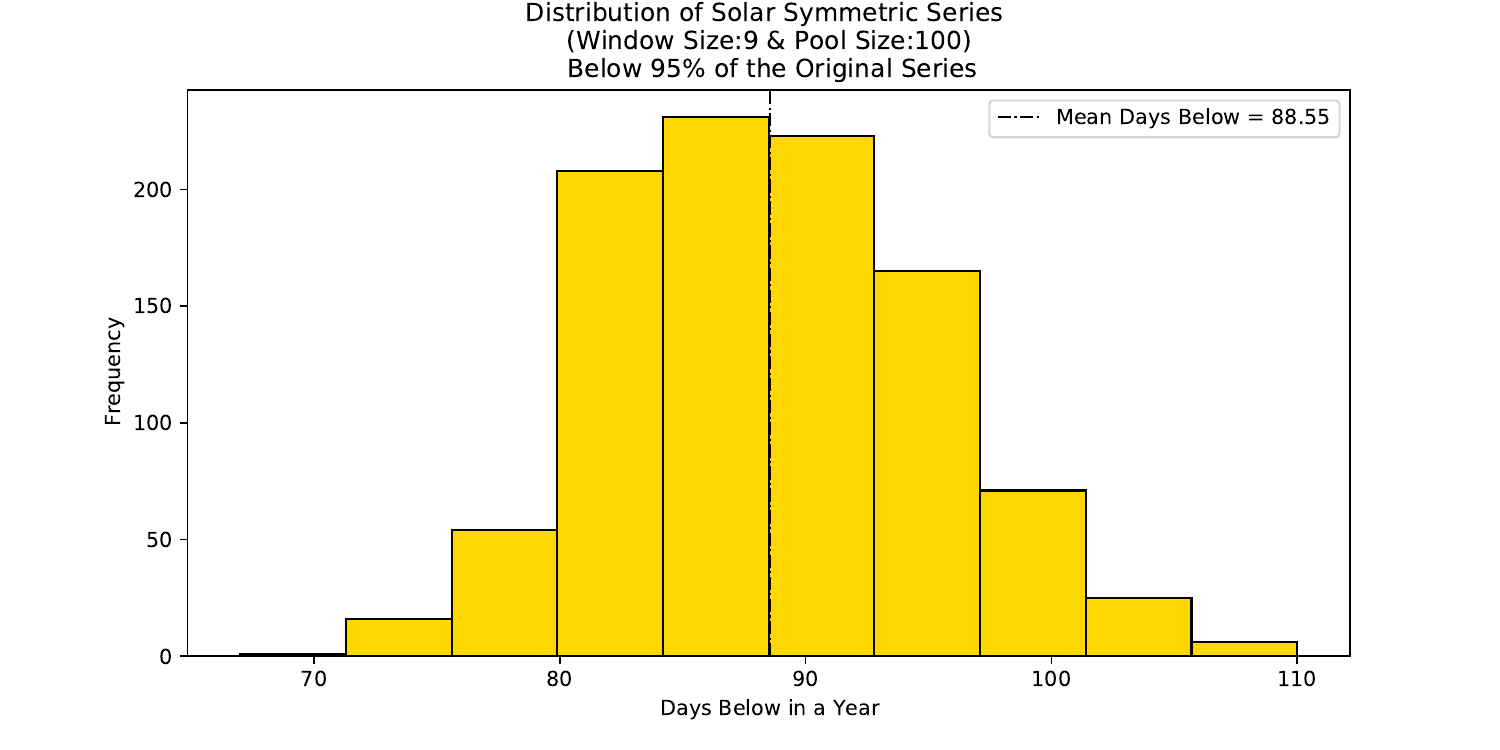}
    \caption{Distribution of the number of days from 1000 solar symmetric series that fall below 95\% of the original PJM data for solar energy.}
    \label{fig_solar_sym_extreme_below}
\end{figure}

Our basic question comes down to this specifically: Given a generated sample of alternate time series, what is the distribution of 24-hour (one day) underage? Looking at solar first, a generated sample of 1000 series, a window size of 5, pool size of 20, Figure \ref{fig_solar_sym_extreme_below} shows the resulting experimental distribution.

We see in Figure \ref{fig_solar_sym_extreme_below} that there is a fair amount of variability in the generated series. On average, compared to the daily solar production in the original, source series, there are about 25 days per year in the generated series in which solar production for the corresponding day is less that 95\% of the original, with a noticeable 30 or more days on the high (i.e., low) end. This suggests that there is a fair chance that another year of generation drawn from the same distribution would have 30 or so drought days. How significant this is operationally is of course conditioned on specific circumstances. We leave this analysis (mostly) to future work.

\begin{figure}[h!]
    \centering
    \includegraphics[width=0.45\textwidth]{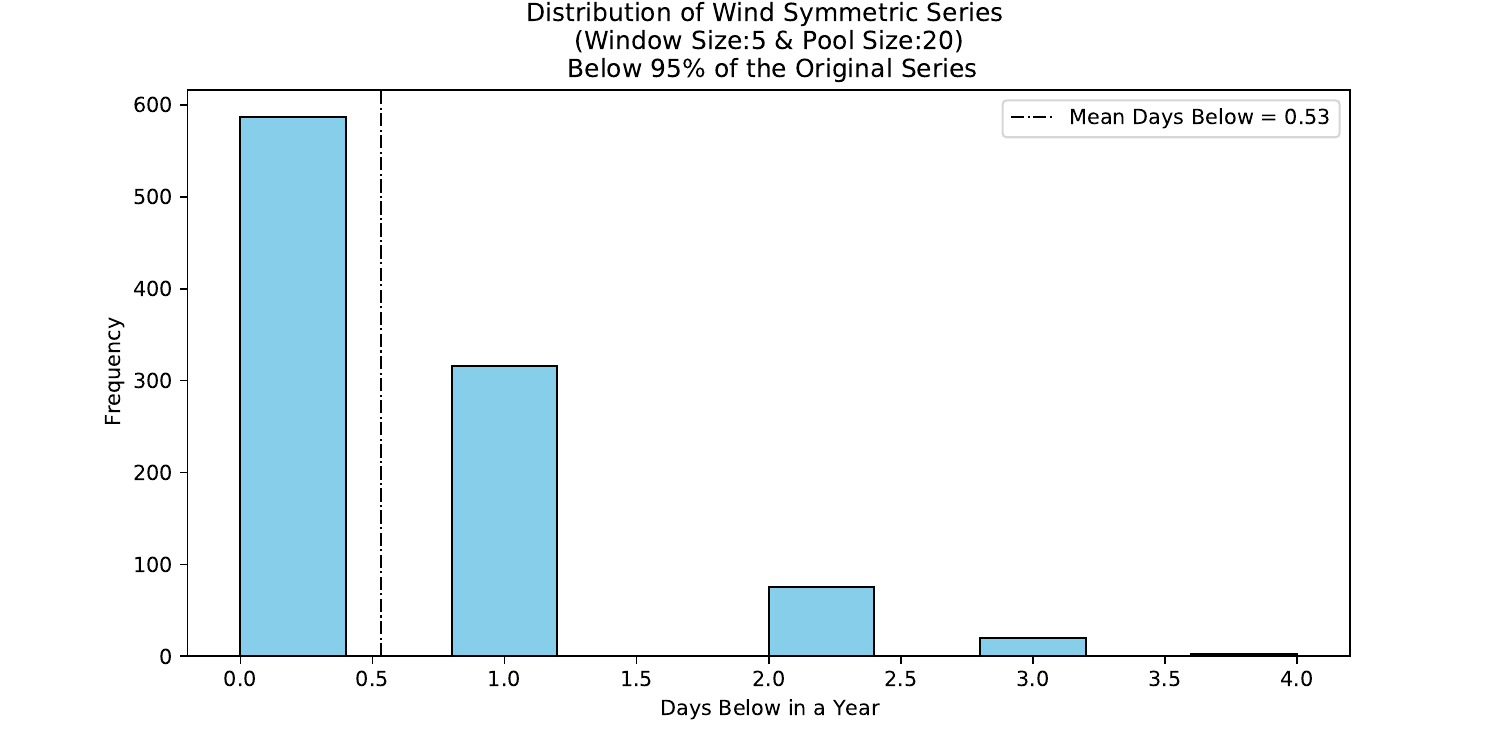}
    \includegraphics[width=0.45\textwidth]{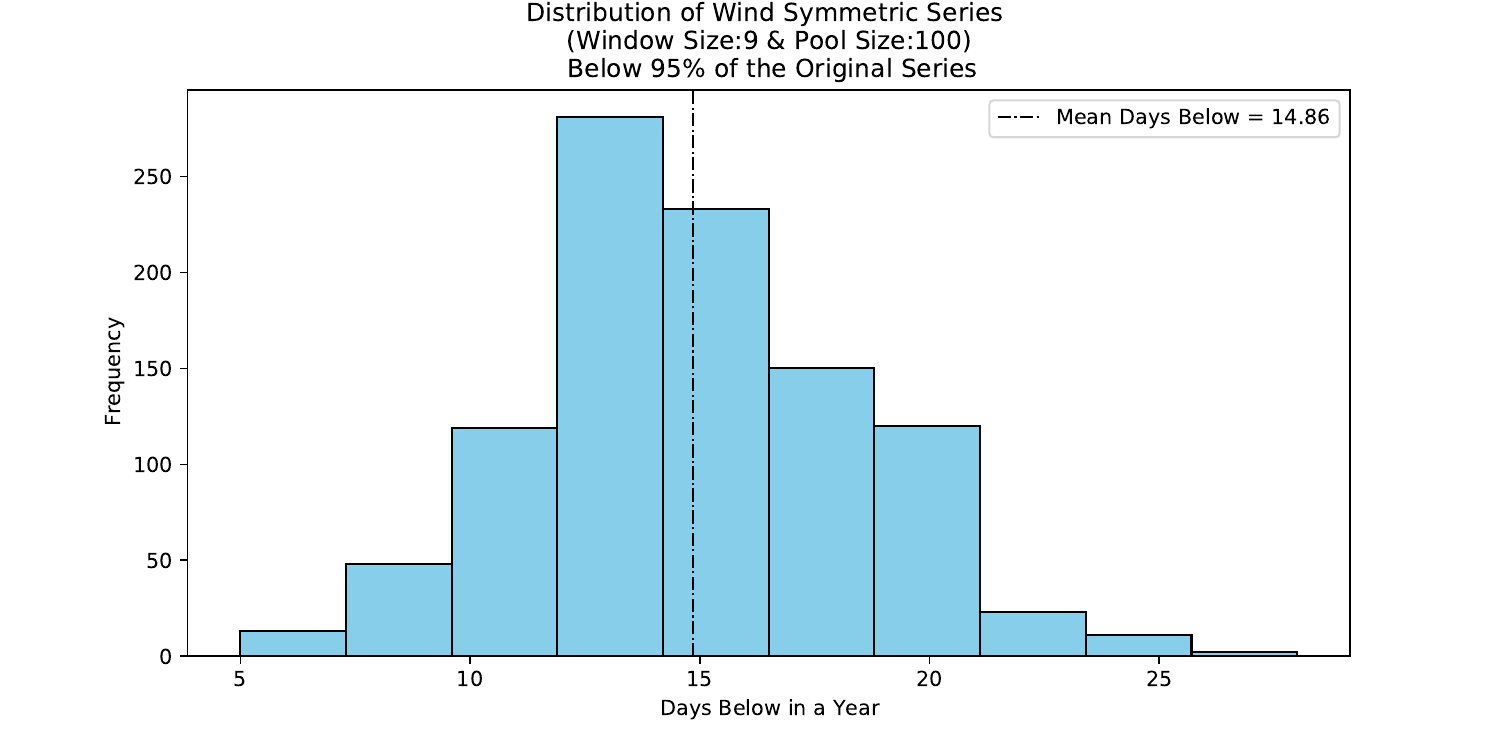}
    \caption{Distribution of the number of days from 1000 wind symmetric series that fall below 95\% of the original PJM data for wind energy.}
    \label{fig_wind_sym_extreme_below}
\end{figure}

Looking at wind with the same parameters, see Figure \ref{fig_wind_sym_extreme_below}.
What we see here is much less variability than is the case with solar. Instead of about 25 days of underage on average, with wind we have about half a day and on the high end 2 or 3. This is surprising in some sense, but is virtually implied by the comparative coefficient of variation, which is 0.67 for wind and 1.34 for solar (see above). If  we change to a window size of 9 (sash 4) and a pool of 100, then the mean for wind is about 15 days, while for solar it is about 89 days. So, the amount of variability to use is in part up to the analyst. In our data, wind does look much less variable than solar, by the present measure and by comparing their coefficient of variations. See the supplemental materials for a more extensive series of graphs and reports.

As a note, load was not showcased because both sets of parameters yielded no underage of 5\%. The explanation follows similar reasoning laid out in \S\ref{sec_NNLB}.

\begin{figure}[h!]
    \centering
    \includegraphics[width=0.45\textwidth]{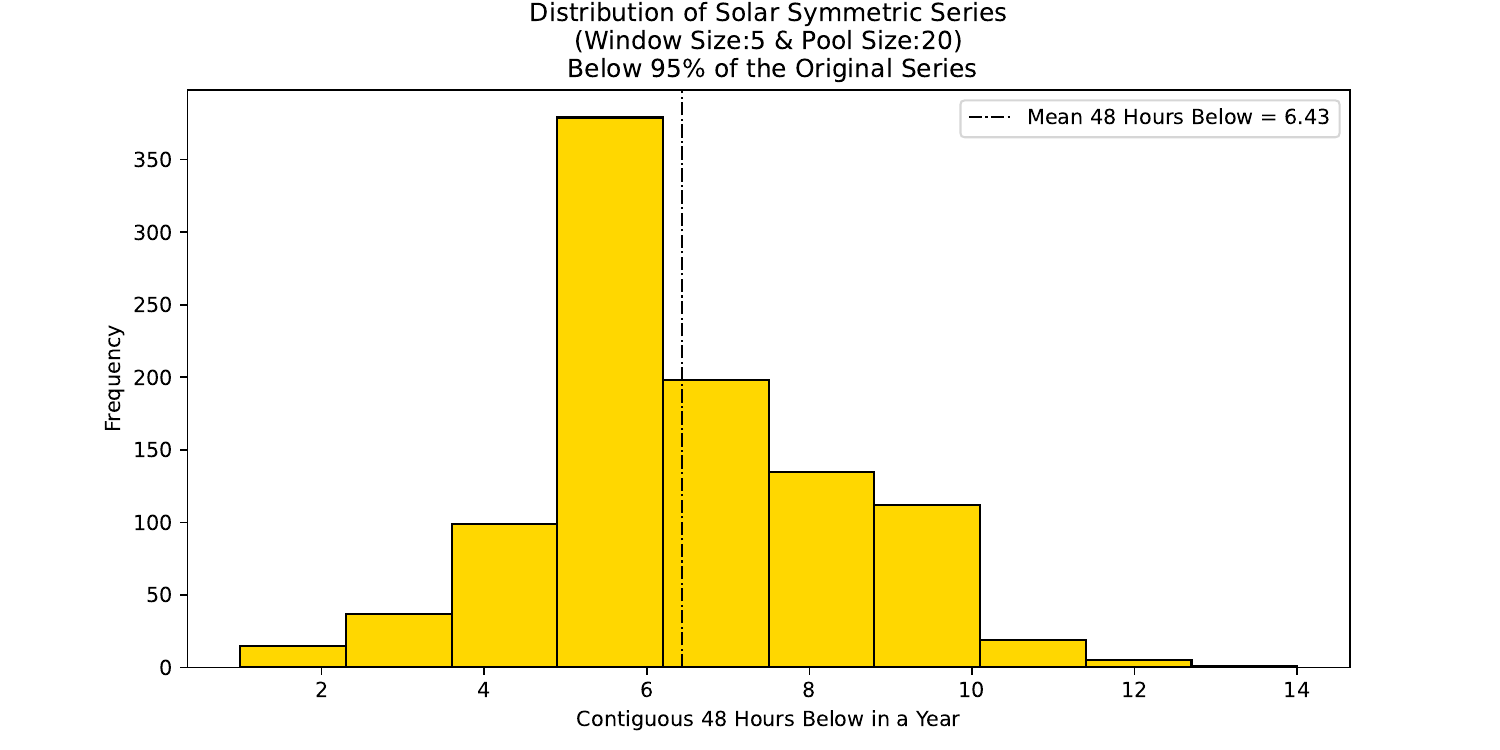}
    \includegraphics[width=0.45\textwidth]{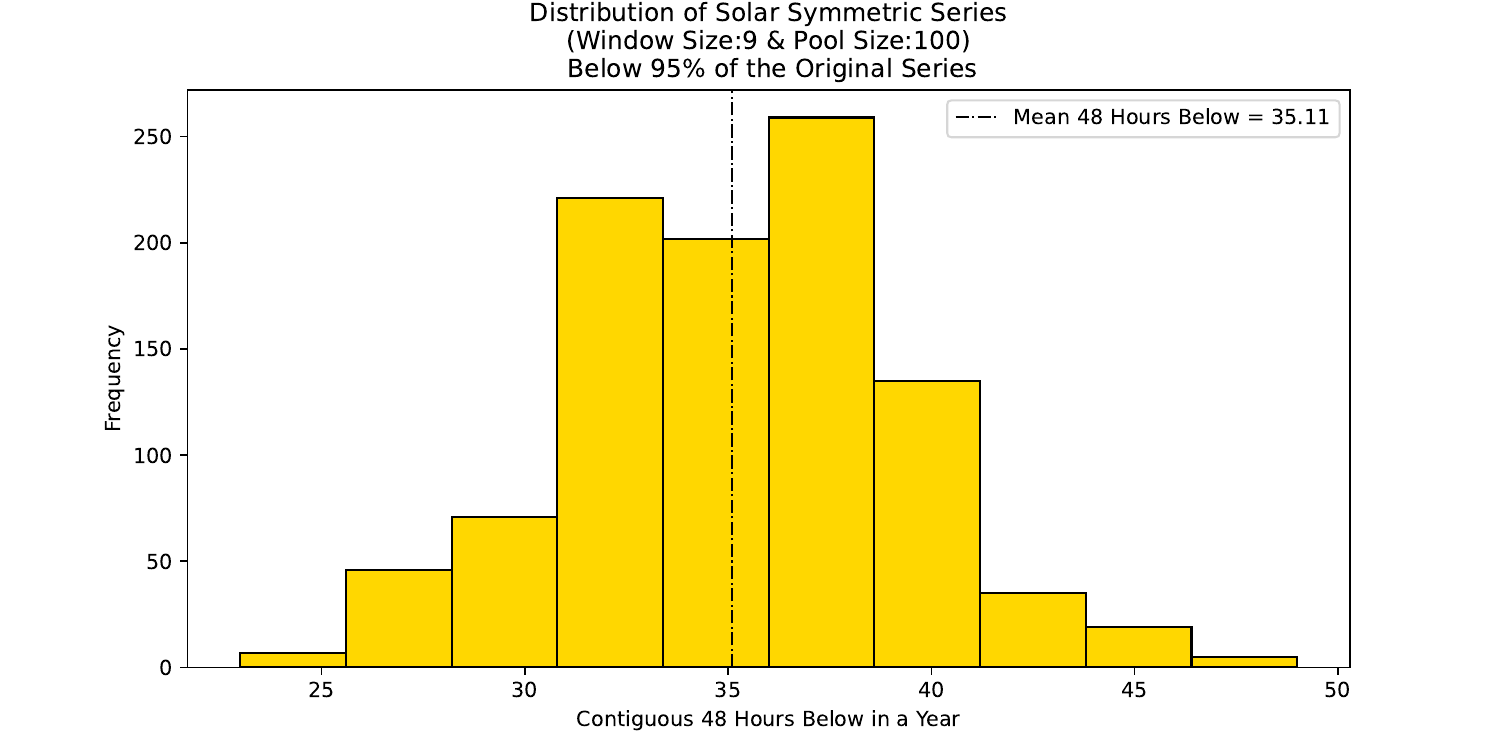}
    \caption{Distribution of the number of contiguous 48 hours from 1000 solar symmetric series that fall below 95\% of the original PJM data for solar energy.}
    \label{fig_solar_sym_extreme_below_48hrs}
\end{figure}

This under-performance statistic is not limited to a 24 hour period. We can consider any number of contiguous hourly periods. For example, in Figure \ref{fig_solar_sym_extreme_below_48hrs}, we can see that for both window sizes there is a decrease in the frequency of contiguous 24 hours to 48 hours below 5\%.
This is to be expected since it is a necessary but not sufficient condition for the 24 hour counts to also be included in the 48 hour counts. This is to illustrate that our code allows us to easily change the time frame parameter in order to visualize different distributions of under-performance measures.






\section{Creating Displaced Time Series} 

The methods we have discussed so far are designed to generate new time series that strongly resemble a reference series, typically observed or forecast. Such generated series are helpful in exploring model robustness, among other things. 

But what about model behavior under conditions systematically changed from existing observations? There is much need to operate models on scenarios far into the future. Given we can state broad directionality presumptions, how should synthetic time series data be produced?

We  propose two methods by which synthetic time series data may be systematically nudged higher or lower, depending on the purposes of the analysis.

\subsection{Incremental Selection \label{sec_inc_select}}

\subsubsection{Description of the Procedure}

Incremental selection is a general method that allows for any desired probability distribution to be used to alter a given series, typically the normal, exponential, or lognormal. Our code can be modified to allow any probability distribution function depending on the user's intentions. We describe the procedure below.



\begin{enumerate}
    \item Given a time series data set $\mathbf{X}$, we want to create a list of residuals $\mathbf{E}$ for each observation $i$ in $\mathbf{X}$ drawing from some probability distribution. Therefore, the elements of $\mathbf{E}$ are $\varepsilon_i \sim \text{Distribution}(\mu,\sigma^2)$ for some mean $\mu$ and standard deviation $\sigma$. We say $\varepsilon_i$ is the 
    offset to observation $i$ from $\mathbf{X}$. 
    
    \item To create $\mathbf{E}$, we first draw a value $z$ from the given distribution with its mean and standard deviation, and then we define the offsets as
    \begin{equation}
        \varepsilon_i = \begin{cases}
        \max_i & \text{if } z \sim \mbox{Distribution}(\mu,\sigma^2) > \max_i \\
        \min_i & \text{if } z \sim \mbox{Distribution}(\mu,\sigma^2) < \min_i \\
        z & \text{otherwise}
        \end{cases}
        \label{equation:incremental_selection}
    \end{equation}
    where $\max_i = \alpha_{\max}\cdot x_i$ and $\min_i = \alpha_{\min}\cdot x_i$ given $x_i$ is the original observation at step $i$ in $\mathbf{X}$ and the $\alpha$'s are parameters such that $\alpha_{\max} > \alpha_{\min}$. This means $|\max_i|$ is the largest acceptable offset from the observation in the original series. We have this condition to set boundaries on the alteration. For example, we will set an $\alpha$ such that no offset will make the data point negative because that would be interpreted as negative energy generation.
    
    \item We take the offset vector and subtract it from our given series to get a newly altered series generated as $\mathbf{X}' = \mathbf{X} - \mathbf{E}$.
    
\end{enumerate}
The following describes how we used the two distributions we examined in this study:
\begin{itemize}
    \item Normal: If the mean is 0, then there is a 50-50 chance a point increases or decreases from its original data point. We have provided an option where using a given probability $p$, we can translate any normal distribution with a given standard deviation to have a probability of $p$ being below the point in the original series. For example, if $p = 0.75$ and our original normal distribution has some mean $\mu$ and standard deviation $\sigma = 1$, then we are essentially pulling offsets from a normal distribution that has a mean $\approx\mu + 0.674$ with standard deviation $\sigma = 1$. 

    \item Exponential: We create a list of residuals $\mathbf{E}$ for each observation $i$ in $\mathbf{X}$ drawing from the exponential distribution where $\varepsilon_i \sim \text{Exp}(\beta)$ for some mean $\beta$. Using the exponential distribution to create offsets is an option of altering time series to always lie below the original series because this distribution is exclusively positive.
\end{itemize}

\subsubsection{Examining the Altered Series}

\begin{figure}[h!]
    \centering
    \includegraphics[width=\textwidth]{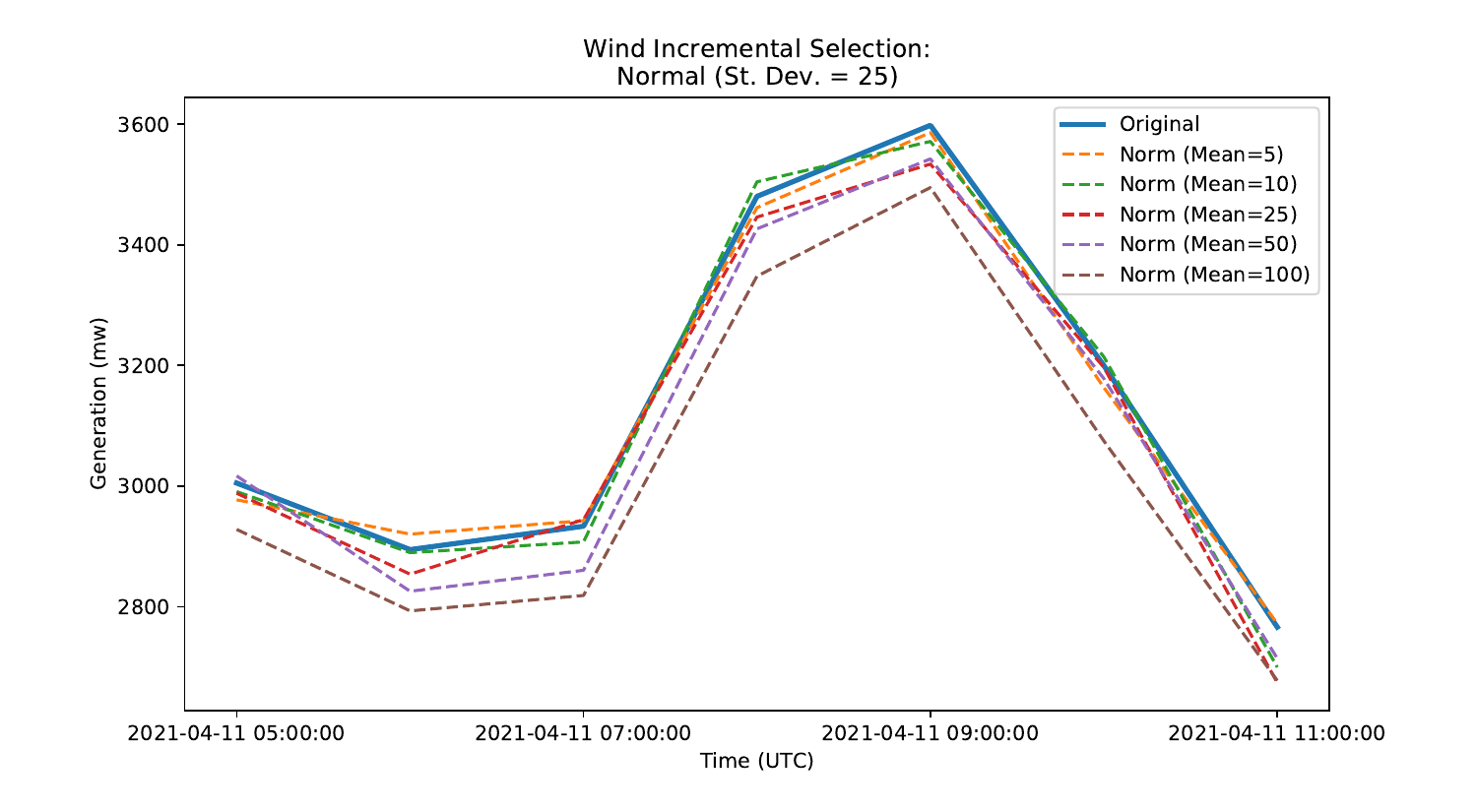}
    \caption{Five alterations of PJM's 2021 wind generation data over a time period of 6 hours using the normal  distribution. Normal, $\sigma=25$.}
    \label{fig_wind_inc_norm}
\end{figure}

\begin{figure}[h!]
    \centering
    \includegraphics[width=\textwidth]{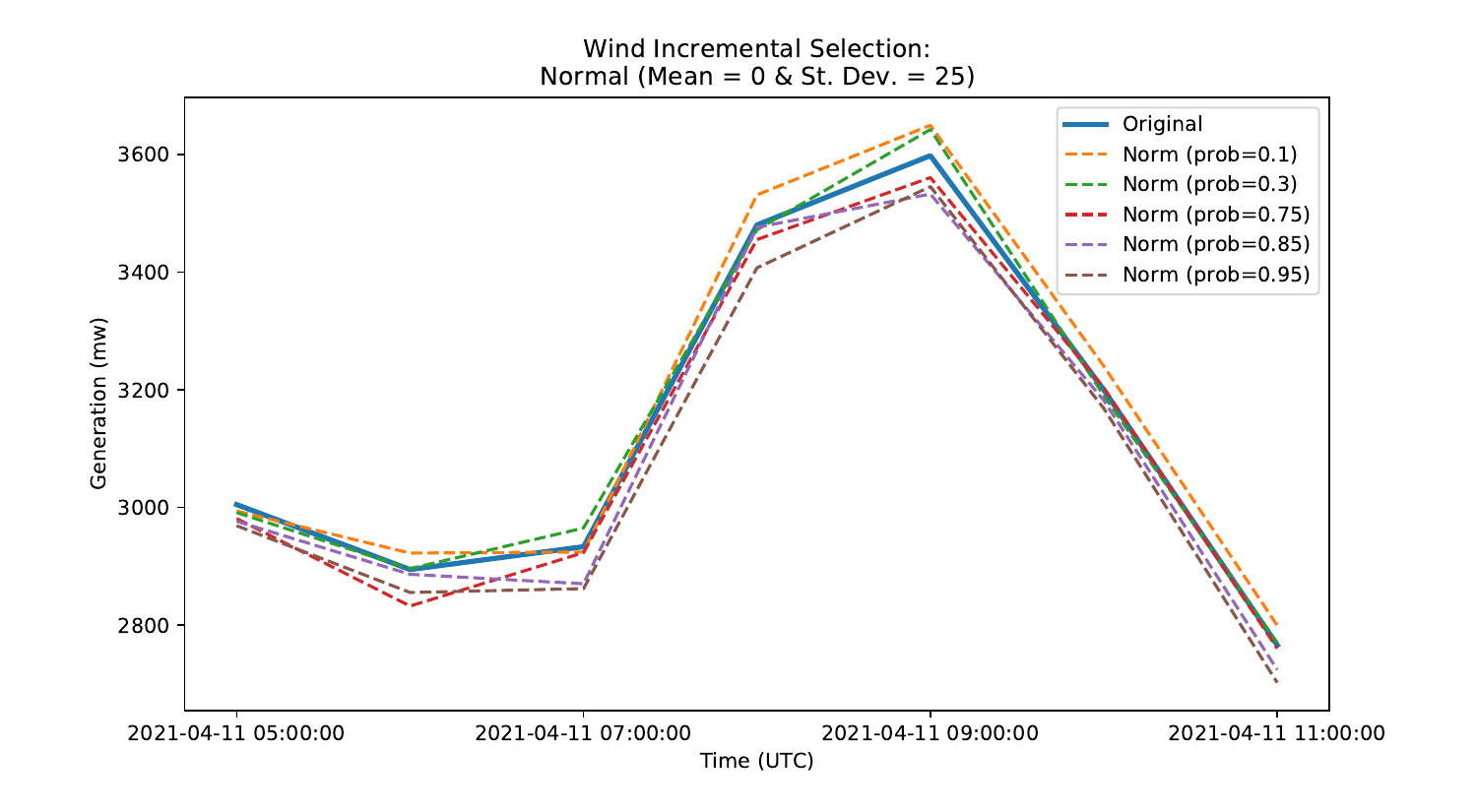}
    \caption{Five alterations of PJM's 2021 wind generation data over a time period of 6 hours using the normal  distribution. Normal, $\sigma=25$.}
    \label{fig_wind_inc_norm_2}
\end{figure}

\begin{figure}[h!]
    \centering
    \includegraphics[width=\textwidth]{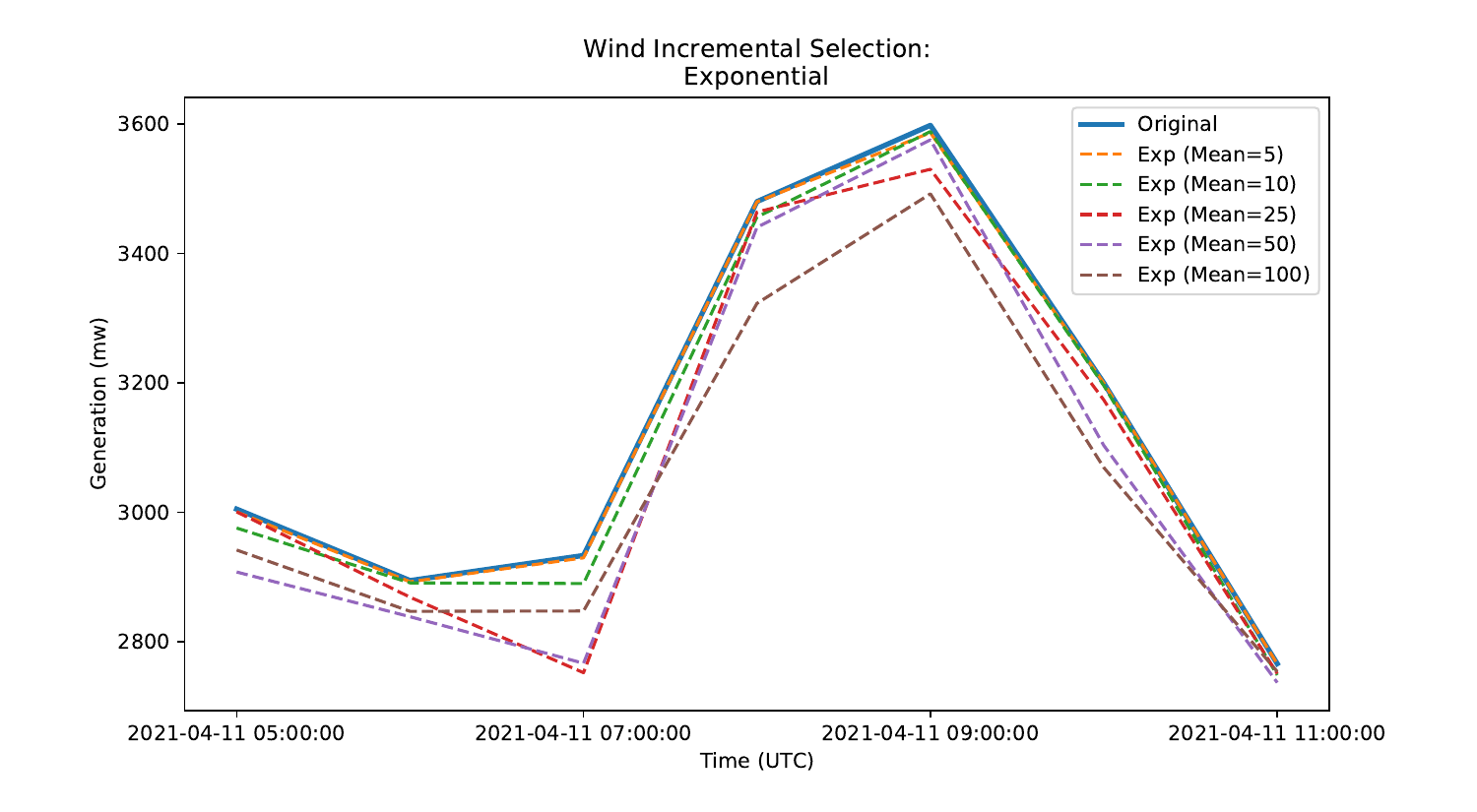}
    \caption{Five alterations of PJM's 2021 Wind generation data over a time period of 6 hours using the exponential distribution.}
    \label{fig_wind_inc_exp}
\end{figure}

In Figures \ref{fig_wind_inc_norm}, \ref{fig_wind_inc_norm_2}, and \ref{fig_wind_inc_exp}, we used the incremental selection method on PJM's 2021 wind generation data. The pictures are a six hour window to better visualize the deviations from the incremental selection. Using the normal distribution, we can see that there are times when the altered versions can produce a higher generation of wind than the original data. For the left graph in Figure \ref{fig_wind_inc_norm}, we fixed the mean to certain positive values, shown  in the legend. Table \ref{tab_wind_inc_norm_means_stats} shows that averages for the wind generation of the altered series are around the mean of the normal distribution less than the original series average. 

The  graph in Figure \ref{fig_wind_inc_norm_2} had two alterations that purposefully only allow on average 10\% and 30\%, respectively, of the altered points to lie below the original point. That is why Table \ref{tab_wind_inc_norm_probs_stats} has an extra row that details the number of days above 105\% of the original data in a year. 

\begin{table}[h!]
    \centering
    \begin{tabular}{|l|r|r|r|r|r|r|}
    \toprule
    Normal &  Original &  Mean=5 &  Mean=10 &  Mean=25 &  Mean=50 &  Mean=100 \\
    \midrule
    Min               &    63.300 &         32.943 &           0.000 &           5.173 &           4.034 &            0.000 \\
    First Quartile    &  1372.900 &       1369.412 &        1360.425 &        1349.594 &        1325.741 &         1270.657 \\
    Median            &  2739.500 &       2739.124 &        2726.149 &        2710.887 &        2682.018 &         2638.829 \\
    Third Quartile    &  4743.150 &       4728.162 &        4724.995 &        4715.413 &        4685.210 &         4638.055 \\
    Max               &  8990.000 &       8986.151 &        9006.112 &        8995.876 &        8927.513 &         8883.403 \\
    Mean              &  3186.568 &       3181.496 &        3176.355 &        3161.458 &        3136.301 &         3086.600 \\
    Standard Dev.     &  2138.397 &       2138.390 &        2138.462 &        2138.180 &        2138.114 &         2138.069 \\
    Coeff. of Var.    &     0.671 &          0.672 &           0.673 &           0.676 &           0.682 &            0.693 \\
    Autocorr. Lag: 24 &     0.437 &          0.437 &           0.437 &           0.437 &           0.437 &            0.436 \\
    Days Below 95\%   &        0 &             0 &              0 &              7 &             45 &             121 \\
    \bottomrule
    \end{tabular}
    \caption{Statistics on altering PJM's 2021 wind generation data using the normal distribution with the given means and a standard deviation of 25.}
    \label{tab_wind_inc_norm_means_stats}
\end{table}
\begin{table}[h!]
    \centering
    \begin{tabular}{|l|r|r|r|r|r|r|}
    \toprule
    Normal &  Original &  prob=0.1 &  prob=0.3 &  prob=0.75 &  prob=0.85 &  prob=0.95 \\
    \midrule
    Min               &    63.300 &          101.451 &           48.437 &            20.983 &             2.353 &             7.071 \\
    First Quartile    &  1372.900 &         1407.977 &         1379.735 &          1353.247 &          1342.418 &          1327.743 \\
    Median            &  2739.500 &         2763.368 &         2750.374 &          2719.955 &          2715.660 &          2691.893 \\
    Third Quartile    &  4743.150 &         4770.020 &         4745.956 &          4724.907 &          4712.031 &          4700.977 \\
    Max               &  8990.000 &         9013.417 &         9036.559 &          8974.119 &          8980.445 &          8957.442 \\
    Mean              &  3186.568 &         3218.822 &         3199.766 &          3169.547 &          3160.507 &          3145.465 \\
    Standard Dev.     &  2138.397 &         2138.322 &         2138.483 &          2138.571 &          2138.743 &          2138.677 \\
    Coeff. of Var.    &     0.671 &            0.664 &            0.668 &             0.675 &             0.677 &             0.680 \\
    Autocorr. Lag: 24 &     0.437 &            0.437 &            0.437 &             0.437 &             0.437 &             0.437 \\
    Days Below 95\%   &        0 &             0 &              0 &              3 &             10 &             32 \\
    Days Above 105\%  &        0 &            20 &              1 &              0 &              0 &             0 \\
    \bottomrule
    \end{tabular}
    \caption{Statistics on altering PJM's 2021 wind generation data using the normal distribution with a mean of 0, a standard deviation of 25, and the given probabilities of being below the original dataset.}
    \label{tab_wind_inc_norm_probs_stats}
\end{table}

As with the normal distribution alteration, using the exponential distribution also results in averages of the altered series being offset by the same amount as the mean we used to generate each exponential distribution. 

Unlike the normal distribution alteration, we can see in Table \ref{tab_wind_inc_exp_stats} that all the altered series lie below the original series in terms of minimum, quartiles, maximum, mean, and median. Our bootstrap estimate also shows the number of days that fell below 95\% of the actual wind generation from PJM in 2021.

\begin{table}[h!]
    \centering
    \begin{tabular}{|l|r|r|r|r|r|r|}
    \toprule
    Exponential &  Original &  Mean=5 &  Mean=10 &  Mean=25 & Mean=50 & Mean=100 \\
    \midrule
    Min               &    63.300 &        63.259 &         54.554 &          0.000 &          0.000 &           0.000 \\
    First Quartile    &  1372.900 &      1368.917 &       1363.091 &       1341.631 &       1325.879 &        1265.014 \\
    Median            &  2739.500 &      2734.657 &       2728.293 &       2714.190 &       2682.897 &        2631.227 \\
    Third Quartile    &  4743.150 &      4737.954 &       4732.544 &       4710.470 &       4688.843 &        4618.897 \\
    Max               &  8990.000 &      8981.927 &       8989.626 &       8975.272 &       8972.591 &        8952.388 \\
    Mean              &  3186.568 &      3181.531 &       3176.531 &       3162.019 &       3136.983 &        3085.197 \\
    Standard Dev.     &  2138.397 &      2138.307 &       2138.461 &       2138.361 &       2138.156 &        2142.348 \\
    Coeff. of Var.    &     0.671 &         0.672 &          0.673 &          0.676 &          0.682 &           0.694 \\
    Autocorr. Lag: 24 &     0.437 &         0.437 &          0.437 &          0.437 &          0.437 &           0.437 \\
    Days Below 95\%   &        0 &             0 &              0 &              6 &             44 &             126 \\
    \bottomrule
    \end{tabular}
    \caption{Statistics on altering PJM's 2021 wind generation data using the exponential distribution with the given means.}
    \label{tab_wind_inc_exp_stats}
\end{table}

Incremental selection is a method to shift the trend of a given series without introducing a significant amount of variability of the observations in relation to each other. This is seen in the statistic tables of this section and the graphs of the altered series. The standard deviations do not noticeably deviate from the original standard deviation for PJM's 2021 wind generation time series, and the coefficients of variation increase only modestly. However, the means and ranges of the altered series have generally shifted down (or up) given the distribution we used to perturb the original series.

\subsection{Altered-Difference Distribution \label{sec_altered_dif}}

\subsubsection{Description of the Procedure\label{sec_description_altered_series}}
This method takes in two time series $\mathbf{X}$ and $\mathbf{Y}$ where we label $\mathbf{X}$ as the series with values that are generally higher than the values in series $\mathbf{Y}$. Either or both of these series may be generated or be an altered version of the other.
\begin{enumerate}
    \item Let $\mathbf{\Delta} = \mathbf{X} - \mathbf{Y}$ (In our code, we have an option where alternately, $\mathbf{\Delta}_i = \max(\mathbf{X}_i-\mathbf{Y}_i, 0), \forall i$, which forces all the delta changes to be greater than or equal to zero) \\
    Note: Our purpose is to create generally lower series, but if the desired result is to create an altered series that is generally higher, let $\mathbf{X}$ be generally lower than $\mathbf{Y}$ when computing $\mathbf{\Delta}$.
    
    \item Let $\mathbf{R} = \mathbf{Y} - \alpha\mathbf{\Delta}$ where $\alpha$ is a scalar chosen to multiply the vector of differences $\mathbf{\Delta}$. If prompted, our code will ensure every entry in $\mathbf{R}$ is a non-negative value by turning all negative values to $0$. \\
    Note: In our implementation, $\alpha$ is a scalar, though other implementations could create a vector of multiples by drawing $\alpha_i \sim \text{Uniform}(0,\max_i)$ for each observation in the series. We chose to not use this approach because we wanted all the differences to be proportional to one another. Using a uniform distribution would introduce random variability for each point, which has the possibility of creating huge jumps between hours, which would lead to an altered series that follows the original distribution less systematically. This could be fixed with a smoothing function. In our case, by using a single scalar, we do not require smoothing of the altered series. 
    
    \item Create---using NNLB, SBB, or some other generating method---new time series $\mathbf{R}_1, \mathbf{R}_2, \ldots$ by using $\mathbf{R}$ as the input for additional perturbation through any systematic time series resampling or alteration methods mentioned in this paper.
\end{enumerate}


\subsubsection{Examining the Altered Series}

We altered PJM's 2021 wind generation data using some of our series that were generated from our resampling bootstrap methods in \S\S\ref{sec_NNLB} and \ref{sec_SBB} and using a few altered series from \S\ref{sec_inc_select}. As mentioned in the ``Description'' section for this alteration method, \S\ref{sec_description_altered_series}, we have two series and designate one to be generally lower than the other. This is intentionally left vague because it is up to the user to decide the statistic they intend to use to compare the two time series.  

\begin{figure}[h!]
    \centering
    \includegraphics[width=\textwidth]{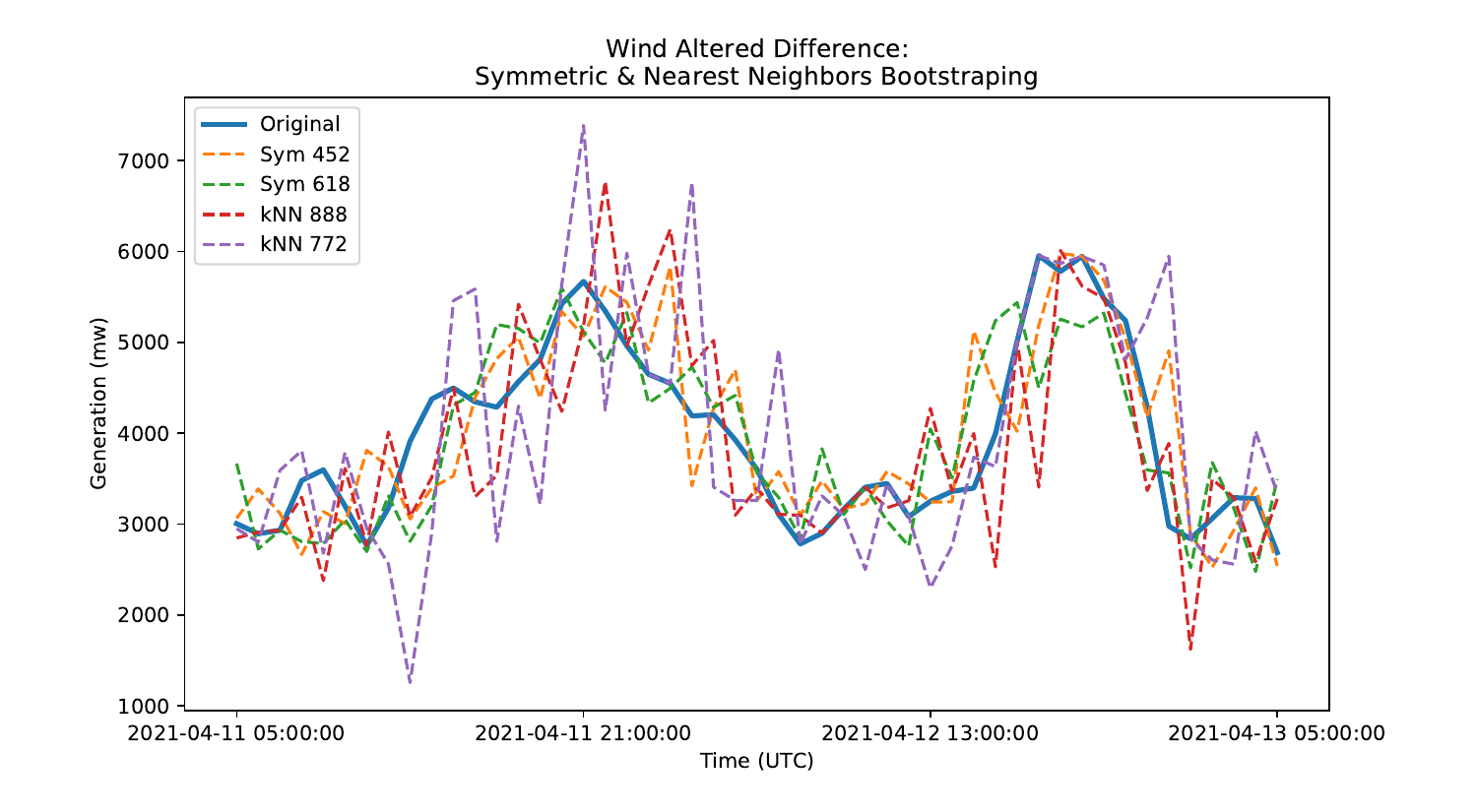}
    \caption{Four alterations of PJM's 2021 Wind generation data over a time period of 24 hours using the altered difference method with $\alpha = 0.5$.}
    \label{fig_wind_alt_dif_sym_knn}
\end{figure}

First, we discuss the resampled series used in Figure \ref{fig_wind_alt_dif_sym_knn}. We took two series from the 1000 Symmetric Block resampled series on PJM's 2021 wind generation using a window size of 9 and a pool size of 100. We also took two series from the 1000 Nearest Neighbors resampled series on PJM's 2021 wind generation using a lag size of 9 and 100 nearest neighbors. The series were taken from these sets because the parameters used for the Symmetric Block and Nearest Neighbors Bootstrap created more variation from the original series. 

``Sym 452'' and ``kNN 888'' were chosen from their respective sets because these two series had the lowest mean. ``Sym 612'' and ``kNN 772'' were chosen from their respective sets because these two series had the most number of days in a year that had generation below 95\% of the original wind generation's mW per day. 

We used the multiple $\alpha = 0.5$ with the series in Figure \ref{fig_wind_alt_dif_sym_knn}, and we can see that the variation from the original wind series has been amplified by the Altered Difference Method. All the means are less than the   averages of original PJM series, but the range of wind generation has become more extreme on the high and low end, which can be seen in Table \ref{tab_wind_altered_dif_sym_knn}. One reason is because we did not make our $\mathbf{\Delta}$s exclusively non-negative. Therefore, if the original data point in the lower series was actually greater than than the value in the generally higher series, the altered point would be increasing from the lower series even more above the original series.

\begin{figure}[h!]
    \centering
    \includegraphics[width=\textwidth]{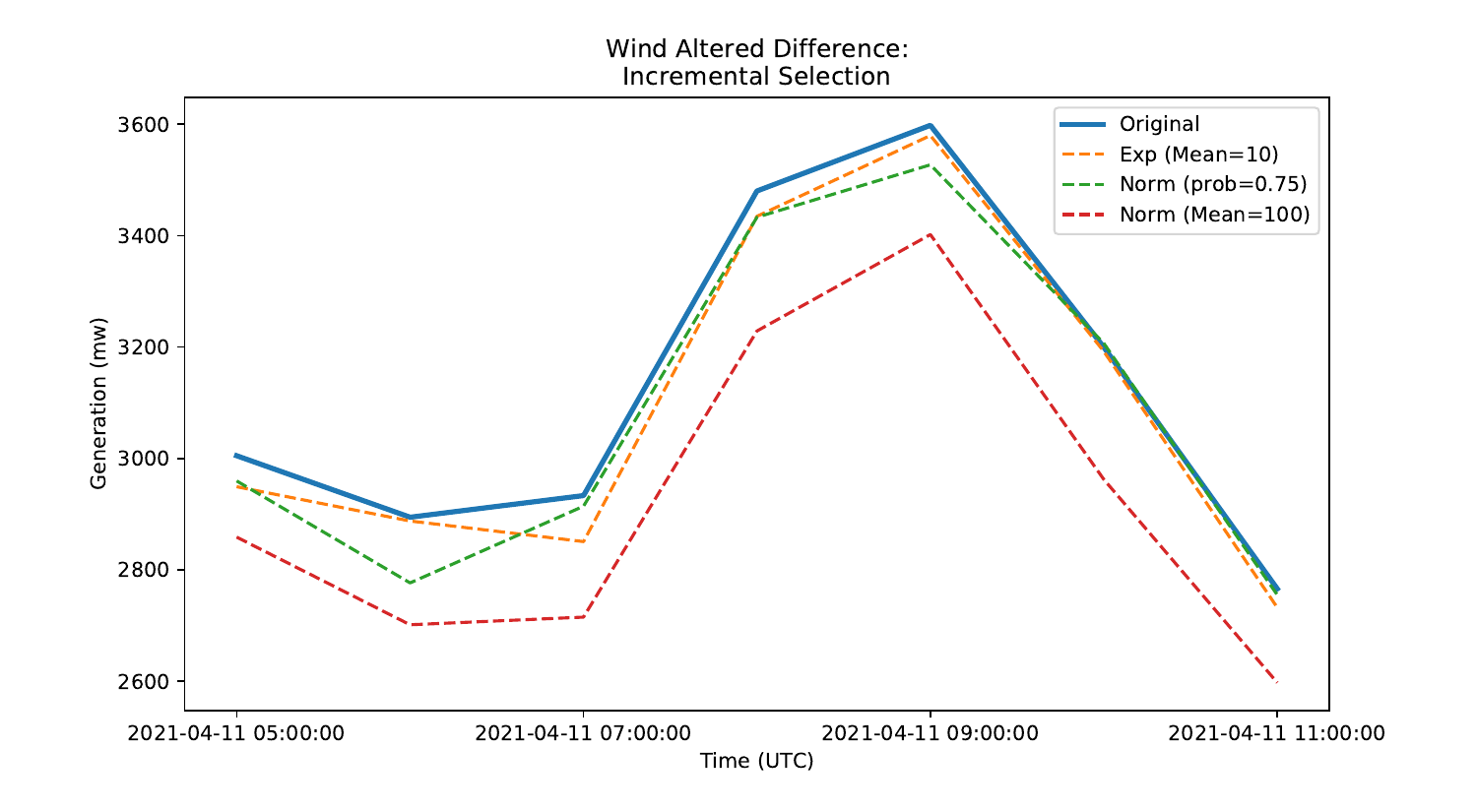}
    \caption{Three alterations of PJM's 2021 Wind generation data over a time period of 6 hours using the altered difference method with $\alpha = 0.9$.}
    \label{fig_wind_alt_dif_inc}
\end{figure}

The other reason is that the SBB and NNLB methods are designed to systematically generate resampled versions of the original series, so they are more similar to the original PJM data than the altered series used in Figure \ref{fig_wind_alt_dif_inc}. The 24-hour time period shown in Figure \ref{fig_wind_alt_dif_sym_knn} shows that even with more variation, the altered series still capture the general trend of the original PJM wind generation data.

\begin{table}[h!]
    \centering
    \begin{tabular}{|l|r|r|r|r|r|}
        \toprule
        Altered Difference &  Original &  Sym 452 &  Sym 618 &  kNN 888 &   kNN 772 \\
        \midrule
        Min               &    63.300 &   11.750 &    1.650 &    0.000 &     0.000 \\
        First Quartile    &  1372.900 & 1363.050 & 1339.275 & 1296.838 &  1318.775 \\
        Median            &  2739.500 & 2714.525 & 2734.050 & 2671.625 &  2683.225 \\
        Third Quartile    &  4743.150 & 4713.100 & 4710.475 & 4663.787 &  4709.962 \\
        Max               &  8990.000 & 9158.400 & 9029.450 & 9561.950 & 10026.450 \\
        Mean              &  3186.568 & 3164.553 & 3170.833 & 3134.828 &  3144.294 \\
        Standard Dev.     &  2138.397 & 2130.048 & 2135.847 & 2165.174 &  2167.364 \\
        Coeff. of Var.    &     0.671 &    0.673 &    0.674 &    0.691 &     0.689 \\
        Autocorr. Lag: 24 &     0.437 &    0.443 &    0.436 &    0.390 &     0.395 \\
        Days Below 95\%   &         0 &      55 &      43 &       94 &       97 \\
        Days Above 105\%  &         0 &      19 &      23 &       74 &       95 \\
        \bottomrule
    \end{tabular}
    \caption{Statistics on altering PJM's 2021 wind generation data using the altered difference method with two symmetric and two nearest neighbors generated series with $\alpha = 0.5$.}
    \label{tab_wind_altered_dif_sym_knn}
\end{table}

Unlike the resampling methods used in Figure \ref{fig_wind_alt_dif_sym_knn}, the series using incremental selection in Figure \ref{fig_wind_alt_dif_inc} create less variability from one observation to the next. This is shown by comparing differences in the standard deviations with the original series' standard deviation from Table \ref{tab_wind_altered_dif_sym_knn} with Table \ref{tab_wind_altered_dif_inc}. That's because the incremental selection used the original series first and altered it with a given distribution with set parameters. Therefore, the graphs we see in Figure \ref{fig_wind_alt_dif_inc} are similar to Figures \ref{fig_wind_inc_norm} and \ref{fig_wind_inc_exp}, but they lie further below the original series, as seen with an increase in the number of days below 95\% of the original generation from wind in Table \ref{tab_wind_altered_dif_inc} with the statistic tables in Section \ref{sec_inc_select}.

\begin{table}[h!]
    \centering
    \begin{tabular}{|l|r|r|r|r|}
        \toprule
        Altered Difference &  Original &  Exponential  &  Normal  &  Normal  \\
                          &           &     (Mean=10)  & (prob=0.75)  &  (Mean=100) \\
        \midrule
        Min               &    63.300 &         46.682 &          0.000 &            0.000 \\
        First Quartile    &  1372.900 &       1353.362 &       1339.904 &         1179.467 \\
        Median            &  2739.500 &       2719.508 &       2705.926 &         2546.122 \\
        Third Quartile    &  4743.150 &       4722.762 &       4707.969 &         4543.598 \\
        Max               &  8990.000 &       8989.290 &       8959.826 &         8802.135 \\
        Mean              &  3186.568 &       3167.498 &       3154.230 &         2996.864 \\
        Standard Dev.     &  2138.397 &       2138.558 &       2138.973 &         2137.692 \\
        Coeff. of Var.    &     0.671 &          0.675 &          0.678 &            0.713 \\
        Autocorr. Lag: 24 &     0.437 &          0.437 &          0.436 &            0.436 \\
        Days Below 95\%   &         0 &             4 &             16 &              237   \\
        Days Above 105\%  &         0 &             0 &              0 &              0  \\
        \bottomrule
    \end{tabular}
    \caption{Statistics on altering PJM's 2021 wind generation data using the altered difference method with three altered series using the incremental selection method with $\alpha = 0.9$.}
    \label{tab_wind_altered_dif_inc}
\end{table}

Overall, the Altered Difference Method is a process to transfer the differences between two similar series onto a the ``lower'' series to retain the differences seen between the two. Depending on what the ``lower'' series is, this method can create more variation within a series and decrease (or increase) the trend of a given series by outputting the altered series.

\subsection{Case Study: Scaling PJM's Renewable Energy\label{sec_pjm_case}}

The gist of the  case study is as  follows.
    \begin{enumerate}[a.]
        \item We use PJM 2021 demand (load) data, and nuclear, solar, and wind production data.
        \item We scale up the 2021 PJM data for solar and wind generation in order to examine a hypothetical system dominated by VRE.
        \item We determined  specific weights for solar and wind production, which that we use to sum the wind and solar series for a given year and to effect a maximum level of supply subject to a constraint on  curtailment level.
        \item We chose a curtailment levels of 10\%, and then  50\%, for wind and solar production.
        \item Taking the 1000 wind and 1000 solar series that we generated with the SBB method, we added together two randomly chosen wind and solar synthetic series for the given weights, creating thereby 1000 combined wind and solar series.
        \item Thereby, we created combined renewable energy series as a weighted sum of the 2021 wind and solar production series, such that renewable production was maximized but no more than 10 or 50\% of production would be curtailed for the year. We found that maximizing for 10\% and 50\% curtailment levels provided in consequence 68\% and 90\%, respectively, of demand by hour over the year.
        \item We also compared the differences in seasonality for these scaled-up series. We find that while overall 90\% of demand is met by the configuration of renewables, the effect is highly seasonal with significant shortfalls during the winter.
        \item We chose as the measure of performance, $\Phi$, the number of 24 hour periods per year in which the total renewable supply was less than 90\% of demand. With the scaled up data that curtailed 50\% of its energy, that number was 64 days.
        \item We then substituted for the scaled series of actual data, the 1000 synthetic series and collected the distribution of our statistic, $\Phi$. This distribution is explained in the case study and can be explored with the code in our supplemental materials.
    \end{enumerate}   
    We now present the case study in narrative form.

Suppose that in the future, we are able to supply the majority of energy demand with renewable sources. We want to analyze this possibility using the current 2021 PJM supply of energy from wind and solar sources. We multiplied each dataset with a weight to meet a certain percent of curtailment. However, as illustrated in the sections above, wind and solar energy generation follow different daily, monthly, and yearly trends, so it would not seem logical to scale up solar and wind generation with the same scalar multiple. We ran a weight analysis to decide the most optimal weight combinations for a given upper bound on the percent of energy to be curtailed that will be applied to the solar and wind datasets respectively. 
\begin{equation}
\text{VRE} = w_s \cdot S + w_w \cdot W
\end{equation}
(The $S$ and $W$ values are solar PV and wind capacities respectively, in megawatts.)
We chose the weights that led to the highest percent supplied by renewable energy, which in our case means the percent of energy supplied by a weighted sum of solar and wind with PJM's nuclear energy for 2021. Our combination of weights $w_s = 45$ and $w_w = 22$ leads to a 10\% curtailment of VRE which supplied around 68\% of demand for the whole year. As shown in the supplemental materials, there are multiple weight combinations that lead to similar outcomes in percent supplied for a given percent curtailed. Therefore, our choice of weights was arbitrary.

After picking a pair of weights, we randomly combined pairs of the synthetic data that was previously generated using the SBB method from Section \ref{sec_SBB} for the solar and wind time series and applied a weighted sum with $w_s = 45$ and $w_w = 22$. 
We compared the percent supplied and curtailed for these combinations and found them to be similar to original data's statistics.  
\begin{table}[!ht]
    \centering
    \begin{tabular}{|l|r|r|r|r|r|r|r|}
        \toprule
        \% Supplied &   mean &    std &    min &    25\% &    50\% &    75\% &    max \\
        \midrule
        Solar ($n=2$) \& Wind ($n=2$) & 0.6801 & 0.0002 & 0.6794 & 0.6799 & 0.6801 & 0.6802 & 0.6808 \\
        Solar ($n=2$) \& Wind ($n=4$) & 0.6800 & 0.0003 & 0.6788 & 0.6797 & 0.6800 & 0.6802 & 0.6811 \\
        Solar ($n=4$) \& Wind ($n=2$) & 0.6792 & 0.0003 & 0.6783 & 0.6790 & 0.6792 & 0.6794 & 0.6801 \\
        Solar ($n=4$) \& Wind ($n=4$) & 0.6791 & 0.0004 & 0.6777 & 0.6788 & 0.6791 & 0.6793 & 0.6803 \\
        \bottomrule
    \end{tabular}
    \caption{Summary statistics for the percent supplied using the weights $w_s = 45$ and $w_w = 22$ for the sash size $n$.}
    \label{tab_vre_percent_supplied}
\end{table}
\begin{table}[!ht]
    \centering
    \begin{tabular}{|l|r|r|r|r|r|r|r|}
        \toprule
        \% Curtailed &   mean &    std &    min &    25\% &    50\% &    75\% &    max \\
        \midrule
        Solar ($n=2$) \& Wind ($n=2$) & 0.0991 & 0.0002 & 0.0984 & 0.0990 & 0.0991 & 0.0993 & 0.0998 \\
        Solar ($n=2$) \& Wind ($n=4$) & 0.0981 & 0.0004 & 0.0968 & 0.0979 & 0.0982 & 0.0984 & 0.0992 \\
        Solar ($n=4$) \& Wind ($n=2$) & 0.0985 & 0.0003 & 0.0975 & 0.0983 & 0.0985 & 0.0987 & 0.0995 \\
        Solar ($n=4$) \& Wind ($n=4$) & 0.0975 & 0.0004 & 0.0962 & 0.0972 & 0.0975 & 0.0978 & 0.0989 \\
        \bottomrule
    \end{tabular}
    \caption{Summary statistics for the percent curtailed using the weights $w_s = 45$ and $w_w = 22$ for the sash size $n$.}
    \label{tab_vre_percent_curtailed}
\end{table}

\begin{figure}[h!]
    \centering
    \includegraphics[width=0.45\textwidth]{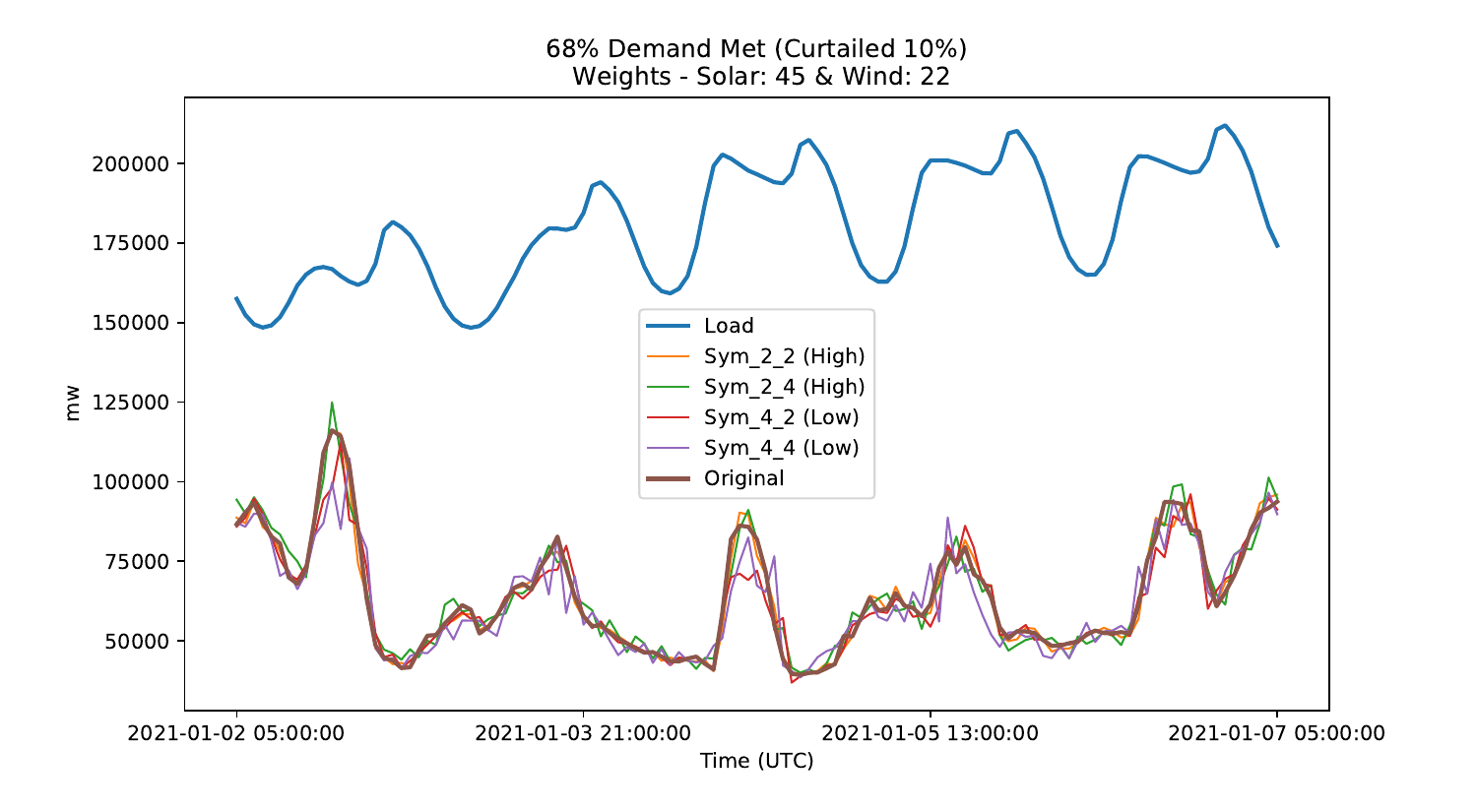}
    \includegraphics[width=0.45\textwidth]{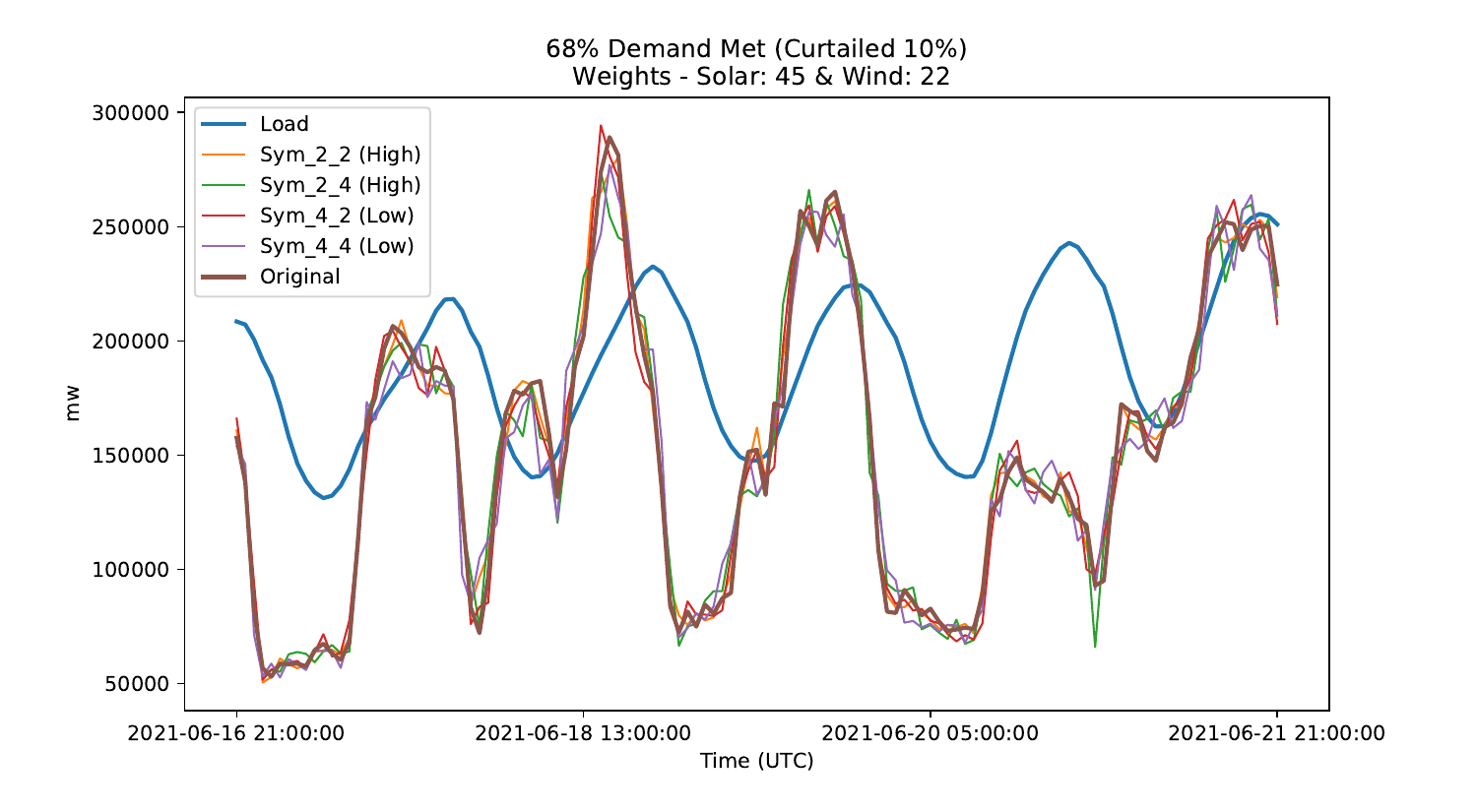}
    \caption{Seasonal 5 day comparisons of the scaled up renewable energy by the weights $w_s = 45$ and $w_w = 22$.}
    \label{fig_vre_curtail_10_seasons}
\end{figure}

It is important to note that the measurements of percent supplied and curtailed apply across the whole year. However, wind and solar energy generation depend on season. We see this in the graphs in Figure \ref{fig_vre_curtail_10_seasons}. In the winter, the scaled up renewable energy fails to meet demand, and is even substantially below demand for energy. In the summer, there are hours where the scaled up renewable energy surpasses the demand at that time. Therefore, additional research is needed for energy storage designs to meet or curtail energy generation.

We use PJM's 2021 nuclear generation as a baseline addition to all the synthetically generated series because we are under the assumption that nuclear energy will be proportionally scaled to meet as much demand that solar and wind energy do not fill at the time.  
We also calculated the percent supplied and curtailed within these 5 day periods in the winter and summer from Figure \ref{fig_vre_curtail_10_seasons}. In the winter, the percent supplied was about 35\% while curtailment was 0\%, which matches the visual shown in the left graph. In comparison, the 5 days in the summer the supplied about 75.5\% while curtailment ranged between 7-8\%.

\begin{figure}[h!]
    \centering
    \includegraphics[width=0.45\textwidth]{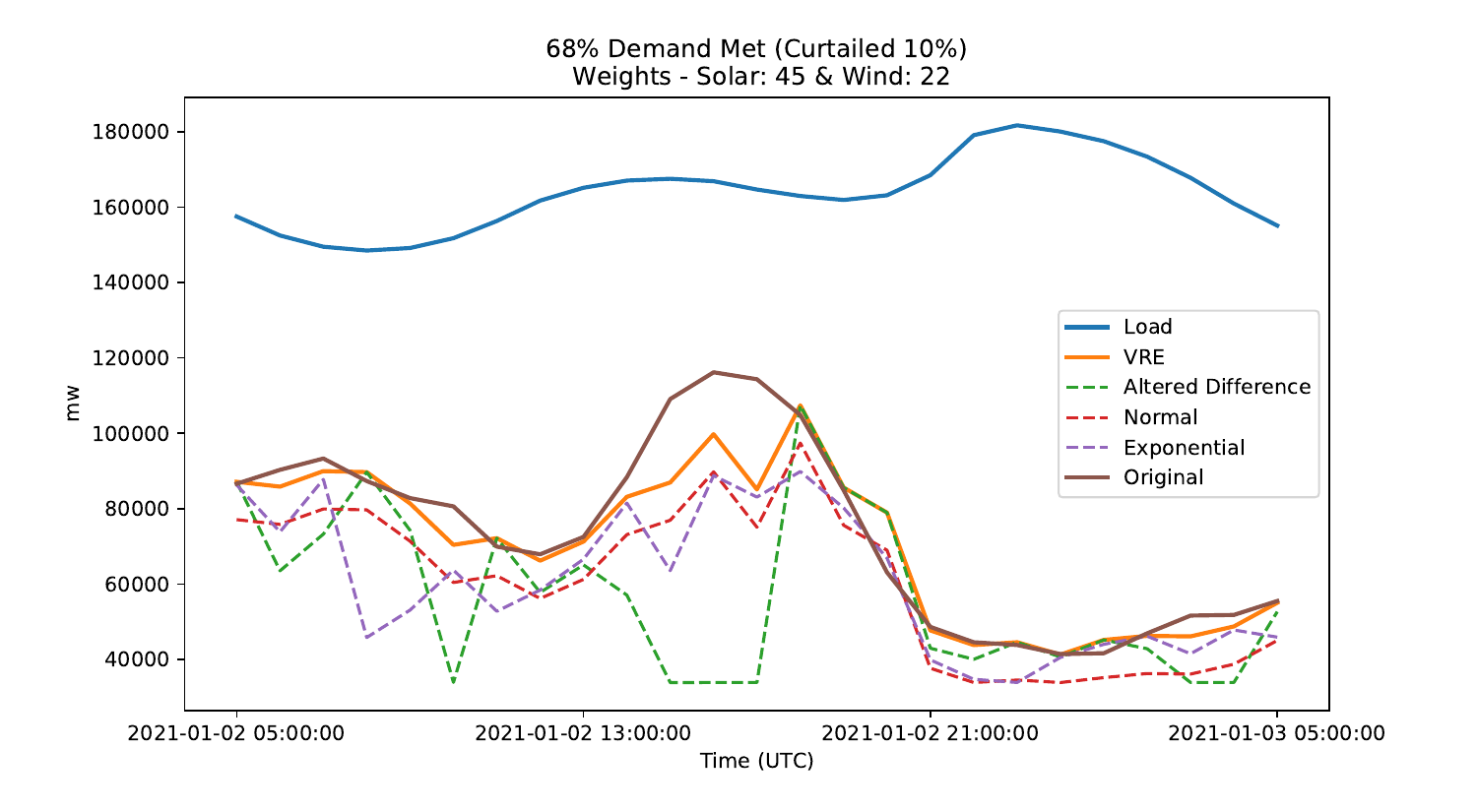}
    \includegraphics[width=0.45\textwidth]{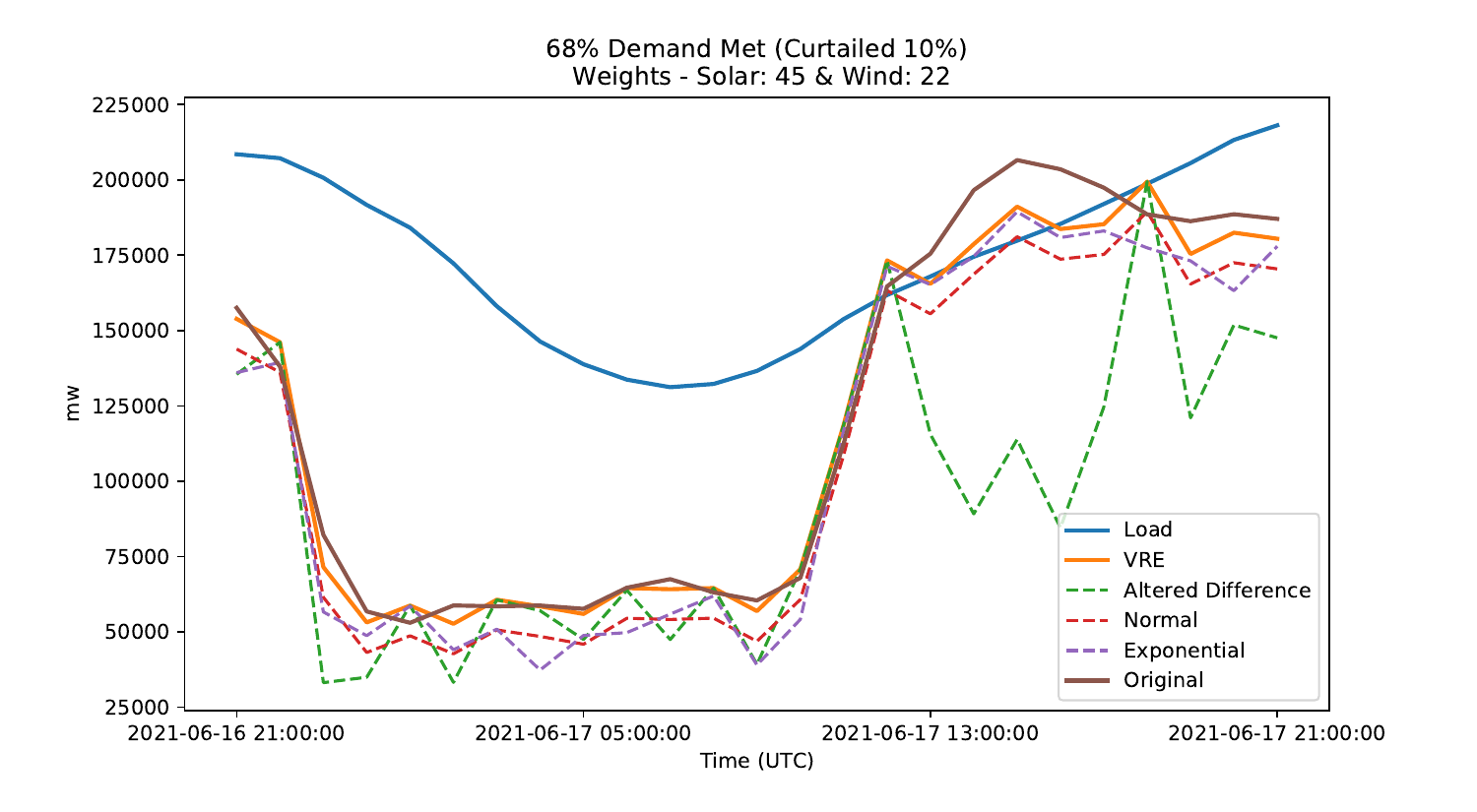}
    \caption{Seasonal 1 day comparisons of altered versions of a scaled up renewable energy dataset by the weights $w_s = 45$ and $w_w = 22$ synthetically generated from the SBB method.}
    \label{fig_vre_curtail_10_seasons_alt}
\end{figure}

Because the purpose of testing the robustness of energy systems requires many different scenarios, we have provided an altered version of the scaled renewable energy data set in Figure \ref{fig_vre_curtail_10_seasons_alt}.
We also considered curtailment levels of 50\%, where we chose $w_s = 84$ and $w_w = 64$ as the scaling weights. 
\begin{figure}[h!]
    \centering
    \includegraphics[width=0.45\textwidth]{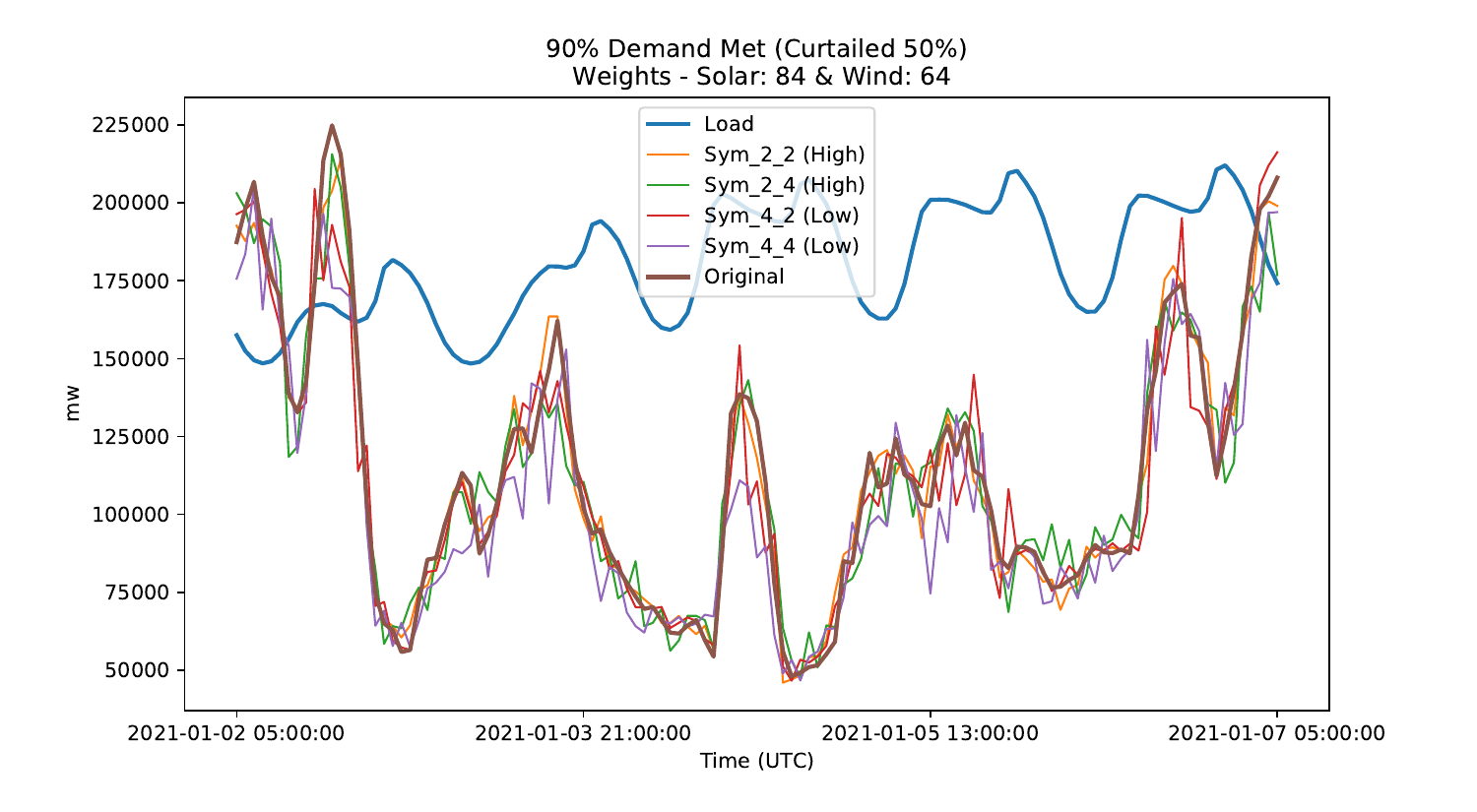}
    \includegraphics[width=0.45\textwidth]{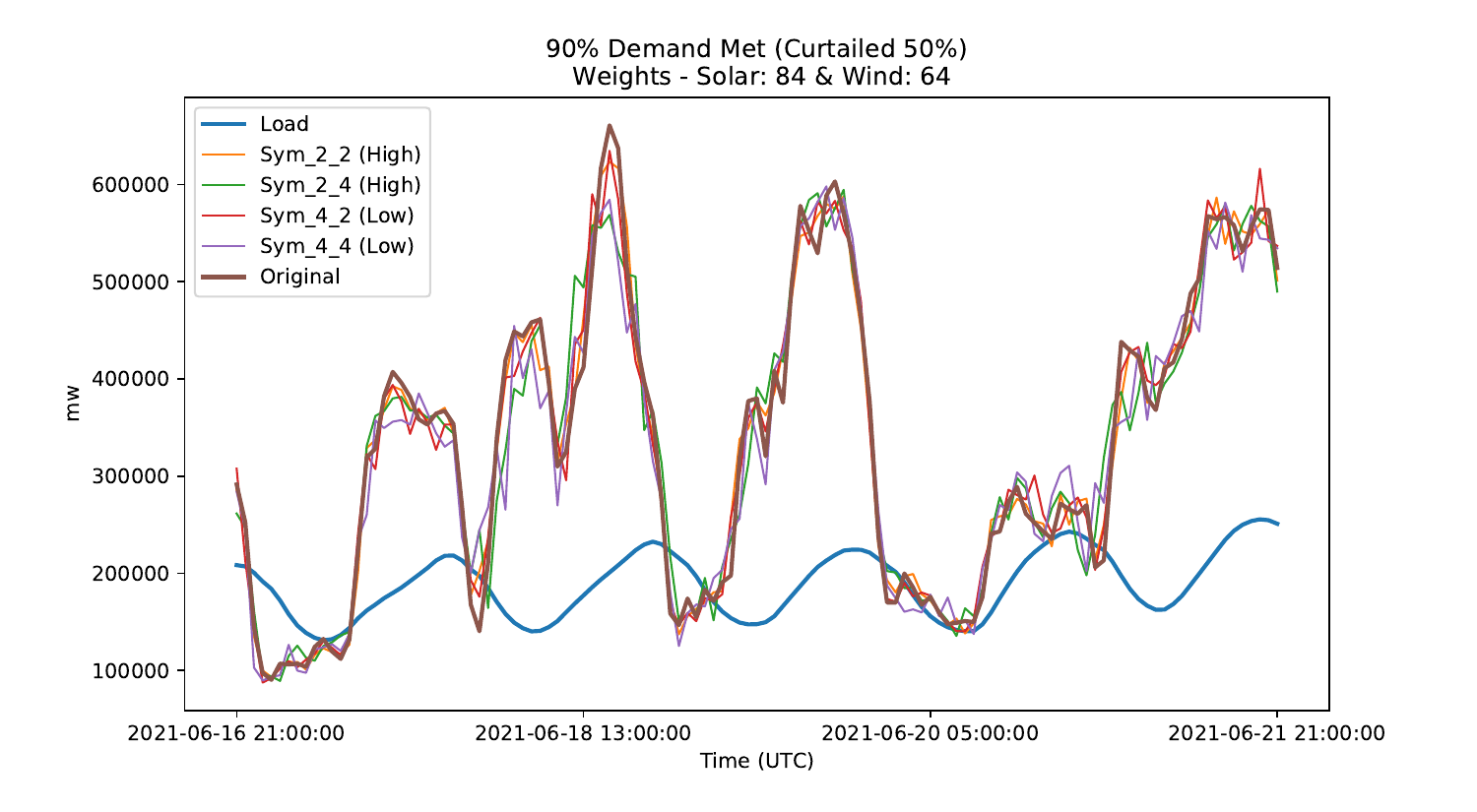}
    \caption{Seasonal 5 day comparisons of the scaled up renewable energy by the weights $w_s = 84$ and $w_w = 64$.}
    \label{fig_vre_curtail_50_seasons}
\end{figure}
In the winter, the percent supplied was about 60\% with curtailment of 4-5.5\%. In comparison, the 5 days in the summer supplied about 97\% while curtailment was 50\%, which aligns with the average curtailment throughout the year (see Figure \ref{fig_vre_curtail_50_seasons}.). These findings demonstrate the need for capable storage designs to help meet energy demand, even when these energy sources have been scaled up. To illustrate this further, we used the bootstrap estimate statistic described in Section \ref{sec_Bootstrap} to calculate the number of 24 hour periods that do not meet a given day's demand. For a cutoff of 50\% curtailment, the original weighted PJM series failed to meet energy demand for 64 days of the year. Using our combined synthetic series, the failure to meet a year's demand ranges from 61--69 days, with the median between 64--65 days.\footnote{See our code in the supplemental materials for purposes of replication.} Therefore, this demonstrates that our synthetic series do closely follow the original series, but can create more variability to allow energy models to test the sensitivity of their systems.


\section{Discussion and Conclusion\label{sec_conclusion}}

\subsection{Discussion\label{sec_discussion}}


To summarize, it is universally agreed that  sensitivity analysis, scenario analysis, and many other forms of model analysis \cite{kimbrough_business_2016} are essential and invaluable tools for effective use of models, particularly those that would inform policy making. Time series data constitute a serious challenge to model analysis, compared to scalar data. The latter affords largely unproblematic alteration for  what-if analyses (a foundational element of much of model analysis), but how to do this with time series data has received scant attention. We have found only two fleeting, albeit promising, forays in this direction, in \cite{amonkar_k-nearest_2022} and in the gray literature
\cite{ray_confronting_2015}. In addition, \cite{turowski_modeling_2022} raises the enticing prospect of generating synthetic anomaly data for energy systems modeling. The methods under discussion in this paper may well prove useful for that purpose as well.

We have found only two practicable methods for generating synthetic time series suitable for post-solution model analysis: NNLB \cite{lall_nearest_1996}, which evidences mild anomalies (predicting solar generation at night), and SBB \cite{kimbrough_symmetric_2021}, which evidences a small bias. Both methods use bootstrap sampling and both we find  fit for the purpose of supporting post-solution model analysis involving time series data, although we prefer SBB to NNLB if only for the latter's faulty treatment of PV production after dark. We then introduced and explored two methods---incremental selection and altered-difference distribution---for principled directional alteration of time series. These methods work with any time series generation method, including NNLB and SBB. They afford exploratory investigation of the consequences of systematic displacements of observed time series. Thus, they might be used to explore generally increasing load on the grid or solar or wind periods of drought. 

We illustrated the application of synthetic time series for model analysis with a mini-study 
of the PJM service area.   The mini-study explored the prevalence of 24 hour shortage periods under multiple scenarios (\S\ref{sec_pjm_case}). An adequately comprehensive demonstration of the synthetic time series methods introduced in this paper and their uses for model analysis would require multiple papers. Our purpose with the PJM mini-study is to illustrate concretely the methods in action and to spark the imagination of the reader. 

The key observation is \emph{in brief} this. It is standard (even inevitable) practice in many areas of system design   to configure a system to perform adequately under a range of conditions. Buildings and civic infrastructure are discussed in this regard in \S\ref{sec_related_applications}. One undertakes model analysis by first specifying a measure of performance (MoP) to be investigated, such as maintaining acceptable temperature (e.g., for a building) or maintaining usage under adverse weather conditions (e.g., for a bridge or roadway). Then one investigates the model's performance with respect to the MoP under varying parameter values. If the parameters are scalars, this is fairly straightforward in principle. In the case of buildings, for example, we might look at the historical distribution of maximum yearly temperatures and assess the performance of the building's cooling system under those conditions. But if the parameters are time series vectors, there is a problem, especially in the case of energy system modeling where we are interested in MoPs regarding behavior over time, such as periods of shortage or surplus. In such cases, we need to examine alternative time series, not simply alternative scalars. The PJM mini-study illustrates how the basic analysis process in the case of scalars can be  duplicated in essence with time series data. Our MoP was the number of 24 hour periods of shortage of electricity production (shortage defined as less than 90\% of demand for the period). We used the SBB method to generate 1000 series of renewable energy production values, each a year in length, and we report the distribution of the MoP under this setup.  Clearly, a complete study would be much more elaborate, but we have demonstrated the core of such studies using synthetic time series generated by the SBB method (with our without directional alterations).

\subsection{Limitations and Future Work\label{sec_limit}}

In concluding, we can identify limitations of the study, along with opportunities and needs for future research. First, we have only hinted in  \S\ref{sec_pjm_case} at the possibilities of direct application of our intended use case: exploring in decision support mode how robust various designs are to secular shocks. Thorough application studies need to be undertaken, well beyond the scope of this paper. 

Second, we implicitly assumed that solar PV and wind capacity were constant during the 2021 data year we report here.  That is, we neglect the consequences of new units coming on line and existing units being taken off line. This is surely not true; there was net expansion of both wind and PV during 2021.  Even so, the error introduced will be small and correctable if needed.

A third limitation is in our view more of a challenge. In scaling up wind and solar PV for the PJM mini-study we altered the means of the time series, as is the point of the exercise, but this did not alter the variability of the series, as measured by coefficients of variation (the appropriate measure, as the means were shifting). It is unlikely that this would happen in a realistic scenario. We would expect that variability would be reduced because of the geographic dispersion of generator farms incumbent on scaling up. They cannot all be where they are today. Whether the variability is increased or reduced is of course an empirical matter. The question for future research is how to modify the methods at hand to accommodate increases or even decreases in variability along with changes in average levels. We find it promising that bootstrap methods will again be useful on this aspect of generating synthetic time series, but the details are not yet available and would likely require a much longer study, so we leave this as a challenge problem for the research community.

A fourth limitation is that we have conducted the discussion in this paper exclusively in terms of single dimensional time series. This can and should be extended to multiple dimension cases, especially for purposes of modeling geographic (2-D) extents. Given the SBB algorithm, the extension is in principle straightforward: the symmetric blocks before and after the focal point are replaced by  symmetric neighborhoods, in any fixed number of dimensions, surrounding the focal point. This issue has begun to be 
explored fruitfully but less straightforwardly  in \cite{amonkar_k-nearest_2022}.

Finally, regarding future work, there are manifold opportunities. One important target  will be to explore how machine learning techniques may produce results that complement or augment the bootstrapping methods on display here. During the past 5--10 years there has been surge of interest  from the machine learning community in synthetic data. Endres et al. \cite{endres_synthetic_2022} provide a useful overview and comparison of about a half dozen projects and software tools, mainly focused on scalar data (but see \cite{dahmen_synsys_2019} for an early but underdeveloped foray into time series generation).  The Synthetic Data Vault (SDV) project \cite{patki_synthetic_2016} is notable among these projects in that it has developed extensions for sequential data (time series is a special case), \cite{zhang_sequential_2022}. The sequential data methods \cite{zhang_sequential_2022} represent interesting and important approaches to generating time series under the learn an accurate model problematic. We do not see, however, application of the methods to the generate ensembles of series problematic that is the focus of this paper. The methods, however, can be used to generate multiple series, as they work by extending existing series using a stochastic process informed by neural net learning on earlier elements in the series. This is promising in the present context. What is less promising is that the method appears to assume that the underlying series are stationary, which is amply false for energy-related data. Future work, well beyond the scope of this paper, could modify the existing methods, compare them with the results of the bootstrapping methods discussed here, and test their efficacy for model analysis. A recent paper on synthetic anomalies in energy time series \cite{turowski_generating_2024} raises intriguing issues that might be addressable with the methods explored in the present study. Recent research has emphasized the importance in planning out how energy grids can integrate renewable energy sources because of the uncertainty of the power source. The methods that were developed in this paper are potentially an additional tool for addressing  uncertainty in  models, as explored in  \cite{fornier_joint_2024} and \cite{haugen_representation_2023}. We sense a set of rich possibilities here for productive interplay of multiple methods.

The most meaningful future work, however, will be successful application of these methods in practice. Much will be learned in the attempt. Besides energy system modeling, building design, and civic works design (all discussed above), we posit that transportation systems design and other forms of the built environment design (e.g., pedestrian accommodation and management in urban areas, theme parks, and sports venues), with their time-varying service demands, can profit from the methods and approach of this study.

\section*{Supplemental Material Repository}

Code and data be placed in a public repository upon acceptance of the paper. Temporarily:

\url{https://www.dropbox.com/scl/fi/fzej4f2y0g7cnqjcitplu/Synthetic-Data-Generation.zip?rlkey=37ji83f6bah1nxywr4qkjqh9d&st=8ui0kjej&dl=0}



\bibliographystyle{ieeetr}
\bibliography{SyntheticData.bib}

\begin{thebibliography}{10}

\bibitem{kimbrough_business_2016}
S.~O. Kimbrough and H.~C. Lau, {\em Business {Analytics} for {Decision}
  {Making}}.
\newblock Boca Ratan, FL: CRC Press, 2016.

\bibitem{hansen_status_2019}
K.~Hansen, C.~Breyer, and H.~Lund, ``Status and perspectives on 100\% renewable
  energy systems,'' {\em Energy}, vol.~175, pp.~471 -- 480, 2019.

\bibitem{yilmaz_power--gas_2022}
H.~{\"U}. Yilmaz, S.~O. Kimbrough, C.~van Dinther, and D.~Keles,
  ``Power-to-gas: {Decarbonization} of the {European} electricity system with
  synthetic methane,'' {\em Applied Energy}, vol.~323, p.~119538, Oct. 2022.
\newblock https://www.sciencedirect.com/science/article/pii/S0306261922008546.

\bibitem{pjm_pjm_2024}
PJM, ``{PJM} {Website},'' 2024.
\newblock https://pjm.com/.

\bibitem{ashrae_handbook_2024}
{ASHRAE}, ``Handbook,'' 2024.
\newblock https://www.ashrae.org/technical-resources/ashrae-handbook.

\bibitem{nrel_tmy_2024}
{NREL}, ``{TMY} - {NSRDB},'' 2024.
\newblock https://nsrdb.nrel.gov/about/tmy.html.

\bibitem{usgs_100-year_2024}
{USGS}, ``The 100-{Year} {Flood},'' 2024.
\newblock
  https://www.usgs.gov/special-topic/water-science-school/science/100-year-flood?qt-science\_center\_objects=0\#qt-science\_center\_objects.

\bibitem{turowski_generating_2024}
M.~Turowski, B.~Heidrich, L.~Weing{\"a}rtner, L.~Springer, K.~Phipps,
  B.~Sch{\"a}fer, R.~Mikut, and V.~Hagenmeyer, ``Generating synthetic energy
  time series: {A} review,'' {\em Renewable and Sustainable Energy Reviews},
  vol.~206, p.~114842, Dec. 2024.

\bibitem{papaefthymiou_mcmc_2008}
G.~Papaefthymiou and B.~Klockl, ``{MCMC} for {Wind} {Power} {Simulation},''
  {\em IEEE Transactions on Energy Conversion}, vol.~23, no.~1, pp.~234--240,
  2008.

\bibitem{ettoumi_statistical_2003}
F.~Y. Ettoumi, H.~Sauvageot, and A.-E.-H. Adane, ``Statistical bivariate
  modelling of wind using first-order {Markov} chain and {Weibull}
  distribution,'' {\em Renewable energy}, vol.~28, no.~11, pp.~1787--1802,
  2003.
\newblock Publisher: Elsevier.

\bibitem{brokish_pitfalls_2009}
K.~Brokish and J.~Kirtley, ``Pitfalls of modeling wind power using {Markov}
  chains,'' in {\em 2009 {IEEE}/{PES} {Power} {Systems} {Conference} and
  {Exposition}}, pp.~1--6, IEEE, 2009.

\bibitem{pesch_new_2015}
T.~Pesch, S.~Schr{\"o}ders, H.-J. Allelein, and J.-F. Hake, ``A new
  {Markov}-chain-related statistical approach for modelling synthetic wind
  power time series,'' {\em New journal of physics}, vol.~17, no.~5, p.~055001,
  2015.
\newblock Publisher: IOP Publishing.

\bibitem{bertolina_exploring_2024}
R.~M. Bertolina, E.~S. Costa, M.~M. Nunes, R.~N. Silva, M.~Guimar{\~a}es, T.~F.
  Oliveira, and A.~C.~P. Brasil~Junior, ``Exploring wind energy for small
  off-grid power generation in remote areas of {Northern} {Brazil},'' {\em
  Energy Systems}, May 2024.

\bibitem{billinton_time-series_1996}
R.~Billinton, H.~Chen, and R.~Ghajar, ``Time-series models for reliability
  evaluation of power systems including wind energy,'' {\em Microelectronics
  Reliability}, vol.~36, no.~9, pp.~1253--1261, 1996.
\newblock Publisher: Elsevier.

\bibitem{chen_synthetic_2017}
J.~Chen and C.~Rabiti, ``Synthetic wind speed scenarios generation for
  probabilistic analysis of hybrid energy systems,'' {\em Energy}, vol.~120,
  pp.~507--517, 2017.
\newblock Publisher: Elsevier.

\bibitem{chen_arima-based_2009}
P.~Chen, T.~Pedersen, B.~Bak-Jensen, and Z.~Chen, ``{ARIMA}-based time series
  model of stochastic wind power generation,'' {\em IEEE transactions on power
  systems}, vol.~25, no.~2, pp.~667--676, 2009.
\newblock Publisher: IEEE.

\bibitem{usaola_synthesis_2014}
J.~Usaola, ``Synthesis of hourly wind power series using the {Moving} {Block}
  {Bootstrap} method,'' in {\em 2014 {International} {Conference} on
  {Probabilistic} {Methods} {Applied} to {Power} {Systems} ({PMAPS})},
  pp.~1--6, IEEE, 2014.

\bibitem{veall_boostrapping_1987}
M.~R. Veall, ``Boostrapping the {Probability} {Distribution} of {Peak}
  {Electricity} {Demand},'' {\em International Economic Review}, vol.~28,
  no.~1, pp.~203--212, 1987.
\newblock Publisher: [Economics Department of the University of Pennsylvania,
  Wiley, Institute of Social and Economic Research, Osaka University].

\bibitem{al-sahlawi_forecasting_1990}
M.~A. Al-Sahlawi, ``Forecasting the {Demand} for {Electricity} in {Saudi}
  {Arabia},'' {\em The Energy Journal}, vol.~11, no.~1, pp.~119--125, 1990.
\newblock Publisher: International Association for Energy Economics.

\bibitem{efron_introduction_1993}
B.~Efron and R.~J. Tibshirani, {\em An {Introduction} to the {Bootstrap}}.
\newblock New York, NY: Chapman \&{\textbackslash} Hall, 1993.

\bibitem{lahiri_bootstrap_2006}
S.~N. Lahiri, ``Bootstrap methods: {A} review,'' {\em Frontiers in statistics},
  pp.~231--255, 2006.
\newblock Publisher: World Scientific.

\bibitem{hardle_bootstrap_2003}
W.~H{\"a}rdle, J.~Horowitz, and J.-P. Kreiss, ``Bootstrap {Methods} for {Time}
  {Series},'' {\em International Statistical Review / Revue Internationale de
  Statistique}, vol.~71, no.~2, pp.~435--459, 2003.
\newblock Publisher: [Wiley, International Statistical Institute (ISI)].

\bibitem{hongyi_li_bootstrapping_1996}
G.~S. Hongyi~Li and {Maddala}, ``Bootstrapping time series models,'' {\em
  Econometric reviews}, vol.~15, no.~2, pp.~115--158, 1996.
\newblock Publisher: Taylor \& Francis.

\bibitem{kreiss_bootstrap_2012}
J.-P. Kreiss and S.~N. Lahiri, ``Bootstrap {Methods} for {Time} {Series},'' in
  {\em Handbook of {Statistics}}, vol.~30, pp.~3--26, Elsevier, 2012.

\bibitem{lall_nearest_1996}
U.~Lall and A.~Sharma, ``A {Nearest} {Neighbor} {Bootstrap} for {Resampling}
  {Hydrologic} {Time},'' {\em Water Resources Research}, vol.~32, pp.~679--693,
  Mar. 1996.

\bibitem{kimbrough_symmetric_2021}
S.~O. Kimbrough and H.~{\"U}. Yilmaz, ``A {Symmetric} {Block} {Resampling}
  {Method} to {Generate} {Energy} {Time} {Series} {Data},'' in {\em 2021 22nd
  {IEEE} {International} {Conference} on {Industrial} {Technology} ({ICIT})},
  vol.~1, pp.~546--551, 2021.
\newblock IEEE Xplore.

\bibitem{amonkar_k-nearest_2022}
Y.~Amonkar, D.~J. Farnham, and U.~Lall, ``A k-nearest neighbor space-time
  simulator with applications to large-scale wind and solar power modeling,''
  {\em Patterns}, vol.~3, p.~100454, Mar. 2022.

\bibitem{ray_confronting_2015}
P.~A. Ray and C.~M. Brown, ``Confronting climate uncertainty in water resources
  planning and project design: the decision tree framework,'' tech. rep., World
  Bank, 2015.
\newblock
  https://documents.worldbank.org/en/publication/documents-reports/documentdetail/516801467986326382/Confronting-climate-uncertainty-in-water-resources-planning-and-project-design-the-decision-tree-framework;
  DecisionTreeFramework.pdf.

\bibitem{turowski_modeling_2022}
M.~Turowski, M.~Weber, O.~Neumann, B.~Heidrich, K.~Phipps, H.~K. {\c C}akmak,
  R.~Mikut, and V.~Hagenmeyer, ``Modeling and generating synthetic anomalies
  for energy and power time series,'' in {\em Proceedings of the {Thirteenth}
  {ACM} {International} {Conference} on {Future} {Energy} {Systems}},
  e-{Energy} '22, (New York, NY, USA), pp.~471--484, Association for Computing
  Machinery, June 2022.

\bibitem{endres_synthetic_2022}
M.~Endres, A.~Mannarapotta~Venugopal, and T.~S. Tran, ``Synthetic {Data}
  {Generation}: {A} {Comparative} {Study},'' in {\em Proceedings of the 26th
  {International} {Database} {Engineered} {Applications} {Symposium}}, {IDEAS}
  '22, (New York, NY, USA), pp.~94--102, Association for Computing Machinery,
  Sept. 2022.

\bibitem{dahmen_synsys_2019}
J.~Dahmen and D.~Cook, ``{SynSys}: {A} {Synthetic} {Data} {Generation} {System}
  for {Healthcare} {Applications},'' {\em Sensors}, vol.~19, p.~1181, Jan.
  2019.

\bibitem{patki_synthetic_2016}
N.~Patki, R.~Wedge, and K.~Veeramachaneni, ``The {Synthetic} {Data} {Vault},''
  in {\em 2016 {IEEE} {International} {Conference} on {Data} {Science} and
  {Advanced} {Analytics} ({DSAA})}, pp.~399--410, 2016.

\bibitem{zhang_sequential_2022}
K.~Zhang, K.~Veeramachaneni, and N.~Patki, ``Sequential {Models} in the
  {Synthetic} {Data} {Vault}.'' https://arxiv.org/pdf/2207.14406.pdf, June
  2022.

\bibitem{fornier_joint_2024}
Z.~Fornier, D.~Grosso, and V.~Leclere, ``Joint production and energy supply
  planning of an industrial microgrid,'' {\em Energy Systems}, Jan. 2024.

\bibitem{haugen_representation_2023}
M.~Haugen, H.~Farahmand, S.~Jaehnert, and S.-E. Fleten, ``Representation of
  uncertainty in market models for operational planning and forecasting in
  renewable power systems: a review,'' {\em Energy Systems}, July 2023.

\end{thebibliography}

\vfill
\noindent File: main\_energy\_systems.tex 
\end{document}